\documentclass[letterpaper,usenatbib]{mn2e}
 
\usepackage{inputenc}
\usepackage{amssymb}
\usepackage{float}
\usepackage{bm}
\usepackage{epsfig}
\usepackage{soul}
\usepackage{subcaption}
\usepackage{multirow,balance}
\usepackage{amsmath}
\usepackage{amssymb}
\usepackage{aas_macros}
\usepackage{graphicx}
\usepackage{natbib}
\usepackage{color}
\usepackage{listings}
\usepackage{hyperref}   
\usepackage{cases}
\usepackage{tikz}
\usepackage{xcolor}
\usetikzlibrary{decorations.pathreplacing}
\usepackage{times}
\usepackage{balance}
\hypersetup{dvips, colorlinks=true, linkcolor=black, citecolor=black, filecolor=black, urlcolor=black}

\newcommand{\ri}{\mathrm{i}}
\newcommand{\re}{\mathrm{e}}
\newcommand{\rd}{\mathrm{d}}
\newcommand{\rp}{\mathrm{p}}
\newcommand{\rg}{\mathrm{g}}
\newcommand{\rG}{\mathrm{G}}
\newcommand{\rb}{\mathrm{b}}

\newcommand{\nbpWKBPlane}{100}
\newcommand{\nbpPsiZ}{50}
\newcommand{\nbpFLambda}{104}

\begin{document}

\setcounter{tocdepth}{3}

\title[Resonant thickening of self-gravitating discs]
{ 
Resonant thickening of self-gravitating discs:
\\ imposed or self-induced orbital diffusion in the tightly wound limit
}

\author[J.-B. Fouvry, C. Pichon, P.-H Chavanis \& L. Monk]{Jean-Baptiste Fouvry$^{1,2}$\thanks{Hubble Fellow.}, Christophe Pichon$^{1,3,4}$, Pierre-Henri Chavanis$^{5}$, and Laura Monk$^{1}$
\vspace*{6pt}\\
\noindent
$^{1}$ Institut d'Astrophysique de Paris, and UPMC Univ. Paris 06, (UMR7095), 98 bis Boulevard Arago, 75014, Paris, France\\
$^{2}$ Institute for Advanced Study, Einstein Drive, Princeton, NJ 08540, USA\\
$^{3}$ Institute of Astronomy, University of Cambridge, Madingley Road, Cambridge, CB3 0HA, United Kingdom\\
$^{4}$ Korea Institute for Advanced Study (KIAS) 85 Hoegiro, Dongdaemun-gu, Seoul, 02455, Republic of Korea\\
$^{5}$ Laboratoire de Physique Th\'eorique (IRSAMC), CNRS and UPS, Univ. de Toulouse, F-31062 Toulouse, France
}

\date{\today}
\label{firstpage}
\pagerange{\pageref{firstpage}--\pageref{lastpage}}

\maketitle

\begin{abstract}
The secular thickening of a self-gravitating stellar galactic disc is investigated using the dressed collisionless Fokker-Planck equation and the inhomogeneous multi-component Balescu-Lenard equation. The thick WKB limits for the diffusion fluxes are found using the epicyclic approximation, while assuming that only radially tightly wound transient spirals are sustained by the disc. This yields simple quadratures for the drift and diffusion coefficients, providing a clear understanding of the positions of maximum vertical orbital diffusion within the disc, induced by fluctuations either external or due to the finite number of particles. These thick limits also offer a consistent derivation of a thick disc Toomre parameter, which is shown to be exponentially boosted by the ratio of the vertical to radial scale heights.

Dressed potential fluctuations within the disc statistically induce a vertical bending of a subset of resonant orbits, triggering the corresponding increase in vertical velocity dispersion. When applied to a tepid stable tapered disc perturbed by shot noise, these two frameworks reproduce qualitatively the formation of ridges of resonant orbits towards larger vertical actions, as found in direct numerical simulations, but over-estimates the timescale involved in their appearance. Swing amplification is likely needed to resolve this discrepancy, as demonstrated in the case of razor-thin discs. Other sources of thickening are also investigated, such as fading sequences of slowing bars, or the joint evolution of a population of giant molecular clouds within the disc.
\end{abstract}


\section{Introduction}
\label{sec:introduction}

The problem of explaining the origin of thick discs in our Galaxy and in external galaxies has been with us for some time~\cite[e.g.,][]{GilmoreReid1983,Freeman1987}. Its interest has been revived recently in the context of galactic archeology as probed by the upcoming result of the GAIA mission. 
Star formation typically occurs on circular orbits within such disc, so that young stars should form a very thin disc~\citep{Wielen1977}. On the other hand, chemo-kinematic observations of old stars within our Milky Way~\citep{GilmoreReid1983,Juric2008,Izevic2008,Bovy2012},
 or in other galactic discs~\citep{YoachimDalcanton2006} have shown that thick components are very common.
The formation of thickened stellar discs yet remains a puzzle for galactic formation theory.
Various physical processes, either internal or external, have been proposed as possible drivers of this observed thickening, but their respective impacts are still unclear.
Violent major events could be at the origin of the extended distribution of stars in disc galaxies. These could be due to the accretion of galaxy satellites~\citep{Meza2005,AbadiNavarro2003}, major mergers of gas-rich systems~\citep{Brook2004}, or gravitational instabilities in gas-rich turbulent clumpy discs~\citep{Noguchi1998,Bournaud2009}.
While mergers do have a strong impact on galactic structures, these extreme events may not be required to create a thickened stellar disc, which could originate from the continuous heating of a preexisting thin disc.
Numerous smooth evolution mechanisms have then been investigated.
Galactic discs could be thickened as a result of galactic infall of cosmic origin leading to multiple minor mergers~\citep{Toth1992,Quinn1993,Villalobos2008,DiMatteo2011}, and evidence for such events has been found in the phase-space structure of the Milky Way~\citep[e.g.,][]{Purcell2011}.
Spiral density waves~\citep{SellwoodCarlberg1984,MinchevQuillen2006,Monari2016} are also possible candidates for increasing the velocity dispersion within the disc, which can then be converted into vertical motion through the deflection from giant molecular clouds (GMCs)~\citep{SpitzerSchwarzschild1953,HanninenFlynn2002}.
Radial migration~\citep{LyndenBell1972,SellwoodBinney2002}, the change of angular momentum of a star with no increase in its radial energy, is also believed to be an important mechanism for the secular evolution of galactic discs. This migration may be induced by spiral-bar coupling~\citep{MinchevFamaey2010}, transient spiral structures~\citep{BarbanisWoltjer1967,CarlbergSellwood1985,SellwoodBinney2002,SolwaySellwood2012}, or even perturbations by minor mergers~\citep{Quillen2009,Bird2012}.
\cite{SchonrichBinney2009a,SchonrichBinney2009b} used an analytical model of radial migration to investigate in detail the impact of radial migration on the vertical heating of the disc and recovered the main characteristics of the Milky Way thick and thin discs.
Recent ${N-}$body simulations also focused on the role played by radial migration~\citep[e.g.,][]{Haywood2008,LoebmanRoskar2011,Minchev2014}, but the efficiency of this thickening mechanism was recently shown as limited~\citep{MinchevFamaey2012}.
Finally, large numerical simulations are now in a position to investigate such processes consistently in a cosmological context~\citep{Minchev2015,GrandSpringel2016}, and the developments of these global approaches are expected to offer new clues on the interplay between these various competing thickening mechanisms. All these investigations can be broadly categorised as relying on either an internal (nature), or external (nurture) origin to trigger the orbital restructuration of the disc. Defining the frameworks in which to address either processes is the purpose of the present paper.

The seminal paper of~\cite{BinneyLacey1988} addressed the origin of the thick disc using an orbit-averaged Fokker-Planck formalism in angle-action. Yet, it fell short of accounting for the self-gravity of the disc, which was shown recently~\citep{FouvryPichonMagorrianChavanis2015} to play a very significant role in boosting the amplitude of the diffusion coefficient for razor-thin discs via successive sequences of spiral waves. It is therefore of interest to try and estimate which orbits are involved in that regime, whether the boost remains significant for thickened discs and if the corresponding secular orbital distortion can account for the observed vertical heating. 
 
Indeed, in such discs made of a finite number of stars and giant molecular clouds (GMCs), fluctuations in the potential alone induced by discrete (possibly distant) encounters may be strongly amplified. Resonances will tend to confine and localise the dissipation of these fluctuations, which can then lead to a spontaneous thickening of discs. Quantifying the relative importance of this intrinsically driven evolution w.r.t. that driven by the environment is timely, as the cosmological environment of self-gravitating discs is now firmly established in the context of the $\Lambda$CDM paradigm. While ${N-}$body simulations  offer a flexible and powerful framework in which to Monte-Carlo these processes \citep[e.g.,][]{Minchev2013}, the effect of the disc's intrinsic fluctuations and susceptibility can also be addressed in the context of kinetic theory, which captures  discrete resonant interactions over secular timescales.
 
The kinetic theory of stellar systems was initiated by~\cite{Jeans1929} and 
\cite{Chandrasekhar1942} for elliptical galaxies and globular clusters. In these works, spatial inhomogeneity was taken into account in the advection term (Vlasov) but the collisional term was calculated by making a local approximation as if the system were homogeneous. Furthermore, collective effects were neglected. In plasma physics, where the system is homogeneous,~\cite{Balescu1960} and~\cite{Lenard1960} developed a rigorous kinetic theory, taking collective effects into account, and obtained a kinetic equation which accounts for the system's susceptibility and for Debye shielding. More recently, in the context of stellar dynamics,~\cite{Heyvaerts2010,Chavanis2012} derived the inhomogeneous Balescu-Lenard equation, a kinetic equation written in angle-action variables that describes spatially inhomogeneous multi-periodic systems and takes collective effects into account. This Balescu-Lenard equation accounts for the self-driven orbital diffusion of a self-gravitating system induced by its intrinsic shot noise due to discreteness and the corresponding long-range correlations. The inhomogeneous Balescu-Lenard equation has recently been implemented by~\cite{FouvryPichonChavanis2015,FouvryPichonMagorrianChavanis2015} in ${ 2D }$ for razor-thin discs.
 
In this paper we intend to account for the system's self-gravity while writing down two diffusion equations in the context of tepid galactic discs of finite thickness. The first one considers the system as collisionless and focuses on a forcing induced by external perturbations, while the second one assumes the system to be isolated and collisional and focuses on the role played by the system's intrinsic discreteness.
Both diffusion processes should be considered since it is not known a priori which is most effective at restructuring the orbital distribution of galaxies, i.e. what are the respective roles of nurture (cosmic environment) vs. nature (system's internal properties) in the secular establishment of the observed properties of these systems.
Following~\cite{FouvryPichonPrunet2015} (hereafter FPP15) and~\cite{FouvryPichonChavanis2015} (hereafter FPC15), and relying on the epicyclic approximation, we will for simplicity seek the thick WKB limit of these two diffusion equations while assuming that only radially tightly wound transient spirals are sustained by the disc. We will aim for simple double quadratures for the associated diffusion fluxes, in order to provide a straightforward understanding of the positions of maximum diffusion within the disc. In this cool regime, the self-gravity of the disc can be tracked down radially via a local WKB-like response, while the vertical degree of motion can be partially decoupled. This, in turn, allows us to simplify the a priori ${3D}$ formalism to an effective (non-degenerate) ${1D}$ formalism.
Illustrations of these formalisms will be presented in the context of a shot noise perturbed tepid Toomre-stable tapered thick disc.
We will underline how they recover the formation of vertical ridges of resonant orbits towards larger vertical actions, hence larger heights and vertical velocity dispersions.
Such diffusion processes may capture either the environmentally driven thickening of galactic discs on secular timescales, or the thickening induced by the system's intrinsic graininess.
Our qualitative predictions will be compared to the numerical experiments from~\cite{SolwaySellwood2012} and the intrinsic limitations of the WKB assumptions will be discussed in details.

The paper is organised as follows.
Section~\ref{sec:seculardiffusionequation} briefly presents two diffusion equations: the secular collisionless diffusion equation and the collisional Balescu-Lenard equation.
Section~\ref{sec:WKBlimit} focuses on thick axisymmetric galactic discs within the WKB approximation.
Section~\ref{sec:application} applies these formalisms to the formation of vertical resonant ridges first in an isolated thick self-gravitating Mestel disc driven by its own discreteness, and then in such a disc subject to recurrent decelerating bars or to the joint secular evolution of a population of GMCs.
Section~\ref{sec:conclusion} wraps up.

\section{Secular diffusion}
\label{sec:seculardiffusionequation}

There are two main channels through which a secular evolution of a stable quasi-stationary self-gravitating system can be induced. The system may either be perturbed by its stochastic environment or by its own intrinsic graininess. The first scenario is captured by the secular collisionless diffusion equation and is presented in section~\ref{sec:introFP}, while the second is captured by the inhomogeneous Balescu-Lenard equation and is presented in section~\ref{sec:introBL}.
Such a dichotomy is essential to capture the respective roles of nature and nurture in the secular evolution of these systems.
We will now briefly describe these two diffusion formalisms.

\subsection{Secular collisionless forcing}
\label{sec:introFP}

Let us consider a collisionless self-gravitating system. Let us further assume that the gravitational background $\psi_{0}$, associated with the Hamiltonian $H_{0}$, is stationary and integrable\footnote{We note that in the thickened geometry, integrability is not warranted by symmetry anymore, so that we are effectively assuming that the disc is thin enough so that it can be approximated to be integrable; see~\cite{Weinberg2015I} for a discussion.}, so that one may always remap the physical coordinates ${ (\bm{x} , \bm{v}) }$ to the angle-action coordinates ${ (\bm{\theta} , \bm{J}) }$~\citep{Goldstein1950,Born1960,BinneyTremaine2008}. Along the unperturbed motions, the actions $\bm{J}$ are conserved, while the angles $\bm{\theta}$ are ${ 2\pi-}$periodic. One can then introduce the intrinsic frequencies of the system $\bm{\Omega}$ as
\begin{equation}
\bm{\Omega} = \dot{\bm{\theta}} = \frac{\partial H_{0}}{\partial \bm{J}} \, .
\label{definition_Omega_general}
\end{equation}
Since the collisionless system is assumed to be in a quasi-stationary state, it can be described by a distribution function (DF) ${ F (\bm{J} , t) }$, which depends only on the actions, with the normalisation convention ${ \! \int \!\! \rd \bm{x} \rd \bm{v} F \!=\! M_{\rm tot} }$, where $M_{\rm tot}$ is the total active mass of the system. When perturbed by an external stochastic source of perturbations, such a system may diffuse on secular timescales~\citep{Weinberg2001a,PichonAubert2006,FouvryPichonPrunet2015} via an anisotropic diffusion equation of the form
\begin{equation}
\frac{\partial F}{\partial t} =  \frac{\partial }{\partial \bm{J}} \!\cdot\! \bigg[ \sum_{\bm{m}}  \bm{m} \, D_{\bm{m}} (\bm{J}) \, \bm{m} \!\cdot\! \frac{\partial F}{\partial \bm{J}} \bigg] \, ,
\label{diffusion_equation}
\end{equation}
where the index ${ \bm{m} \!\in\! \mathbb{Z}^{d} }$ corresponds to the Fourier coefficients associated with the Fourier transform w.r.t. the angles $\bm{\theta}$. See FPP15 for a derivation of the secular collisionless diffusion equation~\eqref{diffusion_equation}. Here $d$ is the dimension of the physical space, i.e. ${ d \!=\! 3 }$ for a thick disc. In equation~\eqref{diffusion_equation}, the diffusion coefficients ${ D_{\bm{m}} (\bm{J}) }$ are given by
\begin{equation}
D_{\bm{m}} (\bm{J}) = \frac{1}{2} \! \sum_{p , q} \! \psi_{\bm{m}}^{(p)}  \psi_{\bm{m}}^{(q) *} \, \bigg[ [ \mathbf{I} \!-\! \widehat{\mathbf{M}} ]^{-1} \!\!\cdot\! \widehat{\mathbf{C}} \!\cdot\! [ \mathbf{I} \!-\! \widehat{\mathbf{M}} ]^{-1}  \bigg]_{pq} \, .
\label{Dm_general}
\end{equation}
In equation~\eqref{Dm_general}, the response matrix $\widehat{\mathbf{M}}$ and the cross-power spectra of the external perturbations $\widehat{\mathbf{C}}$ are functions of $\omega$ which should be evaluated at the resonant frequency ${ \bm{m} \!\cdot\! \bm{\Omega} }$. Here $\mathbf{I}$ stands for the identity matrix. Equation~\eqref{Dm_general} for the diffusion coefficients involves potential basis elements $\psi^{(p)}$, which are introduced following Kalnajs matrix method~\citep{Kalnajs1976II}. Indeed, to solve the non-local Poisson's equation, one introduces a biorthonormal basis of potentials and densities ${ \psi^{(p)} (\bm{x}) }$ and ${ \rho^{(p)} (\bm{x}) }$ such that
\begin{equation}
\Delta \psi^{(p)} = 4 \pi G \rho^{(p)} \;\;\; ; \;\;\; \int \!\! \mathrm{d} \bm{x} \, [ \psi^{(p)} (\bm{x}) ]^{*} \, \rho^{(q)} (\bm{x}) = - \delta_{p}^{q} \, .
\label{definition_basis}
\end{equation}
In order to account for the system's self-gravity, i.e. its ability to amplify perturbations, equation~\eqref{Dm_general} involves the system's response matrix $\widehat{\mathbf{M}}$, which reads
\begin{equation}
\widehat{\mathbf{M}}_{pq} (\omega) = (2 \pi)^{d} \sum_{\bm{m}} \!\! \int \!\! \mathrm{d} \bm{J} \, \frac{\bm{m} \!\cdot\! \partial F / \partial \bm{J}}{\omega \!-\! \bm{m} \!\cdot\! \bm{\Omega}} [ \psi_{\bm{m}}^{(p)} (\bm{J}) ]^{*} \psi_{\bm{m}}^{(q)} (\bm{J}) \, ,
\label{Fourier_M}
\end{equation}
where one should note the specific role played by the pole at the intrinsic frequency ${ \omega \!=\! \bm{m} \!\cdot\! \bm{\Omega} }$. In the expressions~\eqref{Dm_general} and~\eqref{Fourier_M}, ${ \psi_{\bm{m}}^{(p)} (\bm{J}) }$ corresponds to the Fourier transform in angles of the basis elements ${ \psi^{(p)} (\bm{x}) }$, defined as
\begin{equation}
\psi_{\bm{m}}^{(p)} (\bm{J}) = \frac{1}{(2 \pi)^{d}} \!\! \int \!\! \mathrm{d} \bm{\theta} \, \psi^{(p)} (\bm{x} (\bm{\theta} , \bm{J})) \, \re^{-\ri \bm{m} \cdot \bm{\theta}} \, .
\label{definition_psi_FT}
\end{equation}
It then finally remains to specify how one should compute $\widehat{\mathbf{C}}$, the autocorrelation of the external perturbations. We assume that the system is stochastically perturbed by an external potential ${ \psi^{\rm e} (\bm{x} , t) }$. Using the basis elements $\psi^{(p)}$, it may be decomposed as
\begin{equation}
\psi^{\rm e} (\bm{x} , t) = \sum_{p} b_{p}(t) \, \psi^{(p)} (\bm{x}) \, .
\label{decomposition_psi_ext}
\end{equation}
If we assume that the ensemble average ${ (\langle \, \cdot \, \rangle) }$ of these perturbations is stationary in time, one can define their temporal autocorrelation matrix $\mathbf{C}$ as
\begin{equation}
\mathbf{C}_{pq} (t_{1} \!-\! t_{2}) = \big< b_{p} (t_{1}) \, b_{q}^{*} (t_{2}) \big> \, .
\label{definition_C_time}
\end{equation}
In frequency space, using the convention ${ \widehat{f} (\omega) \!=\! \! \int \! \mathrm{d} t \, f(t) \, \re^{\ri \omega t} }$, it can equivalently be written as
\begin{equation}
\big< \widehat{b}_{p} (\omega) \, \widehat{b}_{q}^{*} (\omega ') \big> = 2 \pi \, \delta_{\rm D} (\omega \!-\! \omega ') \, \widehat{\mathbf{C}}_{pq} (\omega) \, ,
\label{definition_C_omega}
\end{equation}
where one recovers the autocorrelation matrix $\widehat{\mathbf{C}}$ which enters in the expression~\eqref{Dm_general} of the diffusion coefficients. To emphasise the conservation of the total number of stars, one may finally introduce the total collisionless flux density $\bm{\mathcal{F}}_{\rm tot}$ as
\begin{equation}
\bm{\mathcal{F}}_{\rm tot} = \sum_{\bm{m}} \bm{m} \, D_{\bm{m}} (\bm{J}) \, \bm{m} \!\cdot\! \frac{\partial F}{\partial \bm{J}} \, ,
\label{definition_Ftot}
\end{equation}
so that equation~\eqref{diffusion_equation} takes the shortened form\footnote{With this convention, ${- \bm{\mathcal{F}}_{\rm tot} }$ corresponds to the direction of diffusion of individual particles in action space.}
\begin{equation}
\frac{\partial F}{\partial t} = \text{div} (\bm{\mathcal{F}}_{\rm tot}) \, .
\label{diffusion_equation_Ftot}
\end{equation}
While formally simple, equations~\eqref{diffusion_equation} and~\eqref{Dm_general} capture a wealth of non-linear physical processes: the secular radial and vertical distortion of resonant orbits induced by a spectrum of dressed perturbations (i.e. accounting for gravitational polarisation) corresponding to uncorrelated swing amplified spiral waves. 
We will show in section~\ref{sec:WKBlimit} how one may use this collisionless diffusion formalism  to describe the induced secular evolution of axisymmetric thick discs.

\subsection{The inhomogeneous Balescu-Lenard equation}
\label{sec:introBL}

If the system is now assumed to be isolated but discrete (i.e. made of a finite number of particles), its long-term evolution is described by the inhomogeneous Balescu-Lenard equation~\citep{Heyvaerts2010,Chavanis2012}. This equation aims at describing the evolution on secular timescales of this isolated DF under the effects of discrete resonant ``encounters'' between stars (finite${-N}$ effects). It reads, using the shortened notation ${ \bm{\Omega}_{i} \!=\! \bm{\Omega} (\bm{J}_{i}) }$,
\begin{align}
\frac{\partial F}{\partial t} \!=\! & \, \pi (2 \pi)^{d} \mu \frac{\partial}{\partial \bm{J}_{1}} \!\cdot\! \!\bigg[ \!\! \sum_{\bm{m}_{1} , \bm{m}_{2}} \!\!\!\! \bm{m}_{1} \!\!\! \int \!\!\! \rd \bm{J}_{2} \, \frac{\delta_{\rm D} (\bm{m}_{1} \!\cdot\! \bm{\Omega}_{1} \!-\! \bm{m}_{2} \!\cdot\! \bm{\Omega}_{2})}{|\mathcal{D}_{\bm{m}_{1} , \bm{m}_{2}} (\bm{J}_{1} , \bm{J}_{2} , \bm{m}_{1} \!\cdot\! \bm{\Omega}_{1}) |^{2}}  \nonumber
\\
& \, \times \, \bigg(\! \bm{m}_{1} \!\cdot\! \frac{\partial }{\partial \bm{J}_{1}} \!-\! \bm{m}_{2} \!\cdot\! \frac{\partial }{\partial \bm{J}_{2}} \!\bigg) \, F (\bm{J}_{1} , t) \, F(\bm{J}_{2} , t) \bigg] \, ,
\label{definition_BL}
\end{align}
where ${ \mu \!=\! M_{\rm tot} / N }$ is the mass of the individual particles, ${ 1 / \mathcal{D}_{\bm{m}_{1} , \bm{m}_{2}} (\bm{J}_{1} , \bm{J}_{2} , \omega) }$ are the dressed susceptibility coefficients which quantify the polarisation cloud around each particle which triggers sequences of transient swing amplified spirals~\citep{JulianToomre1966,Toomre1981}, which in the secular timeframe are assumed instantaneous. See FPC15 for a brief derivation of the Balescu-Lenard equation. When collective effects are neglected, equation~\eqref{definition_BL} becomes the inhomogeneous Landau equation~\citep{PolyachenkoShukman1982,Chavanis2013}, see Appendix B in FPC15. The r.h.s. of equation~\eqref{definition_BL} is written as the divergence of a flux, so as to ensure the conservation of the number of stars. One should also note that this r.h.s. involves a resonance condition through the Dirac delta ${ \delta_{\rm D} (\bm{m}_{1} \!\cdot\! \bm{\Omega}_{1} \!-\! \bm{m}_{2} \!\cdot\! \bm{\Omega}_{2}) }$, where ${ \bm{m}_{1} , \bm{m}_{2} \!\in\! \mathbb{Z}^{d} }$ are integer vectors. This condition is the driver of the collisional evolution. Notice also the antisymmetric operator, ${ \bm{m}_{1}^{\rm s} \!\cdot\! \partial / \partial \bm{J}_{1}^{\rm s} \!-\! \bm{m}_{2}^{\rm s} \!\cdot\! \partial / \partial \bm{J}_{2}^{\rm s} }$ in equation~\eqref{definition_BL}, which ``weighs'' the relative number of pairwise resonant orbits caught in this resonant configuration. 
Relying on Kalnajs matrix method, the dressed susceptibility coefficients appearing in equation~\eqref{definition_BL} are given by
\begin{equation}
\frac{1}{\mathcal{D}_{\bm{m}_{1} , \bm{m}_{2}} (\bm{J}_{1} , \bm{J}_{2} , \omega)} \!=\! \sum_{p , q}  \psi_{\bm{m}_{1}}^{(p)} \!(\bm{J}_{1}) \, [ \mathbf{I} \!-\! \widehat{\mathbf{M}} (\omega) ]^{-1}_{p q} \, [ \psi_{\bm{m}_{2}}^{(q)} \!(\bm{J}_{2}) ]^{*} \, ,
\label{definition_1/D}
\end{equation}
where the system's response matrix $\widehat{\mathbf{M}}$ was introduced in equation~\eqref{Fourier_M}. Finally, one may also benefit from rewriting the Balescu-Lenard equation~\eqref{definition_BL} under the form of an anisotropic diffusion equation, by introducing the associated drift and diffusion coefficients. Indeed, equation~\eqref{definition_BL} may be put under the form
\begin{equation}
\frac{\partial F}{\partial t} \!=\! \sum_{\bm{m}_{1}} \frac{\partial }{\partial \bm{J}_{1}} \!\cdot\! \left[\!  \bm{m}_{1} \bigg(\!\! A_{\bm{m}_{1}} ( \bm{J}_{1} ) F ( \bm{J}_{1} ) \!+\!  D_{\bm{m}_{1}}  ( \bm{J}_{1} ) \bm{m}_{1}  \!\cdot\!  \frac{\partial F}{\partial \bm{J}_{1}}  \!\bigg)  \!\right] ,
\label{initial_BL_rewrite}
\end{equation}
where ${ A_{\bm{m}_{1}} (\bm{J}_{1}) }$ and ${ D_{\bm{m}_{1}} (\bm{J}_{1}) }$ are respectively the collisional drift and diffusion coefficients associated with a given resonance $\bm{m}_{1}$. To simplify the notations, we did not write their secular dependence with $F$. The drift coefficients ${ A_{\bm{m}_{1}} (\bm{J}_{1}) }$ are given by
\begin{equation}
A_{\bm{m}_{1}} \!(\!\bm{J}_{1}\!) \! = \! - \pi (2 \pi)^{d} \! \mu \!\!  \sum_{\bm{m}_{2}}  \!\!  \int  \!\!\!  \rd \bm{J}_{2}  \frac{\delta_{\rm D} (\bm{m}_{1}  \!\cdot\!  \bm{\Omega}_{1}  \!-\!  \bm{m}_{2}  \!\cdot\!  \bm{\Omega}_{2}\!)}{|\mathcal{D}_{\bm{m}_{1} , \bm{m}_{2}} \!(\!\bm{J}_{1} , \!\bm{J}_{2}, \!\bm{m}_{1}  \!\!\cdot\!  \bm{\Omega}_{1} \!) |^{2}} \bm{m}_{2}  \!\cdot\!  \frac{\partial F}{\partial \bm{J}_{2}} \, ,
\label{initial_drift}
\end{equation}
while the diffusion coefficients ${ D_{\bm{m}_{1}} (\bm{J}_{1}) }$ are given by
\begin{equation}
D_{\bm{m}_{1}} \!(\!\bm{J}_{1}\!) \! =\! \pi (2 \pi)^{d} \mu \!\! \sum_{\bm{m}_{2}} \!\! \int \!\! \rd \bm{J}_{2} \frac{\delta_{\rm D} (\bm{m}_{1} \!\cdot\! \bm{\Omega}_{1} \!-\! \bm{m}_{2} \!\cdot\! \bm{\Omega}_{2}\!)}{|\mathcal{D}_{\bm{m}_{1} , \bm{m}_{2}} ( \bm{J}_{1} , \bm{J}_{2},\bm{m}_{1} \!\!\cdot\! \bm{\Omega}_{1}) |^{2}} F (\bm{J}_{2}) \, .
\label{initial_diff}
\end{equation}
Finally, let us introduce the total collisional diffusion flux $\bm{\mathcal{F}}_{\rm tot}$ as
\begin{equation}
\bm{\mathcal{F}}_{\rm tot} =  \sum_{\bm{m}} \bm{m} \left( A_{\bm{m}} (\bm{J}) \, F (\bm{J}) \!+\! D_{\bm{m}} (\bm{J}) \, \bm{m} \!\cdot\! \frac{\partial F}{\partial \bm{J}} \right) \, ,
\label{definition_F_tot}
\end{equation}
so as to rewrite the Balescu-Lenard equation~\eqref{definition_BL} and~\eqref{initial_BL_rewrite} as
\begin{equation}
\frac{\partial F}{\partial t} = \text{div} \left( \bm{\mathcal{F}}_{\rm tot} \right) \, .
\label{BL_div_Ftot}
\end{equation}
We will now illustrate how the two previous diffusion formalisms may be used in the context of axisymmetric thick discs.

\section{Thick WKB limit response}
\label{sec:WKBlimit}

In order to compute the collisionless and collisional diffusion fluxes from equations~\eqref{definition_Ftot} and~\eqref{definition_F_tot}, two main difficulties have to be overcome. The first one is to explicitly determine the mapping from the physical phase-space coordinates ${ (\bm{x} , \bm{v}) }$ to the angle-action ones ${ (\bm{\theta} , \bm{J}) }$ for a thick axisymmetric disc. If one assumes the disc to be sufficiently tepid, i.e. assumes that the stars orbits are close from circular orbits, one can rely on the epicyclic approximation to obtain such a mapping, as described in section~\ref{sec:thickepicyclic}. The second difficulty arises from the computation of the response matrix from equation~\eqref{Fourier_M}, which requires the introduction of a biorthogonal basis of potentials and densities. In order to ease the subsequent inversion of ${ \mathbf{I} \!-\! \widehat{\mathbf{M}} }$, one may follow the WKB approximation~\citep{Liouville1837,Toomre1964,Kalnajs1965,LinShu1966,Palmer1989,FouvryPichonPrunet2015}, which amounts to considering only the diffusion of the system sustained by radially tightly wound spirals. Poisson's equation is then transformed into a local equation, which leads to a diagonal response matrix. Such an application of the WKB formalism in the context of secular dynamics was successfully implemented in the context of razor-thin tepid galactic discs~\citep{FouvryPichon2015,FouvryBinneyPichon2015,FouvryPichonChavanis2015}.
While it failed short in predicting the exact amplitude of the response of the disc, as shown in~\cite{FouvryPichonMagorrianChavanis2015} where a full treatment was presented, it captured in that context the physical process of the resonant diffusion, and in particular the loci of the orbital response.
It is therefore interesting to investigate if this formalism can also capture the formation of resonants ridges in the vertical direction, beyond the radial diffusion.
The generalisation of the WKB formalism to thick discs will be detailed in section~\ref{sec:thickbasiselements}.

\subsection{Epicyclic approximation}
\label{sec:thickepicyclic}

If a thick disc is sufficiently cold, i.e. if the radial and vertical excursions of the star are small, one may use the epicyclic approximation to build up a mapping ${ (\bm{x} , \bm{v}) \!\mapsto\! (\bm{\theta} , \bm{J}) }$, as we now detail. 
While it is well known that the vertical motion tends to be anharmonic, we will neglect such complication in the framework of this paper.
We introduce the cylindrical coordinates ${ (R, \phi , z) }$ to describe our thick axisymmetric disc, along with their associated momenta ${ (p_{R} , p_{\phi} , p_{z}) }$, and we assume that the axisymmetric potential ${ \psi_{0} (R , z) }$ is symmetric w.r.t. the equatorial plane ${ z \!=\! 0 }$. The stationary Hamiltonian $H_{0}$ of the system then reads
\begin{align}
H_{0} & \, = \frac{1}{2} \bigg[ p_{R}^{2} \!+\! \frac{L_{z}^{2}}{R^{2}} \!+\! p_{z}^{2} \bigg] + \psi_{0} (R , z) \nonumber
\\
& \, = \frac{1}{2} \bigg[ p_{R}^{2} \!+\! p_{z}^{2} \bigg] \!+\! \psi_{\rm eff} (R,z) \, ,
\label{Hamiltonian_epi}
\end{align}
where we noted as $L_{z}$ the conserved angular momentum of the star and introduced the effective potential ${ \psi_{\rm eff} \!=\! \psi_{0} \!+\! L_{z}^{2}/(2 R^{2})}$. The first action of the system is then straightforwardly the angular momentum $J_{\phi}$ given by
\begin{equation}
J_{\phi} = \frac{1}{2 \pi} \!\! \oint \mathrm{d} \phi \, p_{\phi} = L_{z} = R^{2} \dot{\phi} \, .
\label{definition_Jphi}
\end{equation}
As we are considering a tepid disc, we may place ourselves in the vicinity of circular orbits. We define the guiding radius of an orbit through the implicit relation
\begin{equation}
\frac{\partial \psi_{\rm eff}}{\partial R} \bigg|_{(R_{\rg},0 )} \!\!\!\!\!\! = 0 \, .
\label{definition_Rg}
\end{equation}
Here ${ R_{\rg} (J_{\phi}) }$ corresponds therefore to the radius for which stars with an angular momentum $J_{\phi}$ are on exactly circular orbits. The mapping between $R_{\rg}$ and $J_{\phi}$ is unambiguous (up to the sign of $J_{\phi}$). In addition, this circular orbit is described at the angular frequency $\Omega_{\phi}$ given by
\begin{equation}
\Omega_{\phi}^{2} (R_{\rg}) = \frac{1}{R_{\rg}} \frac{\partial \psi_{0}}{\partial R} \bigg|_{(R_{\rg} , 0)} \, .
\label{definition_Omega}
\end{equation}
In the neighbourhood of circular orbits, one may expand the Hamiltonian from equation~\eqref{Hamiltonian_epi} as
\begin{equation}
H_{0} = \frac{1}{2} [ p_{R}^{2} \!+\! p_{z}^{2} ] \!+\! \psi_{\rm eff} (R_{\rg} , 0) \!+\! \frac{\kappa^{2}}{2} (R \!-\! R_{\rg})^{2} \!+\! \frac{\nu^{2}}{2} z^{2} \, ,
\label{DL_Hamiltonian_epi}
\end{equation}
where we used the symmetry of the potential w.r.t. the plane ${ z \!=\! 0 }$ and introduced the epicyclic frequencies $\kappa$ and $\nu$ as
\begin{equation}
\kappa^{2} (R_{\rg}) = \frac{\partial^{2} \psi_{\rm eff}}{\partial R^{2}} \bigg|_{(R_{\rg} , 0)} \;\;\; ; \;\;\; \nu^{2} (R_{\rg} ) \!=\! \frac{\partial^{2} \psi_{\rm eff}}{\partial z^{2}} \bigg|_{(R_{\rg} , 0)} \, .
\label{definitions_kappa_nu}
\end{equation}
At the level of approximation of the Taylor expansion in equation~\eqref{DL_Hamiltonian_epi}, the radial and vertical motions are decoupled, and correspond to harmonic librations. Therefore, up to initial phases, there exists two amplitudes $A_{R}$ and $A_{z}$ such that ${ R (t) \!=\! R_{\rg} \!+\! A_{R} \cos (\kappa t) }$ and ${ z(t) \!=\! A_{z} \cos(\nu t) }$. The two associated actions $J_{r}$ and $J_{z}$ are then immediately given by
\begin{equation}
J_{r} = \frac{1}{2} \kappa A_{R}^{2} \;\;\; ; \;\;\; J_{z} = \frac{1}{2} \nu A_{z}^{2} \, .
\label{actions_Jr_Jz}
\end{equation}
For ${ (J_{r} , J_{z} ) \!=\! (0,0) }$, the orbit of the star is circular. When one increases $J_{r}$ (resp. $J_{z}$), the amplitude of the radial (resp. vertical) oscillations increases, so that the orbit gets hotter. 
One should also note that within the epicyclic approximation, the intrinsic frequencies ${ \bm{\Omega} \!=\! (\Omega_{\phi} , \kappa , \nu ) }$ only depend on the variable $R_{\rg}$. Such a dynamical degeneracy may impact the system's secular properties.
 Finally, one can explicitly construct the mapping between the physical coordinates ${ (R,\phi,z,p_{R},p_{\phi},p_{z}) }$ and ${ (\theta_{R} , \theta_{\phi} , \theta_{z} , J_{r} , J_{z} , J_{\phi}) }$~\citep{LyndenBell1972,Palmer1994,BinneyTremaine2008}, which at first order takes the form
\begin{equation}
\begin{cases}
\displaystyle R = R_{\rg} \!+\! A_{R} \cos (\theta_{R}) \, ,
\\
\displaystyle \phi = \theta_{\phi} \!-\! \frac{2 \Omega_{\phi}}{\kappa} \frac{A_{R}}{R_{\rg}} \sin(\theta_{R}) \, ,
\\
\displaystyle z = A_{z} \cos (\theta_{z}) \, .
\end{cases}
\label{angles_mapping_epi}
\end{equation}
These relations and equations~\eqref{definition_Jphi} and~\eqref{actions_Jr_Jz}, provide an explicit mapping between the physical phase-space coordinates and the angle-action ones.

Finally, throughout the calculations, it will be assumed that the quasi-stationary DF of the system will initially take the form of a quasi-isothermal DF~\citep{BinneyMcMillan2011} defined as
\begin{equation}
F (R_{\rg} , J_{r} , J_{z}) = \frac{\Omega_{\phi} \Sigma}{\pi \kappa \sigma_{r}^{2}} \exp \!\bigg[\! - \frac{\kappa J_{r}}{\sigma_{r}^{2}} \!\bigg] \, \frac{\nu}{2 \pi \sigma_{z}^{2}} \exp \!\bigg[\! - \frac{\nu J_{z}}{\sigma_{z}^{2}} \!\bigg] \, ,
\label{definition_DF_quasi_isothermal}
\end{equation}
where the functions $\Sigma$, $\Omega_{\phi}$, $\kappa$, $\nu$, $\sigma_{r}$ and $\sigma_{z}$ have to be evaluated at $R_{\rg}$. Here, $\Sigma$ is the projected surface density associated with the system's density $\rho$ so that ${ \Sigma (R) \!=\! \!\int \! \mathrm{d} z \, \rho (R,z) }$, while $\sigma_{r}$ (resp. $\sigma_{z}$) represents the radial (resp. vertical) velocity dispersion of the stars at a given radius, and only depends on the position in the disc. Such a DF becomes the Schwarzschild DF in the epicycle limit~\cite[see equation~(4.153) in][]{BinneyTremaine2008}.

\subsection{Thick WKB basis elements}
\label{sec:thickbasiselements}

In the context of razor-thin discs, FPP15 presented in details how to construct a biorthonormal basis of tightly wound potentials and densities corresponding to a WKB solution of Poisson's equation. This construction of local basis elements led to a diagonal response matrix. One may now generalise this approach to discs of non-zero thickness by accordingly modifying the vertical components of these elements. The detail of some of the upcoming convolved in-plane calculations will not be presented, as they can be found in FPP15. We will focus here on the specifics of the extra vertical degree of freedom.
 Using the cylindrical coordinates ${ (R , \phi , z) }$, let us introduce the basis elements
\begin{equation}
\psi^{[k_{\phi} , k_{r} , R_{0} , n]} (R , \phi , z) = \mathcal{A} \, \psi_{r}^{[k_{\phi} , k_{r} , R_{0}]} (R , \phi) \, \psi_{z}^{[k_{r} , n]} (z) \, ,
\label{definition_psi_p}
\end{equation}
where $\mathcal{A}$ is an amplitude which will be tuned later on to ensure the correct normalisation of the basis elements. Here ${ \psi_{r}^{[k_{\phi} , k_{r} , R_{0}]} (R , \phi) }$ corresponds to the same in-plane dependence as the one introduced in FPP15 for the infinitely thin WKB basis elements and reads
\begin{equation}
\psi^{[k_{\phi} , k_{r} , R_{0}]}_{r} (R , \phi) = \re^{\ri (k_{\phi} \phi + k_{r} R)} \, \mathcal{B}_{R_{0}} (R) \, ,
\label{definition_psi_r}
\end{equation}
where the radial window function ${ B_{R_{0}} (R) }$ is defined as
\begin{equation}
\mathcal{B}_{R_{0}} (R) = \frac{1}{(\pi \sigma^{2})^{1/4}} \, \exp \!\bigg[\! - \frac{(R \!-\! R_{0})^{2}}{2 \sigma^{2}} \!\bigg] \, .
\label{definition_gaussian_window}
\end{equation}
The basis elements from equation~\eqref{definition_psi_p} are indexed by four numbers: $k_{\phi}$ is an azimuthal number which characterises the angular component of the basis elements, $R_{0}$ is the radius in the disc around which the Gaussian window $\mathcal{B}_{R_{0}} $ is centred, $k_{r}$ corresponds to the radial frequency of the basis elements, and finally ${ n \!\geq\! 1 }$ is an integer index, specific to the thick disc case, which numbers the considered vertical dependences, as detailed later on. In equation~\eqref{definition_gaussian_window}, we also introduced a decoupling scale $\sigma$, which ensures the biorthogonality of the basis elements. Figure~\ref{figBasisWKB_Radial} illustrates the radial dependence of these basis elements, while figure~\ref{figBasisWKB_Plane} focuses on their dependence in the ${ (R,\phi,z\!=\!0)-}$plane.
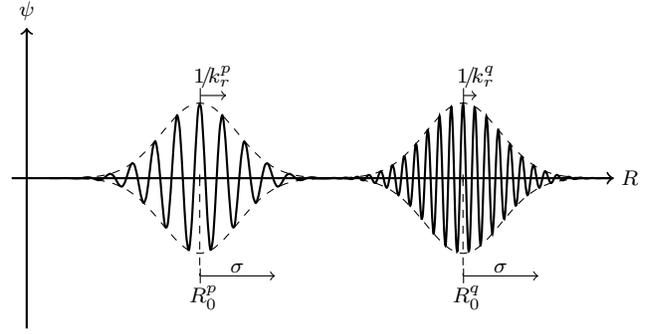
\begin{figure}
\begin{center}
\begin{tikzpicture}
\draw [->] [thick] (-0.5,0) -- (7.5,0) ; \draw (7.5,0) node[font = \small, right]{$R$} ;
\draw [->] [thick] (-0.3,-2) -- (-0.3,2) ; \draw (-0.3,2) node[font = \small, above]{$\psi$} ;
\draw [dashed , smooth] plot [domain=0:4] ( \x,{exp(- (\x-2)*(\x-2)*2)} ) ; \draw [dashed , smooth] plot [domain=0:4] ( \x,{-exp(- (\x-2)*(\x-2)*2)} ) ;
\draw [thick, smooth, samples = 100] plot [domain=0:4] ( \x,{1.005 * exp(- (\x-2)*(\x-2)*2) * cos(1200*(\x-2))} );
\draw [dashed] (2,0) -- (2,-1.3) ; \draw [thin] (2,-0.05) -- (2,0.05) ; \draw (2.05,-1.32) node[font = \small, below]{$R_{0}^{p}$} ;
\draw [|->] [thin] (2,-1.3) -- (3,-1.3) ; \draw (2.5,-1.35) node[font = \small, above]{$\sigma$} ;
\draw [|->] [thin] (2,1.1) -- (2.35,1.1) ; \draw (2.175,1.1) node[font = \small , above]{$ 1 \!/\! k_{r}^{p}$} ;
\draw [dashed , smooth] plot [domain=3.5:7.5] ( \x,{exp(- (\x-5.5)*(\x-5.5)*2)} ) ; \draw [dashed , smooth] plot [domain=3.5:7.5] ( \x,{-exp(- (\x-5.5)*(\x-5.5)*2)} ) ;
\draw [thick, smooth, samples = 200] plot [domain=3.5:7.5] ( \x,{1.005 * exp(- (\x-5.5)*(\x-5.5)*2) * cos(2300*(\x-5.5))} );
\draw [dashed] (5.5,0) -- (5.5,-1.3) ; \draw [thin] (5.5,-0.05) -- (5.5,0.05) ; \draw (5.55,-1.32) node[font = \small, below]{$R_{0}^{q}$} ;
\draw [|->] [thin] (5.5,-1.3) -- (6.5,-1.3) ; \draw (6.0,-1.35) node[font = \small, above]{$\sigma$} ;
\draw [|->] [thin] (5.5,1.1) -- (5.68,1.1) ; \draw (5.675,1.1) node[font = \small , above]{$ 1 \!/\! k_{r}^{q}$} ;
\end{tikzpicture}
\caption{\small{Reproduced from FPC15. Illustration of the radial dependence of two WKB basis elements. Each Gaussian $\mathcal{B}_{R_{0}}$ is centred around a radius $R_{0}$, is modulated at the frequency $k_{r}$, and extends on a region of size given by the decoupling scale $\sigma$.
}}
\label{figBasisWKB_Radial}
\end{center}
\end{figure}
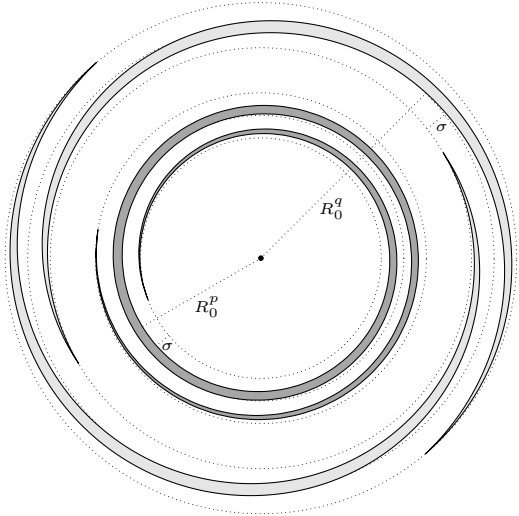
\begin{figure}
\begin{center}
\begin{tikzpicture}
\pgfmathsetmacro{\xmin}{-5} ; \pgfmathsetmacro{\xmax}{5} ;
\pgfmathsetmacro{\ymin}{-5} ; \pgfmathsetmacro{\ymax}{5} ;
\pgfmathsetmacro{\rada}{1.9} ;
\pgfmathsetmacro{\radb}{3.1} ;
\pgfmathsetmacro{\sigb}{0.3} ;
\draw [fill] (0 , 0) circle [radius = 0.03] ;
\draw [thin,dotted,line cap=round,dash pattern=on 0pt off 4 \pgflinewidth] (\rada , 0 ) arc[radius=\rada, start angle=0, end angle=360] ;
\draw [thin,dotted,line cap=round,dash pattern=on 0pt off 4 \pgflinewidth] ({\rada-\sigb} , 0 ) arc[radius={\rada-\sigb}, start angle=0, end angle=360] ;
\draw [thin,dotted,line cap=round,dash pattern=on 0pt off 4 \pgflinewidth] ({\rada+\sigb} , 0 ) arc[radius={\rada+\sigb}, start angle=0, end angle=360] ;
\draw [thin,dotted,line cap=round,dash pattern=on 0pt off 4 \pgflinewidth] (\radb , 0 ) arc[radius=\radb, start angle=0, end angle=360] ;
\draw [thin,dotted,line cap=round,dash pattern=on 0pt off 4 \pgflinewidth] ({\radb-\sigb} , 0 ) arc[radius={\radb-\sigb}, start angle=0, end angle=360] ;
\draw [thin,dotted,line cap=round,dash pattern=on 0pt off 4 \pgflinewidth] ({\radb+\sigb} , 0 ) arc[radius={\radb+\sigb}, start angle=0, end angle=360] ;
\pgfmathsetmacro{\angleRp}{210} ; 
\pgfmathsetmacro{\angleRq}{45} ;
\pgfmathsetmacro{\anglesigmap}{220} ;
\pgfmathsetmacro{\anglesigmaq}{38} ;
\draw [thin,dotted,line cap=round,dash pattern=on 0pt off 4 \pgflinewidth] (0 , 0) -- ({\rada*cos(\angleRp)} , {\rada*sin(\angleRp)}) ;
\draw ({\rada/2.3*cos(\angleRp)} , {\rada/2.3*sin(\angleRp)}) node [below] {${\scriptstyle R}_{\scriptscriptstyle 0}^{\scriptscriptstyle p}$} ; 
\draw [thin,dotted,line cap=round,dash pattern=on 0pt off 4 \pgflinewidth] (0 , 0) -- ({\radb*cos(\angleRq)} , {\radb*sin(\angleRq)}) ; 
\draw ({\radb/2.6*cos(\angleRq)+0.1} , {\radb/2.6*sin(\angleRq)+0.05}) node [below] {${\scriptstyle R}_{\scriptscriptstyle 0}^{\scriptscriptstyle q}$} ; 
\draw [thin,dotted,line cap=round,dash pattern=on 0pt off 4 \pgflinewidth] ({(\rada-\sigb)*cos(\anglesigmap)} , {(\rada-\sigb)*sin(\anglesigmap)}) -- ({\rada*cos(\anglesigmap)} , {\rada*sin(\anglesigmap)}) ;
\draw ({(\rada-\sigb/2)*cos(\anglesigmap)+0.1} , {(\rada-\sigb/2)*sin(\anglesigmap)-0.05}) node {${\scriptstyle \sigma}$} ;
\draw [thin,dotted,line cap=round,dash pattern=on 0pt off 4 \pgflinewidth] ({(\radb-\sigb)*cos(\anglesigmaq)} , {(\radb-\sigb)*sin(\anglesigmaq)}) -- ({\radb*cos(\anglesigmaq)} , {\radb*sin(\anglesigmaq)}) ;
\draw ({(\radb-\sigb/2)*cos(\anglesigmaq)+0.08} , {(\radb-\sigb/2)*sin(\anglesigmaq)-0.08}) node {${\scriptstyle \sigma}$} ;
\pgfmathsetmacro{\agsa}{170} ; \pgfmathsetmacro{\agfa}{200+2*360} ; \pgfmathsetmacro{\dga}{\agfa-\agsa} ;
\pgfmathsetmacro{\ampla}{0.06} ;
\path [draw = black , fill = gray!70 , thin , smooth , samples= \nbpWKBPlane] plot [domain=\agsa:\agfa] ({(\rada+\sigb-(\x-\agsa)*(2*\sigb)/(\dga) - \ampla*(1-((2*\x-\agsa-\agfa)/(\dga))*((2*\x-\agsa-\agfa)/(\dga))))*cos(\x) },{(\rada+\sigb-(\x-\agsa)*(2*\sigb)/(\dga) - \ampla*(1-((2*\x-\agsa-\agfa)/(\dga))*((2*\x-\agsa-\agfa)/(\dga))))*sin(\x)}) -- plot [domain=\agfa:\agsa] ({(\rada+\sigb-(\x-\agsa)*(2*\sigb)/(\dga) + \ampla*(1-((2*\x-\agsa-\agfa)/(\dga))*((2*\x-\agsa-\agfa)/(\dga))))*cos(\x) },{(\rada+\sigb-(\x-\agsa)*(2*\sigb)/(\dga) + \ampla*(1-((2*\x-\agsa-\agfa)/(\dga))*((2*\x-\agsa-\agfa)/(\dga))))*sin(\x)});
\pgfmathsetmacro{\agsb}{-50} ; \pgfmathsetmacro{\agfb}{210} ; \pgfmathsetmacro{\dgb}{\agfb-\agsb} ;
\pgfmathsetmacro{\amplb}{0.08} ;
\path [draw = black , fill = gray!20 , thin , smooth , samples= \nbpWKBPlane] plot [domain=\agsb:\agfb] ({(\radb+\sigb-(\x-\agsb)*(2*\sigb)/(\dgb) - \amplb*(1-((2*\x-\agsb-\agfb)/(\dgb))*((2*\x-\agsb-\agfb)/(\dgb))))*cos(\x) },{(\radb+\sigb-(\x-\agsb)*(2*\sigb)/(\dgb) - \amplb*(1-((2*\x-\agsb-\agfb)/(\dgb))*((2*\x-\agsb-\agfb)/(\dgb))))*sin(\x)}) -- plot [domain=\agfb:\agsb] ({(\radb+\sigb-(\x-\agsb)*(2*\sigb)/(\dgb) + \amplb*(1-((2*\x-\agsb-\agfb)/(\dgb))*((2*\x-\agsb-\agfb)/(\dgb))))*cos(\x) },{(\radb+\sigb-(\x-\agsb)*(2*\sigb)/(\dgb) + \amplb*(1-((2*\x-\agsb-\agfb)/(\dgb))*((2*\x-\agsb-\agfb)/(\dgb))))*sin(\x)}) ;
\path [draw = black , fill = gray!20 , thin , smooth , samples= \nbpWKBPlane] plot [domain=\agsb:\agfb] ({(\radb+\sigb-(\x-\agsb)*(2*\sigb)/(\dgb) - \amplb*(1-((2*\x-\agsb-\agfb)/(\dgb))*((2*\x-\agsb-\agfb)/(\dgb))))*cos(\x-180) },{(\radb+\sigb-(\x-\agsb)*(2*\sigb)/(\dgb) - \amplb*(1-((2*\x-\agsb-\agfb)/(\dgb))*((2*\x-\agsb-\agfb)/(\dgb))))*sin(\x-180)}) -- plot [domain=\agfb:\agsb] ({(\radb+\sigb-(\x-\agsb)*(2*\sigb)/(\dgb) + \amplb*(1-((2*\x-\agsb-\agfb)/(\dgb))*((2*\x-\agsb-\agfb)/(\dgb))))*cos(\x-180) },{(\radb+\sigb-(\x-\agsb)*(2*\sigb)/(\dgb) + \amplb*(1-((2*\x-\agsb-\agfb)/(\dgb))*((2*\x-\agsb-\agfb)/(\dgb))))*sin(\x-180)}) ;
\end{tikzpicture}
\caption{\small{Reproduced from FPC15. Illustration of the dependence of two WKB basis elements in the ${ (R,\phi , z \!=\! 0)-}$plane. Each basis element is located around a central radius $R_{0}$, on a region of size $\sigma$. The winding of the spirals is governed by the radial frequency $k_{r}$, while the number of azimuthal patterns is given by the index $k_{\phi}$, e.g., ${ k_{\phi} \!=\! 1 }$ for the interior dark grey element, and ${ k_{\phi} \!=\! 2 }$ for the exterior light grey one.
}}
\label{figBasisWKB_Plane}
\end{center}
\end{figure}
One should note that the decomposition introduced in equation~\eqref{definition_psi_p} amounts to multiplying the in-plane thin WKB basis elements by a vertical function $\psi_{z}^{[k_{r} , n]}$ which should now be specified.

Starting from the ansatz of equation~\eqref{definition_psi_p}, one now has to solve Poisson's equation~\eqref{definition_basis} to determine the associated density basis elements. Given the assumption of tight-winding (mainly ${ k_{r} R \!\gg\! 1}$, see FPP15), it takes the form
\begin{equation}
- k_{r}^{2} \mathcal{A} \, \psi_{r} \psi_{z} \!+\! \mathcal{A} \, \psi_{r} \frac{\mathrm{d}^{2} \psi_{z}}{\mathrm{d} z^{2}} = 4 \pi G \rho \, ,
\label{Poisson_WKB_thick}
\end{equation}
where the superscripts ${ [ k_{\phi} , k_{r} , R_{0} , n ] }$ have not been written out to shorten the notations. Let us now assume that the density elements satisfy the ansatz of separability 
\begin{equation}
\rho (R , \phi , z) = \frac{\lambda_{\rho}}{4 \pi G} \, \mathcal{A} \, \psi_{r} (R, \phi ) \, \psi_{z} (z) \, w (z) \, ,
\label{ansatz_density}
\end{equation}
where ${ \lambda_{\rho} \!=\! \lambda_{\rho}^{[k_{r} , n]} }$ is a proportionality constant, while ${ w (z) }$ is a cavity function independent of the basis elements' indices. Such a decomposition allows us to rewrite equation~\eqref{Poisson_WKB_thick} as
\begin{equation}
\frac{\mathrm{d}^{2} \psi_{z}}{\mathrm{d} z^{2}} \!-\! k_{r}^{2} \psi_{z} = \lambda_{\rho} \, w (z) \, \psi_{z} \, .
\label{Poisson_WKB_thick_II}
\end{equation}
Notice that equation~\eqref{Poisson_WKB_thick_II} takes the form of a Sturm-Liouville equation~\citep{CourantHilbert1953}, for which one has to determine the eigenfunctions $\psi_{z}^{[k_{r} , n]}$ along with their associated eigenvalues $\lambda_{\rho}^{[ k_{r} , n ]}$. Under sufficient assumptions of regularity the Sturm-Liouville theory states that there exists a discrete spectrum of real eigenvalues ${ \lambda_{1} \!<\! \lambda_{2} \!<\! ... \!<\! \lambda_{n} \!\to\! + \infty }$, with their associated eigenfunctions $\psi_{z}^{1}$, $\psi_{z}^{2}$, ..., $\psi_{z}^{n}$. Moreover, when correctly normalised, the eigenfunctions form a biorthogonal basis such that ${ \int \! \mathrm{d} z \, w (z) \, \psi_{z}^{p} (z) \, \psi_{z}^{q} (z) \!=\! \delta_{p}^{q} }$. 
\\
In order to obtain an explicit expression for our thick basis elements, one now has to specify the considered cavity function ${ w (z) }$. Let us assume that the density basis elements are zero for ${ | z | \!>\! h }$, so that they vanish out of a sharp cavity. This amounts to choosing ${ w (z) }$ such that
\begin{equation}
w (z) = \Theta (z/h) \, ,
\label{definition_w}
\end{equation} 
where ${ \Theta (x) }$ is a door function, equal to $1$ for ${ x \!\in\! [-1 ; 1] }$ and $0$ elsewhere~\citep[see][for a similar ansatz]{GrivGedalin2012}. Since the WKB basis is a local basis, one can adapt the height ${ h \!=\! h (R_{0}) }$ as a function of the position within the disc, so as to better mimic the mean density profile of the disc. Because ${ h (R_{0}) }$ is an ad hoc parameter, one still has to detail how this quantity should be specified as a function of the disc's parameters. The main idea behind equation~\eqref{definition_w} is to approximate the physical cavity of the mean density profile by an approximate sharp cavity of height $h$.
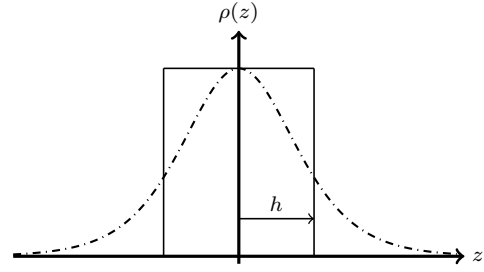
\begin{figure}
\begin{center}
\begin{tikzpicture}
\draw [->] [very thick] (-3,0) -- (3,0) ; \draw (3.,0) node[font = \small, right]{$z$} ;
\draw [->] [very thick] (0.,-0.1) -- (0.,3) ; ; \draw (-0.,3) node[font = \small, above]{$\rho (z)$} ;
\draw [dashdotted , thick, smooth , samples = 100] plot [domain = -3 : 2.9 ] (\x , { 2.5 * (1 / ( cosh ( \x ) ) )^(2) } ) ;
\draw [semithick , smooth] (-1 , 2.5) -- (1 , 2.5) ;
\draw [semithick , smooth] (-1 , 0.) -- (-1 , 2.5) ;
\draw [semithick , smooth] (1 , 0. ) -- (1 , 2.5) ;
\draw [ -> ] [semithick , smooth ] (0 , 0.5) -- (1. , 0.5) ;
\draw ( 0.5 , 0.5 ) node[font = \small , above]{$h$} ; 
\end{tikzpicture}
\caption{\small{Construction of the sharp cavity (solid lines) consistent with the underlying physical vertical density (dotted-dashed lines). We impose the matching of the total volume of the physical and approximate vertical density profiles. The mean density profile corresponds to a Spitzer profile, as introduced in equation~\eqref{definition_rho_star}.
}}
\label{fig_choice_h}
\end{center}
\end{figure}
To do so, as illustrated in figure~\ref{fig_choice_h}, $h$ is chosen to match the volume of the physical and sharp cavities, i.e. one imposes ${ \int \! \mathrm{d} z \, \rho_{\rm tot} (R_{0} , z) \!=\! 2 h (R_{0}) \, \rho_{\rm tot} (R_{0} , 0) }$. When assuming the mean density profile to be a Spitzer profile, as defined later on in equation~\eqref{definition_rho_star}, one can immediately relate $h$ to $z_{0}$ as ${ h (R_{0}) \!=\! 2 \, z_{0} (R_{0}) }$. Therefore, the cavity scale $h$ from equation~\eqref{definition_w} should not be seen as a free parameter of our model, but as imposed by the physical mean density profile of the considered disc.

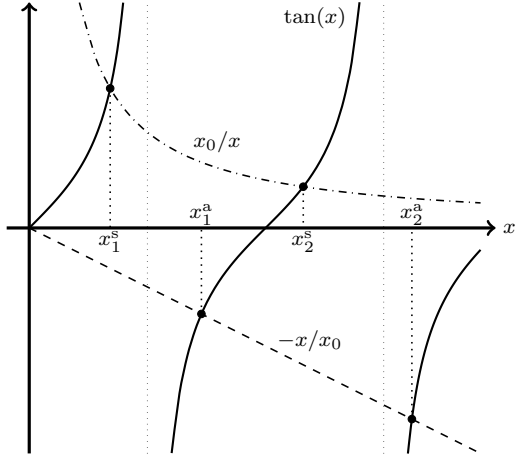
\begin{figure}
\begin{center}
\begin{tikzpicture}
\draw [->] [very thick] (-0.3,0) -- (6.2,0) ; \draw (6.2,0) node[font = \small, right]{$x$} ;
\draw [->] [very thick] (-0.,-3) -- (-0.,3) ; ; \draw (-0.,3) node[font = \small, above]{$$} ;
\draw [dashdotted , semithick, smooth] plot [domain = 0.667 : 6. ] (\x , {2 / \x} ) ;
\draw [dashed , semithick , smooth] plot [domain=0 : 6.] (\x , {-\x / 2});
\draw [thick , smooth] plot[domain=0 : 1.249 ] (\x , { tan (\x * 180 / (3.141592)) }) ;
\draw [thick , smooth] plot[domain = 1.893 : 4.391 ] (\x , { tan (\x * 180 / (3.141592)) }) ;
\draw [thick , smooth] plot[domain = 5.034 : 6.] (\x , { tan (\x * 180 / (3.141592)) }) ;
\draw [very thin , dotted , smooth] (1.571,-3) -- (1.571 , 3) ; \draw [very thin , dotted , smooth] (4.712,-3) -- (4.712 , 3) ;
\draw [fill] (1.076, 1.857) circle [radius = 0.05]; \draw [dotted, semithick] (1.076 , 1.857) -- (1.076, 0) ; \draw (1.076, 0+0.06) node[font = \small, below]{$x_{1}^{\rm s}$} ;
\draw [fill] (3.643, 0.549) circle [radius = 0.05]; \draw [dotted , semithick] (3.643 , 0.549) -- (3.643 , 0); \draw (3.643, 0+0.06) node[font = \small, below]{$x_{2}^{\rm s}$} ;
\draw [fill] (2.289, -1.144) circle [radius = 0.05]; \draw [dotted , semithick] (2.289 , -1.144) -- (2.289 , 0); \draw (2.289, 0-0.06) node[font = \small , above]{$x_{1}^{\rm a}$};
\draw [fill] (5.087, -2.543) circle [radius = 0.05]; \draw [dotted , semithick] (5.087 , -2.543) -- (5.087 , 0); \draw (5.087, 0-0.06) node[font = \small , above]{$x_{2}^{\rm a}$};
\draw (3.8,2.8) node[font = \small]{$ \tan (x) $};
\draw (2.5,1.15) node[font = \small]{$ x_{0} / x $};
\draw (3.2,-1.5) node[font = \small , right]{$- x  / x_{0}$} ;
\end{tikzpicture}
\caption{\small{Illustration of the quantisation relations for the vertical frequency $k_{z}$ induced by the sharp cavity of height $h$. In order to use dimensionless quantities, we introduced ${ x \!=\! k_{z} h }$ and ${ x_{0} \!=\! k_{r} h }$. The top dotted-dashed curve corresponds to the symmetric case from equation~\eqref{quantisation_even} leading to the quantised dimensionless frequencies $x_{1}^{\rm s}$, $x_{2}^{\rm s}$,... The bottom dashed curve corresponds to the antisymmetric case obtained in equation~\eqref{quantisation_odd} associated with $x_{1}^{\rm a}$, $x_{2}^{\rm a}$,... One can note the specific role played by the fundamental symmetric frequency $x_{1}^{\rm s}$, which is the only dimensionless frequency inferior to ${ \pi / 2}$.
}}
\label{fig_quantisation}
\end{center}
\end{figure}

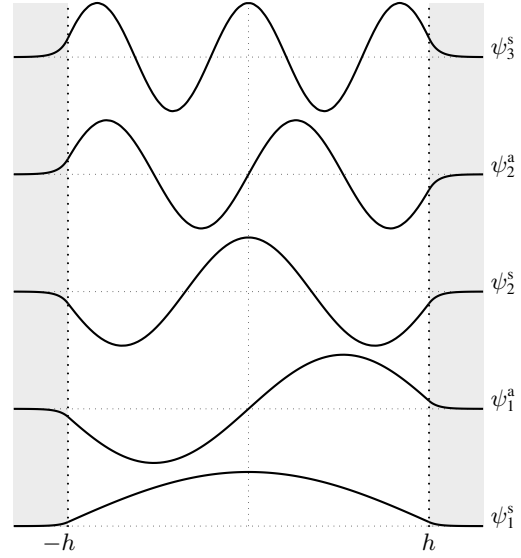
\begin{figure}
\begin{center}
\begin{tikzpicture}[scale = 1.2]
\pgfmathsetmacro{\h}{2} ; \pgfmathsetmacro{\kr}{10} ; 
\pgfmathsetmacro{\eps}{0.001} ; 
\pgfmathsetmacro{\ks}{0.748064} ; \pgfmathsetmacro{\kss}{2.24574} ; \pgfmathsetmacro{\ksss}{3.74771} ; \pgfmathsetmacro{\ka}{1.49652} ; \pgfmathsetmacro{\kaa}{2.99605} ; \pgfmathsetmacro{\kaaa}{4.50092} ; 
\pgfmathsetmacro{\db}{1.3} ; 
\pgfmathsetmacro{\ds}{-2 * \db} ; \pgfmathsetmacro{\da}{- \db} ; \pgfmathsetmacro{\dss}{0} ; \pgfmathsetmacro{\daa}{\db} ; \pgfmathsetmacro{\dsss}{2 * \db}
\pgfmathsetmacro{\alpha}{0.6}
\pgfmathsetmacro{\zmin}{-1.3 * \h} ; \pgfmathsetmacro{\zmax}{1.3 * \h} ; 
\path [fill = gray!15] (\zmin , \ds) rectangle ( - \h , \dsss + \alpha) ;
\path [fill = gray!15] (\h , \ds) rectangle (\zmax , \dsss + \alpha) ; 
\draw [dotted, line cap=round , dash pattern=on 0pt off 4 \pgflinewidth , thick] (-\h , \ds) -- (-\h , \dsss + \alpha) ;
\draw [dotted, line cap=round , dash pattern=on 0pt off 4 \pgflinewidth , thick] (\h , \ds ) -- (\h , \dsss + \alpha) ;
\draw [dotted , very thin] (0 , \ds ) -- ( 0 , \dsss + \alpha) ;
\draw [dotted , very thin] (\zmin , \ds) -- (\zmax , \ds) ;
\draw [dotted , very thin] (\zmin , \dss) -- (\zmax , \dss) ;
\draw [dotted , very thin] (\zmin , \dsss) -- (\zmax , \dsss) ;
\draw [dotted , very thin] (\zmin , \da) -- (\zmax , \da) ;
\draw [dotted , very thin] (\zmin , \daa) -- (\zmax , \daa) ;
\pgfmathsetmacro{\dw}{0.1} ; 
\draw (\zmax, \ds + \dw) node[font = \normalsize, right]{$\psi^{\text{\scriptsize s}}_{\text{\scriptsize 1}}$} ;
\draw (\zmax, \dss + \dw) node[font = \normalsize, right]{$\psi^{\text{\scriptsize s}}_{\text{\scriptsize 2}}$} ; 
\draw (\zmax, \dsss + \dw) node[font = \normalsize, right]{$\psi^{\text{\scriptsize s}}_{\text{\scriptsize 3}}$} ; 
\draw (\zmax, \da + \dw) node[font = \normalsize, right]{$\psi^{\text{\scriptsize a}}_{\text{\scriptsize 1}}$} ; 
\draw (\zmax, \daa + \dw) node[font = \normalsize, right]{$\psi^{\text{\scriptsize a}}_{\text{\scriptsize 2}}$} ; 
\draw (- \h -0.1, \ds) node[font = \normalsize , below]{$- h$} ; \draw ( \h , \ds) node[font = \normalsize , below]{$h$} ;
\draw [thick , smooth] plot[domain=- \h : \h , variable = \z , samples = \nbpPsiZ] (\z , { \alpha * cos ( \ks * \z * 180 / (3.141592)) + \ds }) ;
\draw [thick , smooth] plot[domain=- \h : \h , variable = \z , samples = \nbpPsiZ] (\z , { \alpha * cos ( \kss * \z * 180 / (3.141592)) + \dss}) ;
\draw [thick , smooth] plot[domain=- \h : \h , variable = \z , samples = \nbpPsiZ] (\z , { \alpha * cos ( \ksss * \z * 180 / (3.141592)) + \dsss}) ;
\draw [thick , smooth] plot[domain=- \h : \h , variable = \z , samples = \nbpPsiZ] (\z , { \alpha * sin ( \ka * \z * 180 / (3.141592)) + \da}) ;
\draw [thick , smooth] plot[domain=- \h : \h , variable = \z , samples = \nbpPsiZ] (\z , { \alpha * sin ( \kaa * \z * 180 / (3.141592)) + \daa}) ;
\draw [thick , smooth] plot[domain= \zmin : -\h + \eps , variable = \z , samples = \nbpPsiZ] (\z , { \alpha * exp( \kr * (\h + \z )) * cos(\ks * \h * 180 / (3.141592)) + \ds }) ; 
\draw [thick , smooth] plot[domain= \h - \eps : \zmax , variable = \z , samples = \nbpPsiZ] (\z , { \alpha * exp( \kr * (\h - \z )) * cos(\ks * \h * 180 / (3.141592)) + \ds } ) ; 
\draw [thick , smooth] plot[domain= \zmin : -\h + \eps , variable = \z , samples = \nbpPsiZ] (\z , { \alpha * exp( \kr * (\h + \z )) * cos(\kss * \h * 180 / (3.141592)) + \dss }) ; 
\draw [thick , smooth] plot[domain= \h - \eps : \zmax , variable = \z , samples = \nbpPsiZ] (\z , { \alpha * exp( \kr * (\h - \z )) * cos(\kss * \h * 180 / (3.141592)) + \dss } ) ; 
\draw [thick , smooth] plot[domain= \zmin : -\h + \eps , variable = \z , samples = \nbpPsiZ] (\z , { \alpha * exp( \kr * (\h + \z )) * cos(\ksss * \h * 180 / (3.141592)) + \dsss }) ; 
\draw [thick , smooth] plot[domain= \h - \eps : \zmax , variable = \z , samples = \nbpPsiZ] (\z , { \alpha * exp( \kr * (\h - \z )) * cos(\ksss * \h * 180 / (3.141592)) + \dsss } ) ; 
\draw [thick , smooth] plot[domain= \zmin : -\h + \eps , variable = \z , samples = \nbpPsiZ] (\z , { - \alpha * exp( \kr * (\h + \z )) * sin(\ka * \h * 180 / (3.141592)) + \da }) ; 
\draw [thick , smooth] plot[domain= \h - \eps : \zmax , variable = \z , samples = \nbpPsiZ] (\z , { \alpha * exp( \kr * (\h - \z )) * sin(\ka * \h * 180 / (3.141592)) + \da } ) ; 
\draw [thick , smooth] plot[domain= \zmin : -\h + \eps , variable = \z , samples = \nbpPsiZ] (\z , { - \alpha * exp( \kr * (\h + \z )) * sin(\kaa * \h * 180 / (3.141592)) + \daa }) ; 
\draw [thick , smooth] plot[domain= \h - \eps : \zmax , variable = \z , samples = \nbpPsiZ] (\z , { \alpha * exp( \kr * (\h - \z )) * sin(\kaa * \h * 180 / (3.141592)) + \daa } ) ; 
\end{tikzpicture}
\caption{\small{Illustration of the vertical dependence of the WKB potential basis elements. Here $\psi^{\rm s}$ stands for the symmetric elements from equation~\eqref{full_psi_p_even}, and $\psi^{\rm a}$ for the antisymmetric elements from equation~\eqref{full_psi_p_odd}. As expected from the Sturm-Liouville theory, the basis elements can be ordered via their number of nodes within the cavity.
}}
\label{fig_psi_z_shape}
\end{center}
\end{figure}

Given the cavity function from equation~\eqref{definition_w}, one may then solve Poisson's equation~\eqref{Poisson_WKB_thick_II} -- which takes the simple form of a wave equation -- to obtain an explicit expression for the thick WKB basis elements. It is therefore assumed that $\psi_{z}$ follows the ansatz
\begin{equation}
\psi_{z} (z) = 
\begin{cases}
\begin{aligned}
& A \re^{- k_{r} z }, \;\; & \text{if} & \;\;\; z > h \, ,
\\
& B \re^{\ri k_{z} z} \!+\! C \re^{- \ri k_{z} z}, \;\; & \text{if} & \;\;\; | z | \leq h \, ,
\\
& D \re^{ k_{r} z}, \;\; & \text{if} & \;\;\; z < - h \, ,
\end{aligned}
\end{cases}
\label{ansatz_psi_z}
\end{equation}
where the frequency $k_{z}$ remains to be determined. One immediately obtains ${ \lambda_{\rho} \!=\! -( k_{r}^{2} \!+\! k_{z}^{2} ) }$. In the decomposition from equation~\eqref{ansatz_psi_z}, one must also ensure that both $\psi_{z}$ and ${\mathrm{d} \psi_{z} / \mathrm{d} z }$ are continuous at ${ z \!=\! \pm h }$. At this stage, we will now restrict ourselves to symmetric perturbations, so that ${ \psi_{z} (-z) \!=\! \psi_{z} (z) }$. The very similar antisymmetric case is detailed in Appendix~\ref{sec:appendixantisymbasis}. For even perturbations, one immediately obtains from equation~\eqref{ansatz_psi_z} that ${ A \!=\! D }$ and ${ B \!=\! C }$. The continuity conditions on $\psi_{z}$ and ${ \mathrm{d} \psi_{z} / \mathrm{d} z }$ then take the form
\begin{equation}
\begin{cases}
\displaystyle A \re^{-k_{r} h} = 2 B \cos (k_{z} h) \, ,
\\
\displaystyle k_{r} A \re^{- k_{r} h} = 2 k_{z} B \sin (k_{z} h) \, .
\end{cases}
\label{continuity_conditions_even}
\end{equation}
In order to have a non trivial solution, this requires for $k_{z}$ to satisfy the relation
\begin{equation}
\tan (k_{z} h) = \frac{k_{r}}{k_{z}} \, .
\label{quantisation_even}
\end{equation}
Equation~\eqref{quantisation_even} plays the role of a quantisation relation, which constrains the allowed values for $k_{z}$, once $k_{r}$ and $h$ have been specified. As in the definition of the basis elements from equation~\eqref{definition_psi_p}, we introduce the index ${ n \!\geq\! 1 }$ such that $k_{z}^{n}$ is the ${n-}$th solution of equation~\eqref{quantisation_even}, so that one has
\begin{equation}
k_{z}^{1} \!<\! k_{z}^{2} \!<\! ... \!<\! k_{z}^{n} \!<\! ... \;\;\; \text{and} \;\;\; \tan (k_{z}^{n} h) = \frac{k_{r}}{k_{z}^{n}} \, .
\label{quantised_kz}
\end{equation}
In addition, if one assumes that the disc is sufficiently thin so that ${ k_{z}^{1} h }$ and ${ k_{r} h \!\lesssim\! 1 }$, one can obtain in this limit a simple estimation of the first quantised even $k_{z}^{1}$ which reads
\begin{equation}
k_{z}^{1} \!\simeq\! \sqrt{\frac{k_{r}}{h}} \, .
\label{kz1_even}
\end{equation}
The symmetric quantisation relation~\eqref{quantisation_even} along with its antisymmetric analog from equation~\eqref{quantisation_odd} are illustrated in figure~\ref{fig_quantisation}.
Two important properties of these quantisation relations should be noted. First of all, the fundamental symmetric frequency $k_{z}^{1}$ appears as the only quantised frequency such that ${ k_{z}^{1} h \!<\! \pi/2 }$. Given equation~\eqref{kz1_even}, in the infinitely thin limit where ${ h \!\to\! 0}$, one has ${ k_{z}^{1} h \!\to\! 0 }$, while all the other frequencies are such that ${ k_{z} h }$ remains larger than ${ \pi/2 }$. Such a property already underlines how this fundamental symmetric mode $k_{z}^{1}$ will play a particular role in the razor-thin limit. Moreover, because of the ${ \pi-}$periodicity of the ``$\tan$" function, in the limit of a sufficiently thick disc for which ${ k_{r} h \!\gtrsim\! \pi }$, one may assume that for both symmetric and antisymmetric cases, one has
\begin{equation}
\Delta k_{z} = k_{z}^{n+1} \!-\! k_{z}^{n} \simeq \frac{\pi}{h} \, .
\label{step_distance_kz}
\end{equation}
After some simple algebra, one can finally give a complete definition of the symmetric potential elements as
\begin{align}
\psi^{[k_{\phi} , k_{r} , R_{0} , n]} & \, (R , \phi , z) =  \mathcal{A} \, \psi_{r}^{[k_{\phi} , k_{r} , R_{0}]} (R , \phi) \nonumber
\\
& \;\;\;\;\; \times \,
\begin{cases}
\begin{aligned}
\displaystyle & \! \cos (k_{z}^{n} z) & \!\!\!\! \text{if}& \; |z| \leq h \, ,
\\
\displaystyle & \! \re^{k_{r} h} \! \cos (k_{z}^{n} h) \, \re^{- k_{r} |z|} & \!\!\!\! \text{if}& \; |z| \geq h \, .
\end{aligned}
\end{cases}
\label{full_psi_p_even}
\end{align}
Similarly, the associated density elements are given by
\begin{align}
\rho^{[k_{\phi} , k_{r} , R_{0} , n]}  \, (R ,  \phi , z) =& \, - \frac{k_{r}^{2} \!+\! (k_{z}^{n})^{2}}{4 \pi G} \nonumber
\\
& \times \,  \psi^{[k_{\phi} , k_{r} , R_{0} , n]} (R , \phi , z) \, \Theta \!\bigg[\! \frac{z}{h} \!\bigg] \, .
\label{full_rho_p_even}
\end{align}
The associated antisymmetric basis elements are given in equations~\eqref{full_psi_p_odd} and~\eqref{full_rho_p_odd}. Figure~\ref{fig_psi_z_shape} illustrates the shape of the vertical component of the first basis elements.
As imposed by the definition from equation~\eqref{definition_basis}, one must then ensure that the basis is biorthogonal. As demonstrated in FPP15, we know that for ${ (k_{\phi}^{p} , k_{r}^{p} , R_{0}^{p}) \!\neq\! (k_{\phi}^{q} , k_{r}^{q} , R_{0}^{q}) }$, the orthogonality property is satisfied, under the WKB scale-decoupling assumptions ${ \Delta R_{0} \!\gg\! \sigma \!\gg\! 1/\Delta k_{r} }$, where ${ \Delta R_{0} }$ and ${ \Delta k_{r} }$ stand for the step distances between two successive basis elements, and $\sigma$ is of the width of the radial Gaussian in equation~\eqref{definition_gaussian_window} (see~\cite{FouvryPichonChavanis2015} for details). Moreover, as underlined after equation~\eqref{Poisson_WKB_thick_II}, the Sturm-Liouville theory naturally enforces the orthogonality w.r.t. the $n_{p}$ and $n_{q}$ indices, so that the basis elements from equations~\eqref{full_psi_p_even} and~\eqref{full_rho_p_even} are indeed orthogonal. To finalise the construction of the basis elements, it only remains to correctly normalise them, by determining the value of the amplitude $\mathcal{A}$. One immediately obtains
\begin{equation}
\mathcal{A} = \sqrt{\frac{G}{R_{0} h (k_{r}^{2} \!+\! (k_{z}^{n})^{2})}} \, \alpha_{n} \, ,
\label{amplitude_basis_even}
\end{equation}
where ${1 \!\lesssim\! \alpha_{n} \!\lesssim\! 1.6 }$ is a numerical prefactor given by
\begin{equation}
\alpha_{n} = \sqrt{\frac{2}{1 \!+\! \sin (2 k_{z}^{n} h)/(2 k_{z}^{n} h)}} \, .
\label{alpha_even}
\end{equation}
Using the angle-action mapping from equation~\eqref{angles_mapping_epi}, one may now compute the Fourier transform of the symmetric basis elements as defined in equation~\eqref{definition_psi_FT}. We recall the sum decomposition formula of the Bessel functions of the first kind $\mathcal{J}_{\ell}$ which reads
\begin{equation}
\re^{\ri z \cos(\theta)} = \sum_{\ell} \ri^{\ell} \mathcal{J}_{\ell} (z) \, \re^{\ri \ell \theta} \; ; \; \re^{\ri z \sin (\theta)} = \sum_{\ell} \mathcal{J}_{\ell} (z) \, \re^{\ri \ell \theta} \, ,
\label{sum_decomposition_Bessel}
\end{equation}
along with the property ${ \mathcal{J}_{\ell} (- x) \!=\! (-1)^{\ell} \mathcal{J}_{\ell} (x) }$. Assuming that the vertical excursions of the stars are smaller than $h$, one obtains
\begin{align}
\psi_{\bm{m}}^{[k_{\phi} , k_{r} , R_{0} , n]} (\bm{J}) = & \, \delta_{m_{\phi}}^{k_{\phi}} \, \delta_{m_{z}}^{\rm even} \mathcal{A} \, \re^{\ri k_{r} R_{\rg}} \, \ri^{m_{z} - m_{r}} \mathcal{B}_{R_{0}} (R_{\rg}) \nonumber
\\
& \, \times \, \mathcal{J}_{m_{r}} \!\bigg[\! \sqrt{\tfrac{2 J_{r}}{\kappa}} k_{r} \!\bigg] \, \mathcal{J}_{m_{z}} \!\bigg[\! \sqrt{\tfrac{2 J_{z}}{\nu}} k_{z}^{n} \!\bigg] \, ,
\label{psi_m_even}
\end{align}
while the antisymmetric analog is given in equation~\eqref{psi_m_odd}. Given ${ \psi_{\bm{m}}^{(p)} (\bm{J}) }$, one may now proceed to the evaluation of the response matrix from equation~\eqref{Fourier_M}.

\subsection{Amplification eigenvalues}
\label{sec:amplificationeigenvalues}

A key result of FPP15 was to show that in the infinitely thin limit, the response matrix could be assumed to be diagonal, when computed with WKB basis elements along with the scale-decoupling hypothesis ${ \Delta R_{0} \!\gg\! \sigma \!\gg\! \Delta k_{r} }$. This was a central result allowing for the analytical derivation of the diffusion coefficients. Here, the thick basis elements will have the same radial dependence as in FPP15, but their vertical components might interact and therefore lead to a more complex response matrix. In Appendix~\ref{sec:appendixMdiagonal}, we show that for a thick disc with our thick WKB basis elements, one may still assume the response matrix to be diagonal so as to have
\begin{equation}
 \widehat{\mathbf{M}}_{[k_{\phi}^{p} \!, k_{r}^{p} \!, R_{0}^{p} \!, n_{p} ] , [k_{\phi}^{q} \!, k_{r}^{q} \!, R_{0}^{q} \!, n_{q} ]} \!\!=\! \delta_{k_{\phi}^{p}}^{k_{\phi}^{q}} \delta_{k_{r}^{p}}^{k_{r}^{q}} \delta_{R_{0}^{p}}^{R_{0}^{q}} \delta_{n_{p}}^{n_{q}} \lambda_{[ k_{\phi}^{p} \!, k_{r}^{p} \!, R_{0}^{p} \!, n_{p} ]} \, .
\label{Matrix_diagonal}
\end{equation}
This is a crucial result of the present section.

Let us now estimate the diagonal elements of the response matrix. Compared to FPP15, an additional difficulty in the thick context is to compute the additional integral on $J_{z}$. This can be made using formula $6.615$ from~\cite{Gradshteyn2007}, which reads
\begin{align}
\int_{0}^{+ \infty} \!\!\!\!\!\!\!\! \mathrm{d} J_{z} \, \re^{- a J_{z}} \mathcal{J}_{m_{z}} & \!\big[ b_{p} \sqrt{J_{z}} \big] \mathcal{J}_{m_{z}} \!\big[ b_{q} \sqrt{J_{z}} \big]  \nonumber
\\
& = \frac{1}{a} \mathcal{I}_{m_{z}} \!\bigg[\! \frac{b_{p} b_{q}}{2 a} \!\bigg] \exp \!\bigg[\! -\! \frac{b_{p}^{2} \!+\! b_{q}^{2}}{4 a} \!\bigg] \, ,
\label{formula_Gradshteyn}
\end{align}
where ${ b_{p\!/\!q} \!=\! k_{z}^{p\!/\!q} \sqrt{2/\nu} }$ and ${ a \!=\! \nu / \sigma_{z}^{2} }$. As the disc is supposed to be tepid, in equation~\eqref{Fourier_M}, the contributions from ${ \partial F / \partial J_{\phi} }$ may be neglected w.r.t. ${ \partial F / \partial J_{r} }$ and ${ \partial F / \partial J_{z} }$. Following the same method as in Appendix B of FPP15, and after some algebra, one finally obtains the expression of the symmetric amplification eigenvalues as
\begin{align}
\lambda_{[k_{\phi} , k_{r} , R_{0} , n]}^{\rm sym} = & \, \frac{2 \pi G \Sigma \alpha_{n}^{2}}{h \kappa^{2} (1 \!+\! (k_{z} / k_{r})^{2})} \! \sum_{\ell_{z} {\rm \, even}} \! \frac{\re^{- \chi_{z}} \mathcal{I}_{\ell_{z}}[\chi_{z}] }{(1 \!-\! s_{\ell_{z}}^{2})} \nonumber
\\
& \, \times \, \bigg\{ \mathcal{F} (s_{\ell_{z}} , \chi_{r}) \!-\! \ell_{z} \frac{\nu}{\sigma_{z}^{2}} \frac{\sigma_{r}^{2}}{\kappa} \, \mathcal{G} (s_{\ell_{z}} , \chi_{r}) \bigg\} \, .
\label{lambda_even}
\end{align}
In equation~\eqref{lambda_even}, the dimensionless quantities $\chi_{r}$ and $\chi_{z}$ were defined as
\begin{equation}
\chi_{r} = \frac{\sigma_{r}^{2} k_{r}^{2}}{\kappa^{2}} \;\;\; ; \;\;\; \chi_{z} = \frac{\sigma_{z}^{2} k_{z}^{2}}{\nu^{2}} \, ,
\label{definition_chi}
\end{equation}
and the shifted dimensionless frequency $s_{\ell_{z}}$ as
\begin{equation}
s_{\ell_{z}} = \frac{\omega \!-\! k_{\phi} \Omega_{\phi} \!-\! \ell_{z} \nu}{\kappa} \, .
\label{definition_s}
\end{equation}
Finally, in equation~\eqref{lambda_even}, the (reduction) functions $\mathcal{F}$ and $\mathcal{G}$ were also introduced as
\begin{equation}
\!\!\!\! \begin{cases}
\displaystyle \mathcal{F} (s , \chi) \!=\! 2 (1 \!-\! s^{2}) \frac{\re^{-\chi}}{\chi} \sum_{\ell = 1}^{+ \infty} \frac{\mathcal{I}_{\ell} [\chi]}{1 \!-\! [s/\ell]^{2}} \, ,
\\
\displaystyle \mathcal{G} (s , \chi) \!=\! 2 (1 \!-\! s^{2}) \frac{\re^{-\chi}}{\chi} \bigg[ \frac{1}{2} \frac{\mathcal{I}_{0} [\chi]}{s} \!+\! \frac{1}{s} \sum_{\ell = 1}^{+ \infty} \frac{\mathcal{I}_{\ell} [\chi]}{1 \!-\! [\ell/s]^{2}} \bigg] \, ,
\end{cases}
\label{definitions_F_G}
\end{equation}
where $\mathcal{F}$ is the usual reduction function from the seminal works of~\cite{Kalnajs1965,LinShu1966}.
When considering antisymmetric contributions, thanks to the results from Appendix~\ref{sec:appendixantisymbasis}, one obtains the amplification eigenvalues given by
\begin{align}
\lambda_{[k_{\phi} , k_{r} , R_{0} , n]}^{\rm anti} = & \, \frac{2 \pi G \Sigma \beta_{n}^{2}}{h \kappa^{2} (1 \!+\! (k_{z} / k_{r})^{2})} \! \sum_{\ell_{z} {\rm \, odd}} \! \frac{\re^{- \chi_{z}} \mathcal{I}_{\ell_{z}}[\chi_{z}] }{(1 \!-\! s_{\ell_{z}}^{2})} \nonumber
\\
& \, \times \, \bigg\{ \mathcal{F} (s_{\ell_{z}} , \chi_{r}) \!-\! \ell_{z} \frac{\nu}{\sigma_{z}^{2}} \frac{\sigma_{r}^{2}}{\kappa} \, \mathcal{G} (s_{\ell_{z}} , \chi_{r}) \bigg\} \, ,
\label{lambda_odd}
\end{align}
where the prefactor $\beta_{n}$ has been defined in equation~\eqref{beta_odd}.

Equations~\eqref{lambda_even} and~\eqref{lambda_odd} are also important results of this paper, since they allow us to easily assess the strength of the self-gravitating amplification for a thick disc. When effectively computing the thick amplification eigenvalues from equations~\eqref{lambda_even} and~\eqref{lambda_odd}, in order to obtain physically relevant amplification eigenvalues (i.e. satisyfing ${ 0 \!<\! \lambda \!<\! 1 }$ in their definition domain), one has to enforce two additional restrictions. These amount to neglecting the contributions from the vertical action gradients w.r.t. the radial ones, and restricting the sum on resonance vectors only to closed orbits on resonance. Let us now motivate these two restrictions.

The general expression of the response matrix from equation~\eqref{Fourier_M} involves the gradient of the DF w.r.t. to the actions ${ \partial F / \partial \bm{J} }$. As the disc is supposed to be tepid, one may neglect the contributions from ${ \partial F / \partial J_{\phi} }$ w.r.t. ${ \partial F / \partial J_{r} }$ and ${ \partial F / \partial J_{z} }$ (as was assumed in~\cite{FouvryPichonPrunet2015} in the razor-thin case). In addition, we also neglect the contributions from the vertical action gradients w.r.t. the radial ones, as the radial ones are the only ones which remain in the razor-thin limit. In equations~\eqref{lambda_even} and~\eqref{lambda_odd}, this amounts to neglecting any contributions from the reduction function $\mathcal{G}$ and only conserving contributions from the reduction function $\mathcal{F}$. Let us note that in the razor-thin case, the DF's vertical gradient ${ \partial F / \partial J_{z} }$ becomes infinite and yet does not appear in the razor-thin amplification eigenvalues (see equation~\eqref{limit_lambda_thin}). Our first restriction in the computation of the amplification eigenvalues (i.e. neglecting the ${ \partial F / \partial J_{z} }$ gradients) amounts to propagating this razor-thin property to the thickened case.

Moreover, attention should be paid to the fact that in order to compute the collisionless diffusion coefficients ${ D_{\bm{m}} (\bm{J}) }$ from equation~\eqref{Dm_general} as well as the collisional drift and diffusion coefficients ${ A_{\bm{m}} (\bm{J}) }$ and ${ D_{\bm{m}} (\bm{J}) }$ from equations~\eqref{initial_drift} and~\eqref{initial_diff}, one has to evaluate the amplification eigenvalues at the resonant frequency ${ \omega \!=\! \bm{m} \!\cdot\! \bm{\Omega} }$. Therefore, as noted in equation~\eqref{s_resonance}, the shifted dimensionless frequency $s_{\ell_{z}}^{\bm{m}}$ from equation~\eqref{definition_s}, associated with a resonance $\bm{m}$, takes the form
\begin{equation}
s_{\ell_{z}}^{\bm{m}} = m_{r} \!+\! (m_{z} \!-\! \ell_{z}) \frac{\nu}{\kappa} \!+\! \ri \eta \, ,
\label{s_notinteger}
\end{equation}
where a small imaginary part $\eta$ was added. Since the potential is assumed to be non-degenerate, i.e. ${ \nu / \kappa }$ is not a rational number of low order, $s_{\ell_{z}}^{\bm{m}}$, when evaluated for a resonance $\bm{m}$, is an integer only for ${ \ell_{z} \!=\! m_{z} }$. Here, having an integer $s_{\ell_{z}}^{\bm{m}}$ implies that there exists a rotating frame in which the orbit is closed, i.e. in which the considered stars are exactly on resonance. In the razor-thin ${ 2D }$ case, such a rotating frame always exists (see the razor-thin expression~\eqref{definition_s0}), while in the thickened ${ 3D }$ case this is not always possible. As illustrated in figure~\ref{fig_FandG}, the reduction functions ${ s \!\mapsto\! \mathcal{F} (s , \chi) , \mathcal{G} (s , \chi) }$ diverge in the neighbourhood of integers, but are well defined when evaluated for exactly integer values, provided that one adds a small imaginary part $\eta$ as in equation~\eqref{s_notinteger}. In order to never probe the diverging branches of these reductions functions, one should always evaluate these functions for exactly integer values of $s$. Consequently, because $s_{\ell_{z}}^{\bm{m}}$ is an integer only for ${ \ell_{z} \!=\! m_{z} }$, in the general expressions~\eqref{lambda_even} and~\eqref{lambda_odd} of the amplification eigenvalues, we restrict the sum on $\ell_{z}$ solely to this case. Let us note that in the razor-thin case, the dimensionless frequency ${ s \!=\! (\omega \!-\! k_{\phi} \Omega_{\phi}) / \kappa }$, when evaluated at resonance, is always an integer. Our second restriction in the computation of the amplification eigenvalues (i.e. considering only the term ${ \ell_{z} \!=\! m_{z} }$) amounts to propagating this razor-thin property to the thickened case.

To conclude, given to the two previous critical approximations, the expressions of the amplification eigenvalues from equations~\eqref{lambda_even} and~\eqref{lambda_odd}, when computed for a resonance $\bm{m}$, generically becomes
\begin{equation}
\!\lambda_{\bm{m}} \!( \!J_{\phi} , \!k_{r} , \! k_{z} \!) \!=\! \frac{2 \pi G \Sigma \gamma_{\bm{m}}^{2}}{h \kappa^{2} \! (1 \!\!+\!\! (k_{z} / k_{r}\!)^{2}\!)} \frac{ \re^{- \chi_{z}} \mathcal{I}_{m_{z}} \![\chi_{z}\!]}{(1 \!-\! m_{r}^{2})} \mathcal{F} (\!m_{r} , \! \chi_{r}\!) \, ,
\label{lambda_simple}
\end{equation}
where we introduced the numerical prefactor $\gamma_{\bm{m}}$ as
\begin{equation}
\gamma_{\bm{m}} (J_{\phi} , k_{r} , k_{z})= 
\begin{cases}
\begin{aligned}
\displaystyle & \alpha (J_{\phi} , k_{r} , k_{z}) && \text{if} \;\;\; m_{z} \; \text{even} \, ,
\\
\displaystyle & \beta (J_{\phi} , k_{r} , k_{z}) && \text{if} \;\;\; m_{z} \; \text{odd} \, .
\end{aligned}
\end{cases}
\label{definition_gamma}
\end{equation}
The general rewriting from equation~\eqref{lambda_simple} applies in the same manner to both symmetric and antisymmetric vertical resonances. Notice that the approximated amplification eigenvalues from equation~\eqref{lambda_simple} remain fully compatible with the discussion from Appendix~\ref{sec:appendixMdiagonal}, where we justified that the system's response matrix can be assumed to be diagonal. Finally, let us note that these restrictions on the computation of the amplification eigenvalues were used in all the numerical applications presented in section~\ref{sec:application}.

\subsection{A thickened $Q$ factor}
\label{sec_thickQ}

Before evaluating the collisionless and collisional diffusion fluxes, let us now illustrate how the previous amplification eigenvalues allow us to recover the razor-thin WKB amplification eigenvalues obtained in FPP15 and the known WKB dispersion relations for stellar discs~\citep{Kalnajs1965,LinShu1966}. As a second step, we will emphasise how equation~\eqref{lambda_simple} allows for a generalisation of Toomre's $Q$ parameter~\citep{Toomre1964} to thick discs.

\begin{figure}
\begin{center}
\epsfig{file=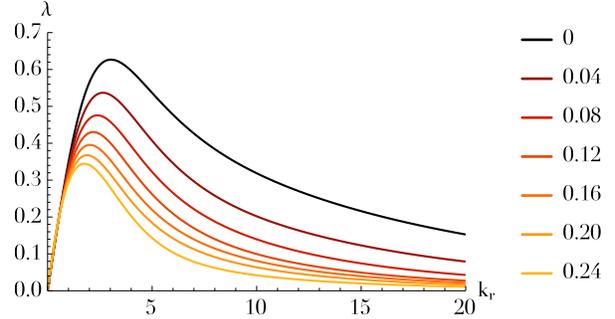,width=0.45\textwidth}
\caption{\small{Illustration of the effect of the disc thickness on the amplification eigenvalues. The disc is the thickened Mestel disc introduced in section~\ref{sec:discmodel}, for the resonance ${ \bm{m} \!=\! \bm{m}_{\rm COR} }$ at the location ${ J_{\phi} \!=\! 2 }$. The different curves correspond to different values of the scale thickness $z_{0}$ from equation~\eqref{definition_rho_star}. For ${ z_{0} \!\neq\! 0 }$, we computed ${ \lambda (k_{r} , k_{z}^{\rm min} (k_{r})) }$ thanks to equation~\eqref{lambda_simple}, while for ${ z \!=\! 0 }$, i.e for the razor-thin case, we computed ${ \lambda_{\rm thin} (k_{r}) }$ following equation~\eqref{limit_lambda_thin}. As expected, one recovers that the thickening of the disc tends to reduce its gravitational susceptibility.
}}
\label{figLambda_thicktothin}
\end{center}
\end{figure}

In the infinitely thin limit, one can only consider resonances associated with ${ m_{z} \!=\! 0 }$, so that only the symmetric basis elements may play a role. Thanks to the quantisation relation illustrated in figure~\ref{fig_quantisation}, notice that except for the fundamental symmetric mode $k_{z, \mathrm{s}}^{1}$, one always has ${ k_{z,\mathrm{s}}^{n} \!>\! \pi / (2h) }$. In the infinitely thin limit, for which ${ h \!\to\! 0 }$, only the fundamental symmetric mode will be relevant for the amplification eigenvalue. In this thin limit, in equation~\eqref{lambda_even}, one can get rid of the degree of freedom w.r.t. $k_{z}^{n}$ and evaluate the symmetric amplification eigenvalue in ${ (k_{r} , k_{z , \rm s}^{1} (k_{r} , h)) \!\simeq\! (k_{r} ,\! \sqrt{k_{r} / h})}$, thanks to equation~\eqref{kz1_even}. Equation~\eqref{lambda_simple} then reads
\begin{equation}
\lambda (\omega , k_{\phi} , k_{r} , h) = \frac{2 \pi G \alpha_{1}^{2} \Sigma k_{r}}{\kappa^{2} (1 \!+\! k_{r} h)} \frac{\re^{- \chi_{z}} \mathcal{I}_{0} [\chi_{z}]}{(1 \!-\! s^{2})} \mathcal{F} (s , \chi_{r}) \, ,
\label{lambda_thin_even_II}
\end{equation}
where the prefactor $\alpha_{1}$ was introduced in equation~\eqref{alpha_even} and is a function of ${ k_{z , \rm s}^{1} h \!=\! \sqrt{k_{r} h} }$, so that ${ \lim_{\rm thin} \alpha_{1} \!=\! 1 }$. In equation~\eqref{lambda_thin_even_II}, we also introduced the dimensionless frequency $s$ as
\begin{equation}
s = \frac{\omega \!-\! k_{\phi} \Omega_{\phi}}{\kappa} \, .
\label{definition_s0}
\end{equation}
Finally, $\chi_{z}$, defined in equation~\eqref{definition_chi}, is only a function of $k_{r}$ and $h$, and reads ${ \chi_{z} \!=\! (\sigma_{z}^{2} k_{r})/(\nu^{2} h) }$. When studying the infinitely thin limit, remember that the physical height ${ \sigma_{z} / \nu }$ and the cavity size $h$ are directly related. Indeed, as detailed in equation~\eqref{link_sigmaz_z0_nu}, given Jeans equation, one has
\begin{equation}
\frac{\sigma_{z}}{\nu} = c_{2} \, h \, ,
\label{definition_c2}
\end{equation}
where $c_{2}$ is a dimensionless constant. For the mean Spitzer density profile introduced in equation~\eqref{definition_rho_star}, one immediately has ${ c_{2} \!=\! 1/\sqrt{2} }$. One can write ${ \chi_{z} \!=\! c_{2}^{2} k_{r} h }$, and has ${ \lim_{\rm thin} \chi_{z} \!=\! 0 }$. Starting from equation~\eqref{lambda_thin_even_II}, since ${ \lim_{\rm thin} \alpha_{1} \!=\! 1 }$ and ${ \lim_{\rm thin} \chi_{z} \!=\! 0 }$, one immediately recovers in the limit of an infinitely thin disc the known amplification eigenvalues of razor-thin discs (see FPP15) as
\begin{equation}
\lim\limits_{\rm thin} \lambda_{\rm sym}  =  \frac{2 \pi G \Sigma |k_{r}|}{\kappa^{2} (1 \!-\! s^{2})} \mathcal{F} (s , \chi_{r}) \, .
\label{limit_lambda_thin}
\end{equation}
This result demonstrates how the thick WKB basis introduced in equation~\eqref{definition_psi_p} is fully consistent with the known razor-thin results. Using the numerical values from the thickened Mestel disc introduced in section~\ref{sec:discmodel}, this property is illustrated in figure~\ref{figLambda_thicktothin}.

\begin{figure}
\begin{center}
\epsfig{file=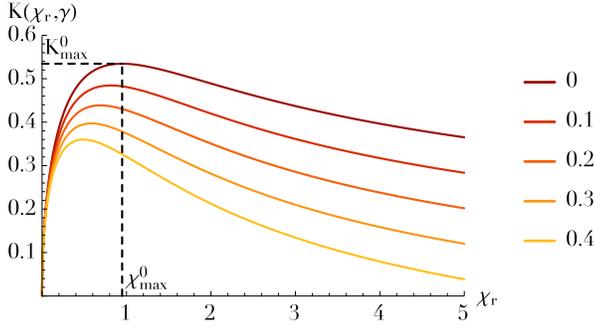,width=0.45\textwidth}
\caption{\small{Behaviour of the structure function ${ \chi_{r} \!\mapsto\! K (\chi_{r} , \gamma) }$ from equation~\eqref{definition_K} for various values of $\gamma$, as defined in equation~\eqref{definition_K}. The razor-thin case corresponds to ${ \gamma \!=\! 0 }$.
}}
\label{figQthick_K}
\end{center}
\end{figure}

Equation~\eqref{lambda_thin_even_II} can now be used to study how Toomre's Q factor~\citep{Toomre1964} gets modified by the thickening of the disc, i.e. by a non-zero value of $h$. Let us recall that $Q$ is a parameter such that ${ Q \!>\! 1 }$ ensures the stability of the disc w.r.t. local axisymmetric tightly wound perturbations. As only stability w.r.t. tightly wound axisymmetric modes is considered, we may first impose ${ k_{\phi} \!=\! 0 }$. Here, we place ourselves at the stability limit given by ${ \omega \!=\! 0 }$, so that ${ s \!=\! 0 }$, and seek a criterion on the disc's parameters such that there exists no ${ k_{r} \!>\! 0 }$ for which ${ \lambda (k_{r} , h) \!=\! 1 }$, i.e. such that the disc is stable. In this context, equation~\eqref{lambda_thin_even_II} immediately takes the form
\begin{align}
\lambda (k_{r} , h) & \, = \frac{2 \pi G \Sigma k_{r}}{\kappa^{2}} \mathcal{F} (0 , \chi_{r}) \bigg\{ \frac{\alpha_{1}^{2} }{1 \!+\! k_{r} h } \re^{-\chi_{z}}  \mathcal{I}_{0} [\chi_{z}]\bigg\} \nonumber 
\\
& \, = \frac{2 \pi G \Sigma k_{r}}{\kappa^{2}} \mathcal{F} (0 , \chi_{r}) \bigg\{ 1 \!-\! \bigg[ \frac{2}{3} \!+\! c_{2}^{2} \bigg] k_{r} h \bigg\} \nonumber 
\\
& \, = \frac{2 \pi G \Sigma}{\kappa \sigma_{r}} \, K (\chi_{r} , \gamma) \, ,
\label{lambda_thin_Q}
\end{align}
where the second line of the previous equation has been obtained using a series development at first order w.r.t. ${ k_{r} h \!\ll\! 1 }$, by expressing $\alpha_{1}$ and $\chi_{z}$ as function of ${ k_{r} h }$. As expected, one recovers that adding a finite thickness to the disc tends to reduce the amplification eigenvalues. In equation~\eqref{lambda_thin_Q}, to shorten the notations, we introduced the parameter ${ \gamma \!=\! [ \frac{2}{3} \!+\! c_{2}^{2} ] (h /\kappa) / \sigma_{r} }$, and defined the structure function ${ K (\chi_{r} , \gamma) }$ as
\begin{equation}
K (\chi_{r} , \gamma) = \frac{1}{\sqrt{\chi_{r}}} \bigg[ 1 \!-\! \re^{- \chi_{r}} \mathcal{I}_{0} [\chi_{r}] \bigg] \bigg[ 1 \!-\! \gamma \sqrt{\chi_{r}} \bigg] \, .
\label{definition_K}
\end{equation}
The shape of the function ${ \chi_{r} \!\mapsto\! K (\chi_{r} , \gamma) }$ is illustrated in figure~\ref{figQthick_K}.
In order to obtain a simple asymptotic expression of a thick stability parameter, one must then study ${ K_{\rm max} (\gamma) }$, the maximum of the function ${ \chi_{r} \!\mapsto\! K (\chi_{r} , \gamma) }$ as a function of $\gamma$. For ${ \gamma \!=\! 0 }$, i.e. for the razor-thin case, ${ K_{\rm max}^{0} \!\simeq\! 0.534 }$ reached for ${ \chi_{\rm max}^{0} \!\simeq\! 0.948 }$. A first order expansion in $\gamma$ yields
\begin{align}
K_{\rm max} (\gamma) & \, \simeq K_{\rm max}^{0} \bigg[ 1 \!-\! \gamma \sqrt{\chi_{\rm max}^{0}} \bigg] \nonumber
\\
& \, \simeq K_{\rm max}^{0} \, \re^{- \gamma \sqrt{\chi_{\rm max}^{0} }} = K_{\rm max}^{\rm approx.} (\gamma) \, ,
\label{Kmax_gamma_Q}
\end{align}
which is found to approximate well ${ K_{\rm max} (\gamma) }$ for ${ \gamma \!\lesssim\! 2 }$. Given the maximum ${ K_{\rm max}^{\rm approx.} (\gamma) }$, the expression of the thickened $Q_{\rm thick}$ parameter follows from equation~\eqref{lambda_thin_Q} and reads
\begin{equation}
Q_{\rm thick} \, = Q_{\rm thin} \, \re^{- \gamma \sqrt{\chi_{\rm max}^{0}}} = Q_{\rm thin} \exp \!\bigg[\! 1.61 \frac{\sigma_{z} / \nu}{\sigma_{r}/\kappa} \!\bigg] \, .
\label{Qthick}
\end{equation}
where the relation~\eqref{definition_c2} was used to rewrite $h$ as a function of ${ \sigma_{z} / \nu }$, given the value ${ c_{2} \!=\! 1/\sqrt{2} }$. Equation~\eqref{Qthick} involves the razor-thin stability parameter $Q_{\rm thin}$~\citep{Toomre1964} reading
\begin{equation}
Q_{\rm thin} = \frac{\sigma_{r} \kappa}{3.36 \, G \, \Sigma} \, .
\label{Qthin}
\end{equation}
As expected, increasing the thickness of the disc leads to larger $Q$ values, and therefore to more stable discs, via an exponential boost in the ratio of the vertical to radial scale heights. Note that expression~\eqref{Qthick} which was obtained through the computation of the response matrix eigenvalues using thick WKB basis elements is fairly general and is not specific to the Spitzer mean density profile from equation~\eqref{definition_rho_star}. When considering a different mean density profile, one only has to change accordingly the value of the constant $c_{2}$ from equation~\eqref{definition_c2}, which relates the size of the mean density profile to the sharp cavity introduced in equation~\eqref{definition_w}.
Let us now discuss how this relates to previous results.
A few authors have tackled the question of characterising the stability of thickened stellar discs (see~\cite{Romeo1992} and references therein). The most reliable and self-consistent analysis is the one of~\cite{Vandervoort1970}, which investigates density waves in thickened stellar discs. This approach is based on the collisionless Boltzmann equation limited to even vertical perturbations, and relies on the assumption of the existence of an adiabatic invariant $J_{z}$, which allows for the description of the vertical motion of the stars. Written with our current notations,~\citet[][his equation~77]{Vandervoort1970} obtains amplification eigenvalues of the form
\begin{equation}
\lambda_{\rm V} = \frac{2 \pi G \Sigma |k_{r}|}{\kappa^{2} (1 \!-\! s^{2})} \mathcal{F} (s , \chi_{r}) \, Q_{\rm V}^{-1} ( k_{r} h ) \, ,
\label{lambda_Vandervoort}
\end{equation}
where figure~\ref{fig_choice_h} was used to relate $h$ and $z_{0}$. 
In equation~\eqref{lambda_Vandervoort}, ${ Q_{\rm V} (k_{r} h) }$ is a non trivial function, which may be computed via variational principles. Similarly, in our present formalism, starting from equation~\eqref{lambda_thin_even_II}, equation~\eqref{lambda_Vandervoort} takes the form
\begin{equation}
\lambda_{\rm F} = \frac{2 \pi G \Sigma |k_{r}|}{\kappa^{2} (1 \!-\! s^{2})} \mathcal{F} (s , \chi_{r}) \, Q_{\rm F}^{-1} (k_{r} h) \, ,
\label{lambda_like_Vandervoort}
\end{equation}
where the function ${ Q_{\rm F} (k_{r} h) }$ is defined as
\begin{equation}
Q_{\rm F} (k_{r} h) = \frac{1 \!+\! k_{r} h}{\alpha_{1}^{2} \, \re^{- \chi_{z}} \, \mathcal{I}_{0} [\chi_{z}] } \, ,
\label{definition_QF}
\end{equation}
where one should pay attention to the fact that $Q_{\rm F}$ only depends on the value of ${ k_{r} h }$. Let us note that he expression of the correction function $Q_{\rm F}$ is explicit and was obtained here by directly estimating the eigenvalues of the response matrix from equation~\eqref{Fourier_M} using the thickened WKB basis elements from equation~\eqref{definition_psi_p}. Thanks to the values from Table~$1$ in~\cite{Vandervoort1970}, which provides approximate values for the function ${ x \!\mapsto\! Q_{\rm V} (x) }$, the behaviours of the functions $Q_{\rm V}$ and $Q_{\rm F}$ can straightforwardly be compared. These functions are found to agree well on the range ${ 0 \!\leq\! k_{r} h \!\leq\! 5 }$.

\subsection{Collisionless orbital diffusion}
\label{sec:FPWKB}

Thanks to the estimation of the system's amplification eigenvalues, one may now estimate the collisionless diffusion coefficients from equation~\eqref{Dm_general}.
In order to shorten the notations, the WKB basis elements from equation~\eqref{definition_psi_p} will be written as
\begin{equation}
\psi^{(p)} = \psi^{[k_{\phi}^{p} , k_{r}^{p} , R_{0}^{p} , n_{p}]} \, .
\label{short_notations_psi}
\end{equation}
Assuming, as in equation~\eqref{Matrix_diagonal}, that the response matrix is diagonal, we may rewrite it under the form ${ \widehat{\mathbf{M}}_{pq} \!=\! \lambda_{p} \delta_{p}^{q} }$. The diffusion coefficients from equation~\eqref{Dm_general} are then given by
\begin{equation}
D_{\bm{m}} (\bm{J}) \!=\! \frac{1}{2}  \sum_{p , q} \!\psi_{\bm{m}}^{(p)} \!(\bm{J}) \, \psi_{\bm{m}}^{(q) *} \!(\bm{J}) \frac{1}{1 \!\!-\!\! \lambda_{p}} \frac{1}{1 \!\!-\!\! \lambda_{q}} \widehat{\mathbf{C}}_{pq} (\bm{m} \!\cdot\! \bm{\Omega}) \, ,
\label{Dm_initial}
\end{equation}
where ${ \widehat{\mathbf{C}}_{pq} }$, as defined in equation~\eqref{definition_C_omega}, corresponds to the cross-correlation between the basis coefficients $\widehat{b}_{p}$ and $\widehat{b}_{q}$. One should note that the Fourier transformed basis elements from equation~\eqref{psi_m_even} involve a $\delta_{m_{z}}^{\rm even}$ (resp. $\delta_{m_{z}}^{\rm odd}$) for the symmetric (resp. antisymmetric) elements. Therefore, in equation~\eqref{Dm_initial}, since $\psi_{\bm{m}}^{(p)}$ and $\psi_{\bm{m}}^{(q)}$ are evaluated for the same resonant vector $\bm{m}$, the diffusion coefficients do not couple the symmetric and antisymmetric basis elements. To estimate $D_{\bm{m}}$, depending on whether $m_{z}$ is even (resp. odd), one only has to consider the symmetric (resp. antisymmetric) basis elements. As was done in section~\ref{sec:thickbasiselements}, let us now restrict ourselves to the symmetric case, while the very similar antisymmetric case is detailed in Appendix~\ref{sec:appendixantisymDm}. Following the same approach as in FPP15, one may first express the basis coefficients $\widehat{b}_{p}$ as a function of the external perturbation $\psi^{\rm e}$. After some calculation, one obtains
\begin{equation}
\widehat{b}_{p} (\omega) = \frac{(k_{r}^{p})^{2} \!+\! (k_{z}^{p})^{2}}{4 \pi G} \frac{\mathcal{A}_{p} R_{0}^{p}}{(\pi \sigma)^{1/4}} (2 \pi)^{2} \widehat{\psi^{\rm e}}_{\! m_{\phi} , k_{r}^{p} , k_{z}^{p}} [R_{0}^{p} , \omega] \, ,
\label{bp_psi_ext}
\end{equation}
where we used the shortening notation ${ k_{z}^{p} \!=\! k_{z}^{n_{p}} }$. In equation~\eqref{bp_psi_ext}, the external potential $\widehat{\psi^{\rm e}}$ has undergone three transformations: (i) an azimuthal Fourier transform of indice $m_{\phi}$, (ii) a local radial Fourier transform centred around $R_{0}^{p}$ at the frequency $k_{r}^{p}$, and (iii) an even-restricted vertical Fourier transform on the scale $h$ at the frequency $k_{z}^{p}$. These three transforms are defined as
\begin{align}
\!\!\! & {\rm (i)\!:} && \!\!\!\!\! f_{m_{\phi}} \!=\! \frac{1}{2 \pi} \!\! \int \!\! \mathrm{d} \phi \, f [\phi] \, \re^{- \ri m_{\phi} \phi} \, , \nonumber
\\
\!\!\! & {\rm (ii)\!:} && \!\!\!\!\! f_{k_{r}} [R_{0}] \!=\! \frac{1}{2 \pi} \!\! \int \!\! \mathrm{d} R \, \re^{- \ri k_{r} (R - R_{0})} \exp \!\!\bigg[\!\! -\! \frac{(\!R \!-\! R_{0}\!)^{2}}{2 \sigma^{2}} \!\bigg] f[R]  \, , \nonumber
\\
\!\!\! & {\rm (iii)\!:} && \!\!\!\!\! f_{k_{z}} \!=\! \!\! \int_{-h}^{+ h} \!\!\!\!\! \mathrm{d} z \, \cos (k_{z} z) \, f [z] \, . 
\label{definition_FT}
\end{align}
By disentangling the sums on $p$ and $q$ in equation~\eqref{Dm_initial}, one may rewrite the diffusion coefficients as
\begin{equation}
D_{\bm{m}}^{\rm sym} (\bm{J}) = \delta_{m_{z}}^{\rm even} \bigg< \frac{1}{2 \pi} \!\! \int \!\! \mathrm{d} \omega' \, g (\bm{m} \!\cdot\! \bm{\Omega}) \, g^{*} (\omega ') \bigg> \, ,
\label{Dm_short}
\end{equation}
where the function ${ g(\omega) }$ is defined as
\begin{equation}
\!\! g (\omega) \!=\! \frac{2 \pi}{2 h} \!\!\!\!\! \sum_{k_{r}^{p} , R_{0}^{p} , n_{p}} \!\!\!\!\!\! g_{s} (k_{r}^{p} , R_{0}^{p} , k_{z}^{p} , \omega \!) \, \re^{\ri k_{r}^{p} (R_{\rg} - R_{0}^{p})} \mathcal{G}_{r} \!( R_{\rg} \!-\! R_{0}^{p}) \, .
\label{definition_g_FP}
\end{equation}
In equation~\eqref{definition_g_FP}, one should note that the sum on $k_{\phi}^{p}$ has been executed thanks to the Kronecker delta from equation~\eqref{psi_m_even}. In equation~\eqref{definition_g_FP}, ${ \mathcal{G}_{r} (R) \!=\! {1}/{\sqrt{2 \pi \sigma^{2}}} \re^{-R^{2}\!/( 2 \sigma^{2} ) } }$ is a normalised Gaussian of width $\sigma$, and $g_{s}$ encompasses all the slow dependences of the diffusion coefficients w.r.t. the radial position so that
\begin{align}
g_{s} (k_{r}^{p} , R_{0}^{p} , k_{z}^{p} , \omega) =  \, & \mathcal{J}_{m_{r}} \!\bigg[\! \sqrt{\tfrac{2 J_{r}}{\kappa}} k_{r}^{p} \!\bigg] \mathcal{J}_{m_{z}} \!\bigg[\! \sqrt{\tfrac{2 J_{z}}{\nu}} k_{z}^{p} \!\bigg] \nonumber
\\
& \, \times \, \frac{\alpha_{p}^{2}}{1 \!-\! \lambda_{p}} \widehat{\psi^{\rm e}}_{\! m_{\phi} , k_{r}^{p} , k_{z}^{p}} [R_{0}^{p} , \omega] \, .
\label{definition_gs}
\end{align}
One can note that in the discrete sums from equation~\eqref{definition_g_FP}, the basis elements are separated by step distances ${ \Delta k_{r} }$ and ${ \Delta R_{0} }$, so that ${ \Delta k_{r} \!=\! n_{k} \Delta k_{r} }$ and ${ R_{0}^{p} \!=\! R_{\rg} \!+\! n_{r} \Delta R_{0} }$. As in FPP15, in order to cancel out the rapidly evolving complex exponential from equation~\eqref{definition_g_FP}, one can straightforwardly show that the basis elements must satisfy a critical sampling condition~\citep{Gabor1946,Daubechies1990} of the form
\begin{equation}
\Delta R_{0} \Delta k_{r} = 2 \pi \, .
\label{critical_sampling}
\end{equation}
Given these step distances, and using Riemann sum formula, equation~\eqref{definition_g_FP} may be rewritten with continuous integrals w.r.t. the $R_{0}^{p}$ and $k_{r}^{p}$ variables. As the Gaussian ${ \mathcal{G}_{r} (R_{\rg} \!-\! R_{0}^{p}) }$ is sufficiently peaked and correctly normalised, it may be replaced by ${ \delta_{\rm D} (R_{\rg} \!-\! R_{0}^{p}) }$. Therefore, equation~\eqref{definition_g_FP} becomes
\begin{equation}
g (\omega) = \frac{1}{2 h} \! \sum_{n_{p}} \!\!\!\ \int \!\! \mathrm{d} k_{r}^{p} \, g_{s} (k_{r}^{p} , R_{\rg} , k_{z}^{p} , \omega) \, ,
\label{g_continuous}
\end{equation}
where one must note that there remains a sum on the index $n_{p}$. At this stage, there are two strategies. On the one hand, one can either assume the disc to be sufficiently thick so as to replace the sum on $n_{p}$ in equation~\eqref{g_continuous} by a continuous integral over $k_{z}$. On the other hand, in the limit of a thin disc, one should keep the discrete sum from equation~\eqref{g_continuous}. In the upcoming calculations, we will follow the first approach. Appendix~\ref{sec:fromthicktothin} details how one should proceed with the second approach, shows that these two approaches are fully consistent one with another, and also fully recovers the razor-thin limit from FPP15. As noted in equation~\eqref{step_distance_kz}, for a sufficiently thick disc, one may assume the distance between successive quantised $k_{z}$ frequencies to be of the order ${ \Delta k_{z} \!\simeq\! \pi / h }$. Provided that ${ \Delta k_{z} }$ is small compared to the typical scale of variation of the function ${ k_{z} \!\mapsto\! g_{s} (k_{z}) }$, one may use once again Riemann sum formula, to rewrite equation~\eqref{g_continuous} as
\begin{equation}
g (\omega) = \!\frac{1}{2 \pi} \!\! \int \!\! \mathrm{d} k_{r}^{p} \mathrm{d} k_{z}^{p} \, g_{s} (k_{r}^{p} , R_{\rg} , k_{z}^{p} , \omega) \, .
\label{g_ultra_continuous}
\end{equation}
Let us now define the autocorrelation $\widehat{C}_{\psi^{\rm e}}$ of the external perturbations as
\begin{align}
\widehat{C}_{\psi^{\rm e}}  & [ m_{\phi} , \omega , R_{\rg} , k_{r}^{p} , k_{r}^{q} , k_{z}^{p} , k_{z}^{q} ] = \nonumber
\\
& \, \;\;\; \frac{1}{2 \pi} \!\! \int \!\! \mathrm{d} \omega ' \, \big< \widehat{\psi^{\rm e}}_{\! m_{\phi} , k_{r}^{p} , k_{z}^{p}} [R_{\rg} , \omega] \, \widehat{\psi^{\rm e}}^{*}_{\! m_{\phi} , k_{r}^{q} , k_{z}^{q}} [ R_{\rg} , \omega '] \big> \, . 
\label{definition_general_autocorrelation}
\end{align}
One can then rewrite the general expression of the symmetric diffusion coefficients from equation~\eqref{Dm_initial} as
\begin{align}
D_{\bm{m}}^{\rm sym} (\bm{J}) & \, =  \delta_{m_{z}}^{\rm even} \frac{1}{(2 \pi)^{2}} \nonumber
\\
& \times \, \!\! \int \!\! \mathrm{d} k_{r}^{p} \mathrm{d} k_{z}^{p} \, \mathcal{J}_{m_{r}} \!\bigg[\! \sqrt{\tfrac{2 J_{r}}{\kappa}} k_{r}^{p} \!\bigg] \mathcal{J}_{m_{z}} \!\bigg[\! \sqrt{\tfrac{2 J_{z}}{\nu}} k_{z}^{p} \!\bigg] \frac{\alpha_{p}^{2}}{1 \!-\! \lambda_{p}} \nonumber
\\
& \times \, \int \!\! \mathrm{d} k_{r}^{q} \mathrm{d} k_{z}^{q} \mathcal{J}_{m_{r}} \!\bigg[\! \sqrt{\tfrac{2 J_{r}}{\kappa}} k_{r}^{q} \!\bigg] \mathcal{J}_{m_{z}} \!\bigg[\! \sqrt{\tfrac{2 J_{z}}{\nu}} k_{z}^{q} \!\bigg] \frac{\alpha_{q}^{2}}{1 \!-\! \lambda_{q}} \nonumber
\\
& \times \, \widehat{C}_{\psi^{\rm e}} [m_{\phi} , \bm{m} \!\cdot\! \bm{\Omega} , R_{\rg} , k_{r}^{p} , k_{r}^{q} , k_{z}^{p} , k_{z}^{q} ] \, .
\label{Dm_autocorrelation_even}
\end{align}
The antisymmetric equivalent of equation~\eqref{Dm_autocorrelation_even} is derived in equation~\eqref{Dm_autocorrelation_odd}. Assuming some stationarity properties on the stochasticity of the external perturbations, one may then further simplify equation~\eqref{Dm_autocorrelation_even}.
As in FPP15, we suppose that the external perturbations are also spatially quasi-stationary so that
\begin{align}
\big< \psi^{\rm e}_{m_{\phi}} & \, [R_{1} ,  z_{1} , t_{1}] \, \psi^{\rm e *}_{m_{\phi}} [R_{2} , z_{2} , t_{2}] \big> = \nonumber
\\
& \;\; \mathcal{C} [m_{\phi} , t_{1} \!\!-\!\! t_{2} ,  (R_{1} \!\!+\!\! R_{2})/2 , R_{1} \!\!-\!\! R_{2} , z_{1} \!\!+\!\! z_{2} , z_{1} \!\!-\!\! z_{2} ] \, ,
\label{assumption_autocorrelation}
\end{align}
where the dependences w.r.t. ${ (R_{1} \!+\! R_{2})/2 }$ and ${ z_{1} \!+\! z_{2} }$ are supposed to be slow. As demonstrated in Appendix~\ref{sec:appendixautocorrelation}, one can then show that
\begin{align}
\bigg<\!  & \, \widehat{\psi^{\rm e}}_{\! m_{\phi}, k_{r}^{1} , k_{z}^{1}}  [R_{\rg} , \omega_{1}] \, \widehat{\psi^{\rm e}}_{\! m_{\phi} , k_{r}^{2} , k_{z}^{2}}^{*} [R_{\rg} , \omega_{2}] \!\bigg> =  2 \pi^{2} \delta_{\rm D} (\omega_{1} \!-\! \omega_{2}) \nonumber
\\
& \;\;\;\;\;\;\;  \times \, \delta_{\rm D} (k_{r}^{1} \!-\! k_{r}^{2}) \, \delta_{\rm D} (k_{z}^{1} \!-\! k_{z}^{2}) \, \widehat{\mathcal{C}} [m_{\phi} , \omega_{1} , R_{\rg} , k_{r}^{1} , k_{z}^{1}] \, .
\label{diagonal_autocorrelation}
\end{align}
Thanks to this autocorrelation diagonalised both in $\omega$, $k_{r}$ and $k_{z}$, the expression of the symmetric diffusion coefficients from equation~\eqref{Dm_autocorrelation_even} becomes
\begin{align}
D_{\bm{m}}^{\rm sym} (\bm{J}) = & \, \delta_{m_{z}}^{\rm even} \frac{\pi}{(2 \pi)^{2}} \!\! \int \!\! \mathrm{d} k_{r}^{p} \mathrm{d} k_{z}^{p} \, \mathcal{J}_{m_{r}}^{2} \!\!\bigg[\! \sqrt{\!\tfrac{2 J_{r}}{\kappa}} k_{r}^{p} \!\bigg] \mathcal{J}_{m_{z}}^{2} \!\!\bigg[\! \sqrt{\!\tfrac{2 J_{z}}{\nu}} k_{z}^{p} \!\bigg] \nonumber
\\
& \times \, \bigg[\! \frac{\alpha_{p}^{2}}{1 \!-\! \lambda_{p}} \!\bigg]^{2} \widehat{\mathcal{C}} [m_{\phi} , \bm{m} \!\cdot\! \bm{\Omega} , R_{\rg} , k_{r}^{p} , k_{z}^{p}] \, .
\label{Dm_diagonalised_even}
\end{align}
Equation~\eqref{Dm_diagonalised_even}, along with its antisymmetric equivalent from equation~\eqref{Dm_diagonalised_odd}, are the main results of this section.
 As in FPP15, equation~\eqref{Dm_diagonalised_even} may be further simplified thanks to the so-called approximation of the small denominators. This amounts to focusing on the contributions from the waves that yield the maximum amplification. One therefore assumes that the function ${ (k_{r} , k_{z}) \!\mapsto\! \lambda (k_{r} , k_{z}) }$, in its allowed domain (i.e. ${ k_{z} \!\geq\! k_{z}^{1} (k_{r}) }$, see figure~\ref{fig_quantisation}) reaches a well-defined maximum ${ \lambda_{\rm max} (R_{\rg} , \omega) }$ for ${ (k_{r} , k_{z}) \!=\! (k_{r}^{\rm max} , k_{z}^{\rm max}) }$. One may then define the neighbouring region $\mathcal{V}_{\rm max} \!=\! \{ (k_{r} , k_{z} ) \, \big| \, \lambda(k_{r} , k_{z}) \!\geq\! \lambda_{\rm max}/2 \}$, and its area $| \mathcal{V}_{\rm max} |$. The previous expression of the diffusion coefficients can then straightforwardly be approximated as
\begin{align}
D_{\bm{m}}^{\rm sym} (\bm{J}) = & \, \delta_{m_{z}}^{\rm even} \frac{\pi |\mathcal{V}_{\rm max}|}{(2 \pi)^{2}} \mathcal{J}_{m_{r}}^{2} \!\!\bigg[\! \sqrt{\!\tfrac{2 J_{r}}{\kappa}} k_{r}^{\rm max} \!\bigg] \mathcal{J}_{m_{z}}^{2} \!\!\bigg[\! \sqrt{\!\tfrac{2 J_{z}}{\nu}} k_{z}^{\rm max} \!\bigg] \nonumber
\\
& \,  \!\!\!\!\! \times \, \bigg[\! \frac{\alpha_{\rm max}^{2}}{ 1 \!-\!  \lambda_{\rm max}} \!\bigg]^{2}\,  \widehat{\mathcal{C}} [m_{\phi} , \bm{m} \!\cdot\! \bm{\Omega} , R_{\rg} , k_{r}^{\rm max} , k_{z}^{\rm max} ] \, .
\label{Dm_diagonalised_ASD_even}
\end{align}
One can improve the previous approximation by performing the integrations from equation~\eqref{Dm_diagonalised_even} only for ${ (k_{r} , k_{z} ) \!\in\!  \mathcal{V}_{\rm max} }$. Such a calculation is more numerically demanding but does not alter the principal conclusions drawn in this paper, while ensuring a better estimation of the diffusion flux.

\subsection{Collisional orbital diffusion}
\label{sec:BLWKB}

Relying similarly on the amplification eigenvalues obtained in section~\ref{sec:amplificationeigenvalues}, we may now proceed to the evaluation of the collisional drift and diffusion coefficients from equations~\eqref{initial_drift} and~\eqref{initial_diff}.

\subsubsection{Estimation of the susceptibility coefficients}
\label{sec:estimationsusceptibilitycoefficients}

Let us first estimate the dressed susceptibility coefficients from equation~\eqref{definition_1/D}. Using the shortened notation from equation~\eqref{short_notations_psi}, they read
\begin{equation}
\frac{1}{\mathcal{D}_{\bm{m}_{1} , \bm{m}_{2}} (\bm{J}_{1} , \bm{J}_{2} , \omega)} = \sum_{p} \psi_{\bm{m}_{1}}^{(p)} (\bm{J}_{1}) \bigg[ \frac{1}{1 \!-\! \lambda_{p} (\omega)} \bigg] \, \psi_{\bm{m}_{2}}^{(p) *} (\bm{J}_{2}) \, .
\label{simple_1/D}
\end{equation}
Separating the contributions from symmetric and antisymmetric basis elements, equation~\eqref{simple_1/D} can be rewritten as
\begin{equation}
\frac{1}{\mathcal{D}_{\bm{m}_{1} , \bm{m}_{2}} (\bm{J}_{1} , \bm{J}_{2} , \omega)} \!=\! \!\sum_{p} \!\bigg[\! \frac{\psi_{\bm{m}_{1}}^{\mathrm{s} , (p)} \, \psi_{\bm{m}_{2}}^{\mathrm{s} , (p) *}}{1 \!-\! \lambda_{p}^{\mathrm{s}}}  +  \!\frac{\psi_{\bm{m}_{1}}^{\mathrm{a} , (p)} \, \psi_{\bm{m}_{2}}^{\mathrm{a} , (p) *}}{1 \!-\! \lambda_{p}^{\mathrm{a}}} \!\bigg] \, ,
\label{simple_1/D_2}
\end{equation}
where the superscripts $``\mathrm{s}"$ and $``\mathrm{a}"$ respectively correspond to symmetric and antisymmetric basis elements. It was shown in equations~\eqref{psi_m_even} and~\eqref{psi_m_odd} that a Fourier transformed basis element $\psi_{\bm{m}}^{(p)}$ involves an azimuthal Kronecker symbol $\delta_{m_{\phi}}^{k_{\phi}^{p}}$. Moreover, in the symmetric (resp. antisymmetric) case, it also involves a $\delta_{m_{z}}^{\rm even}$ (resp. $\delta_{m_{z}}^{\rm odd}$). As a consequence, in equation~\eqref{simple_1/D_2}, in order to have non zero susceptibility coefficients, one must necessarily have
\begin{equation}
m_{1}^{\phi} \!=\! m_{2}^{\phi} = k_{\phi}^{p} \;\;\; \text{and} \;\;\; (m_{1}^{z} \!-\! m_{2}^{z}) \; \text{even} \, .
\label{condition_m_1/D}
\end{equation}
Since $m_{1}^{z}$ and $m_{2}^{z}$ must have the same parity, when computing the susceptibility coefficients from equation~\eqref{simple_1/D_2}, depending on the parity of $m_{1}^{z}$, one has to consider the symmetric elements only or the antisymmetric ones only. Before proceeding with the evaluation of the susceptibility coefficients from equation~\eqref{simple_1/D_2}, we will first emphasise a crucial consequence of the localised thick WKB basis from equation~\eqref{definition_psi_p}, which is the restriction to local resonances. This is the matter of the next section.

\subsubsection{Restriction to local resonances}
\label{sec:localresonances}

The Balescu-Lenard drift and diffusion coefficients from equation~\eqref{initial_drift} and~\eqref{initial_diff} involve an integration over the dummy variable $\bm{J}_{2}$. For a given value of $\bm{J}_{1}$, $\bm{m}_{1}$ and $\bm{m}_{2}$, this should be seen as a scan of the entire action space, searching for regions where the resonant condition ${ \bm{m}_{1} \!\cdot\! \bm{\Omega}_{1} \!-\! \bm{m}_{2} \!\cdot\! \bm{\Omega}_{2} \!=\! 0 }$ is satisfied. Because the epicyclic approximation was assumed, the intrinsic frequencies ${ (\Omega_{\phi} , \kappa , \nu) }$ from equations~\eqref{definition_Omega} and~\eqref{definitions_kappa_nu} only depend on the action $J_{\phi}$, which makes the resonance condition simpler. For fixed values of ${ R_{1} \!=\! R_{\rg} (\bm{J}_{1}) }$, $\bm{m}_{1}$ and $\bm{m}_{2}$, one therefore has to find the resonant radii $R_{2}^{r}$ such that the resonance condition ${ f (R_{2}^{r}) \!=\! 0 }$ is satisfied, where ${ f (R_{2}^{r}) }$ is defined as
\begin{equation}
f (R_{2}^{r}) = \bm{m}_{1} \!\cdot\! \bm{\Omega} (R_{1}) \!-\! \bm{m}_{2} \!\cdot\! \bm{\Omega} (R_{2}^{r}) \, .
\label{definition_f_resonance}
\end{equation}
Once these resonant radii have been identified, one may finally rely on the rule for the composition of a Dirac delta and a function which reads
\begin{equation}
\delta_{\rm D} (f (x)) = \sum_{y \in Z_{f}} \!\! \frac{\delta_{\rm D} (x \!-\! y)}{| f' (y) |} \, ,
\label{composition_delta}
\end{equation}
where ${ Z_{f} \!=\! \big\{ y \, \big| \, f (y) \!=\! 0 \big\} }$. In order to use the expression~\eqref{composition_delta}, one also has to assume that the poles of $f$ are non-degenerate so that
\begin{equation}
\frac{\rd (\bm{m}_{2} \!\cdot\! \bm{\Omega})}{\rd R} \bigg|_{R_{2}^{r}} \!\!\! \neq 0 \, .
\label{non_degenerate_poles}
\end{equation}
As noted in equation~\eqref{condition_m_1/D}, one has ${ m_{1}^{\phi} \!=\! m_{2}^{\phi} }$. As a consequence, the resonance condition from equation~\eqref{definition_f_resonance} takes the form
\begin{equation}
m_{1}^{\phi} \Omega_{\phi}^{1} \!+\! m_{1}^{r} \kappa^{1} \!+\! m_{1}^{z} \nu^{1} \!=\! m_{1}^{\phi} \Omega_{\phi}^{r} \!+\! m_{2}^{r} \kappa^{r} \!+\! m_{2}^{z} \nu^{r} \, ,
\label{rewrite_resonance}
\end{equation}
where we used the shortening notation ${ \Omega_{\phi}^{1} \!=\! \Omega_{\phi}^{1} (R_{1}) }$ and ${ \Omega_{\phi}^{r} \!=\! \Omega_{\phi} (R_{2}^{r}) }$.
Because the Fourier transformed basis elements from equations~\eqref{psi_m_even} and~\eqref{psi_m_odd} involve the narrow radial Gaussian $\mathcal{B}_{R_{0}}$, the relevant resonant radii $R_{2}^{r}$ must necessarily be close to $R_{1}$, so that ${ \Delta R \!=\! R_{2}^{r} \!-\! R_{1} }$ is such that ${ |\Delta R| \!\leq\! \text{(few)} \, \sigma }$. Equation~\eqref{rewrite_resonance} may then be rewritten as
\begin{equation}
\bigg[\! m_{2}^{\phi} \frac{\rd \Omega_{\phi}}{\rd R} \!+\! m_{2}^{r} \frac{\rd \kappa}{\rd R} \!+\! m_{2}^{z} \frac{\rd \nu}{\rd R} \!\bigg] \Delta R \!=\! \bigg[\! m_{1}^{r} \!-\! m_{2}^{r} \!\bigg] \kappa^{1} \!+\! \bigg[\! m_{1}^{z} \!-\! m_{2}^{z} \!\bigg] \nu^{1} \, .
\label{rewrite_resonance_II}
\end{equation}
In the l.h.s. of equation~\eqref{rewrite_resonance_II}, the terms within brackets is non-zero thanks to the assumption from equation~\eqref{non_degenerate_poles} that the resonant poles are simple. Notice that ${ \Delta R }$ is small because of the scale-decoupling approach used in the construction of the WKB basis elements. The r.h.s. of equation~\eqref{rewrite_resonance_II} is discrete in the sense that it is the sum of a multiple of $\kappa$ and of $\nu$. As the disc is supposed to be not too thick, it may be assumed that ${ \nu \!\gg\! \kappa }$. Moreover, as was also shown in equation~\eqref{condition_m_1/D} ${ (m_{1}^{z} \!-\! m_{2}^{z}) }$ is an even number. As a consequence, if ${ (m_{1}^{z} \!-\! m_{2}^{z}) \!\neq\! 0 }$, then
\begin{equation}
\big| (m_{1}^{z} \!-\! m_{2}^{z}) \, \nu (R_{1}) \big| \geq 2 \nu (R_{1}) \gg \big| m_{1}^{r} \!-\! m_{2}^{r} \big| \, \kappa (R_{1})\,,
\label{nu_vs_kappa}
\end{equation}
provided that the resonance vectors $\bm{m}_{1}$ and $\bm{m}_{2}$ are of small order. The l.h.s. of equation~\eqref{rewrite_resonance_II} is therefore small, while its r.h.s. is of the order of ${ \nu (R_{1}) }$. As a consequence, equation~\eqref{rewrite_resonance_II} necessarily implies that ${ m_{1}^{z} \!=\! m_{2}^{z} }$. Equation~\eqref{rewrite_resonance_II} then takes the form
\begin{equation}
\frac{\rd (\bm{m}_{2} \!\cdot\! \bm{\Omega})}{\rd R} \Delta R = \bigg[ m_{1}^{r} \!-\! m_{2}^{r}  \bigg] \, \kappa (R_{1}) \, .
\label{rewrite_resonance_III}
\end{equation}
Similarly, the l.h.s. of equation~\eqref{rewrite_resonance_III} is small because of ${ \Delta R }$, while its r.h.s. is either zero or of the order ${ \kappa (R_{1}) }$. This immediately imposes that both sides of equation~\eqref{rewrite_resonance_III} have to be zero. As a conclusion, the use of the thick WKB basis implies that only local resonances are allowed so that
\begin{equation}
R_{2}^{r} \!=\! R_{1} \;\;\; ; \;\;\; m_{1}^{r} \!=\! m_{2}^{r} \;\;\; ; \;\;\; m_{1}^{z} \!=\! m_{2}^{z} \, .
\label{local_resonances}
\end{equation}
This a crucial consequence of the restriction to the thick WKB basis from equation~\eqref{definition_psi_p}.

\subsubsection{Asymptotic continuous limit}
\label{sec:asymptoticlimit}

One may now evaluate the susceptibility coefficients from equation~\eqref{simple_1/D_2} by restricting ourselves to the cases ${ R_{2} \!=\! R_{1} }$ and ${ \bm{m}_{2} \!=\! \bm{m}_{1} }$. As noted in equation~\eqref{simple_1/D_2}, the symmetric case (i.e. $m_{1}^{z}$ even) and the antisymmetric one (i.e. $m_{1}^{z}$ odd) can be treated separately. The upcoming calculations will be made for the symmetric case, from which the antisymmetric expressions are straightforward to deduce. When writing explicitly the sum on the basis elements, and using the expression~\eqref{psi_m_even} of the Fourier transformed basis elements, equation~\eqref{simple_1/D_2} becomes
\begin{align}
\frac{1}{\mathcal{D}_{\bm{m}_{1} , \bm{m}_{1}} } = & \,  \!\!\! \sum_{k_{r}^{p}, R_{0}^{p} , n_{p}} \!\!\! \frac{G}{R_{0}^{p} h} \frac{1}{(k_{r}^{p})^{2} \!+\! (k_{z}^{p})^{2}}  \frac{1}{\sqrt{\pi \sigma^{2}}} \exp \!\bigg[\! - \frac{(R_{1} \!\!-\! R_{0}^{p})^{2}}{\sigma^{2}} \!\bigg]  \nonumber
\\
& \times \, \frac{\alpha_{p}^{2}}{1 \!-\! \lambda_{p} (\omega)} \,  \mathcal{J}_{m_{1}^{r}} \bigg[\!\! \sqrt{\tfrac{2 J_{r}^{1}}{\kappa_{1}}} \, k_{r}^{p} \!\bigg] \, \mathcal{J}_{m_{1}^{r}} \bigg[\!\! \sqrt{\tfrac{2 J_{r}^{2}}{\kappa_{1}}} \, k_{r}^{p} \!\bigg] \nonumber
\\
& \, \times \,  \mathcal{J}_{m_{1}^{z}} \bigg[\!\! \sqrt{\tfrac{2 J_{z}^{1}}{\nu_{1}}} \, k_{z}^{p} \!\bigg] \, \mathcal{J}_{m_{1}^{z}} \bigg[\!\! \sqrt{\tfrac{2 J_{z}^{2}}{\nu_{1}}} \, k_{z}^{p} \!\bigg] \, .
\label{1/D_full_discrete}
\end{align}
In equation~\eqref{1/D_full_discrete}, the shortened notations ${ 1/\mathcal{D}_{\bm{m}_{1} , \bm{m}_{1}} }$ was introduced for ${ 1 \!/\! \mathcal{D}_{\bm{m}_{1} , \bm{m}_{1}} \!(R_{1} , \!J_{r}^{1} , \!J_{z}^{1} , \!R_{1} , \!J_{r}^{2} , \!J_{z}^{2} , \!\omega) }$, as well as ${ \kappa_{1} \!=\! \kappa (R_{1}) }$, ${ \nu_{1} \!=\! \nu (R_{1}) }$ and ${ k_{z}^{p} \!=\! k_{z}^{n_{p}} }$ One should also note that the sum on $k_{\phi}^{p}$ was executed thanks to the constraint from equation~\eqref{condition_m_1/D}. As in section~\ref{sec:FPWKB}, the next step of the calculation is to replace the discrete sums on $k_{r}^{p}$ and $R_{0}^{p}$ by continuous expressions. To do so, we rely on the step distance from equation~\eqref{critical_sampling} and replace the Gaussian in ${ ( R_{1} \!-\! R_{0}^{p} ) }$ in equation~\eqref{1/D_full_discrete} by ${ \delta_{\rm D} (R_{1} \!-\! R_{0}^{p}) }$. The integration on $R_{0}^{p}$ may then be performed, and equation~\eqref{1/D_full_discrete} becomes
\begin{align}
& \, \frac{1}{\mathcal{D}_{\bm{m}_{1} , \bm{m}_{1}} } \!=\! \frac{G}{2 \pi R_{1} h} \sum_{n_{p}} \! \int \!\!\! \rd k_{r} \frac{1}{k_{r}^{2} \!+\! (k_{z}^{p})^{2}} \frac{\alpha_{p}^{2}}{1 \!-\! \lambda_{p} (\omega)}  \label{1/D_discrete}
\\
& \times \mathcal{J}_{m_{1}^{r}} \bigg[\!\! \sqrt{\!\tfrac{2 J_{r}^{1}}{\kappa_{1}}} \, k_{r} \!\bigg]  \mathcal{J}_{m_{1}^{r}} \bigg[\!\! \sqrt{\!\tfrac{2 J_{r}^{2}}{\kappa_{1}}} \, k_{r} \!\bigg]
\mathcal{J}_{m_{1}^{z}} \bigg[\!\! \sqrt{\!\tfrac{2 J_{z}^{1}}{\nu_{1}}} k_{z}^{p} \!\bigg] \, \mathcal{J}_{m_{1}^{z}} \bigg[\!\! \sqrt{\!\tfrac{2 J_{z}^{2}}{\nu_{1}}} \, k_{z}^{p} \!\bigg] \, .  \nonumber
\end{align}
One must note that in equation~\eqref{1/D_discrete} there still remains a sum on the vertical index $n_{p}$. At this stage of the calculation, there are two possible strategies to complete the evaluation of the susceptibility coefficients. If one assumes the disc to be sufficiently thick, one may replace the sum on $n_{p}$ by a continuous integral over $k_{z}$. Conversely, in the limit of a thin disc, one should keep the discrete sum in equation~\eqref{1/D_discrete}. In the following calculations, the first continuous approach will be pursued. In Appendix~\ref{sec:fromthicktothin}, the second approach is investigated: it is shown that these two approaches are fully consistent one with another, and how the razor-thin limit from FPC15 may be recovered. As noted in equation~\eqref{step_distance_kz}, the distance between two successive quantised $k_{z}$ can be approximated by ${ \Delta k_{z} \!\simeq\! \pi / h }$. Provided that the function present in the r.h.s. of equation~\eqref{1/D_discrete} varies on scales larger than ${ \Delta k_{z} }$, one may use once again the Riemann sum formula to rewrite equation~\eqref{1/D_discrete} as
\begin{align}
& \, \frac{1}{\mathcal{D}_{\bm{m}_{1} , \bm{m}_{1}} } \!=\! \frac{G}{2 \pi^{2} R_{1}} \!\! \int \!\! \rd k_{r} \rd k_{z} \frac{1}{k_{r}^{2} \!+\! k_{z}^{2}} \frac{\alpha_{k_{r} , k_{z}}^{2}}{1 \!-\! \lambda_{k_{r},k_{z}} (\omega)}   \label{1/D_continuous}
\\
& \times \mathcal{J}_{m_{1}^{r}} \bigg[\!\! \sqrt{\!\tfrac{2 J_{r}^{1}}{\kappa_{1}}} \, k_{r} \!\bigg] \mathcal{J}_{m_{1}^{r}} \bigg[\!\! \sqrt{\!\tfrac{2 J_{r}^{2}}{\kappa_{1}}} \, k_{r} \!\bigg] \mathcal{J}_{m_{1}^{z}} \bigg[\!\! \sqrt{\!\tfrac{2 J_{z}^{1}}{\nu_{1}}} \, k_{z} \!\bigg] \mathcal{J}_{m_{1}^{z}} \bigg[\!\! \sqrt{\!\tfrac{2 J_{z}^{2}}{\nu_{1}}} \, k_{z} \!\bigg] \, .  \nonumber
\end{align}
This explicit expression of the dressed susceptibility coefficients is the main result of the present section:
it relates the gravitational susceptibility of the disc to known analytic functions of its actions via a simple regular quadrature.
Following equation~\eqref{Dm_diagonalised_ASD_even}, we may further simplify equation~\eqref{1/D_continuous} by using the so-called approximation of the small denominators, so that it becomes
\begin{align}
& \, \frac{1}{\mathcal{D}_{\bm{m}_{1} , \bm{m}_{1}} } \!=\! \frac{G}{2 \pi^{2} R_{1}} \frac{|\mathcal{V}_{\rm max}|}{(k_{r}^{\rm max})^{2} \!+\! (k_{z}^{\rm max})^{2}} \frac{\alpha_{\rm max}^{2}}{1 \!-\! \lambda_{\rm max}} \mathcal{J}_{m_{1}^{r}} \!\bigg[\!\! \sqrt{\!\tfrac{2 J_{r}^{1}}{\kappa_{1}}}  k_{r}^{\rm max} \!\bigg] \nonumber
\\
&  \times  \mathcal{J}_{m_{1}^{r}} \!\bigg[\!\! \sqrt{\!\tfrac{2 J_{r}^{2}}{\kappa_{1}}} k_{r}^{\rm max} \!\bigg] \mathcal{J}_{m_{1}^{z}} \!\bigg[\!\! \sqrt{\!\tfrac{2 J_{z}^{1}}{\nu_{1}}}  k_{z}^{\rm max} \!\bigg] \mathcal{J}_{m_{1}^{z}} \!\bigg[\!\! \sqrt{\!\tfrac{2 J_{z}^{2}}{\nu_{1}}}  k_{z}^{\rm max} \!\bigg] \, .
\label{1/D_ASD}
\end{align}
One can improve this approximation by rather performing the integrations in equation~\eqref{1/D_continuous} for ${ (k_{r} , k_{z}) \!\in\! \mathcal{V}_{\rm max} }$. This approach is more numerically demanding but allows for a more precise determination of the diffusion flux. Using this improved approximation does not alter the principal conclusions drawn in this paper. Finally, for ${ m_{1}^{z} }$ odd, the antisymmetric analogs of the previous expressions of the susceptibility coefficients are straightforward to obtain through the substitution ${ \alpha \!\to\! \beta }$, introduced in equation~\eqref{beta_odd}, and by considering the antisymmetric amplification eigenvalues from equation~\eqref{lambda_odd}.

\subsubsection{Estimation of the drift and diffusion coefficients}
\label{sec:estimationcoefficients}

The final step of the collisional  calculation is to determine the Balescu-Lenard drift and diffusion coefficients from equations~\eqref{initial_drift} and~\eqref{initial_diff}. Thanks to the restriction to local resonances justified in equation~\eqref{local_resonances}, the sum on $\bm{m}_{2}$ in equations~\eqref{initial_drift} and~\eqref{initial_diff} is only limited to ${ \bm{m}_{2} \!=\! \bm{m}_{1} }$, and using the formula~\eqref{composition_delta}, one may immediately perform the integration on $J_{\phi}^{2}$, which adds a prefactor of the form ${ 1/|\partial (\bm{m}_{1} \!\cdot\! \bm{\Omega}_{1}) / \partial J_{\phi}| }$. Using the shortened notation
\begin{equation}
\frac{1}{(\bm{m}_{1} \!\cdot\! \bm{\Omega}_{1})'} = \frac{1}{\big| \tfrac{\partial}{\partial J_{\phi}} [\bm{m}_{1} \!\cdot\! \bm{\Omega}_{1} ] \big|_{J_{\phi}^{1}}} \, ,
\label{short_omega_derivative}
\end{equation}
one can write the expression of the drift coefficients as
\begin{equation}
A_{\bm{m}_{1}} \!(\!\bm{J}_{1}\!)  \!=\! - \frac{8 \pi^{4} \mu}{(\bm{m}_{1} \!\cdot\! \bm{\Omega}_{1})'}\!\!\! \int \!\! \rd J_{r}^{2} \rd J_{z}^{2} \, \frac{\bm{m}_{1} \!\cdot\! \partial F / \partial \bm{J} (J_{\phi}^{1} , J_{r}^{2} , J_{z}^{2})}{|\mathcal{D}_{\bm{m}_{1} , \bm{m}_{1}} \!(\bm{J}_{1} , \!\bm{J}_{2}  , \!\bm{m}_{1} \!\cdot\! \bm{\Omega}_{1} \!)|^{2}} \, .
\label{final_drift}
\end{equation}
Similarly the diffusion coefficients are given by
\begin{equation}
D_{\bm{m}_{1}} \!(\!\bm{J}_{1}\!)  \!=\! \frac{8 \pi^{4} \mu}{(\bm{m}_{1} \!\cdot\! \bm{\Omega}_{1})'} 
 \!\! \int \!\! \rd J_{r}^{2} \rd J_{z}^{2} \, \frac{F (J_{\phi}^{1} , J_{r}^{2} , J_{z}^{2})}{| \mathcal{D}_{\bm{m}_{1} , \bm{m}_{1}} \!(\bm{J}_{1} , \!\bm{J}_{2}  , \!\bm{m}_{1} \!\cdot\! \bm{\Omega}_{1}\!) |^{2}} \, .
\label{final_diff}
\end{equation}
In equations~\eqref{final_drift} and~\eqref{final_diff}, the susceptibility coefficients are given by equation~\eqref{1/D_continuous}, or equation~\eqref{1/D_ASD} within the approximation of the small denominators (or their antisymmetric analogs depending on the parity of $m_{1}^{z}$).
In particular, they have to be evaluated for ${ J_{\phi}^{2} \!=\! J_{\phi}^{1} }$.
In the case where the DF takes the form of a quasi-isothermal DF as in equation~\eqref{definition_DF_quasi_isothermal} and where the susceptibility coefficients are obtained via the approximation of the small denominators from equation~\eqref{1/D_ASD}, the integrations on $J_{r}^{2}$ and $J_{z}^{2}$ in equations~\eqref{final_drift} and~\eqref{final_diff} may be explicitly computed (see Appendix~C of FPC15 for an illustration in the razor-thin limit). To do so, in addition to equation~\eqref{formula_Gradshteyn}, one relies on the integration formula
\begin{align}
\!\! \int_{0}^{+ \infty} \!\!\!\!\!\! & \, \rd J \, J \, \re^{- a J} \mathcal{J}_{m}^{2} \!\big[ b \sqrt{J} \big] = \frac{1}{a^{2}} \exp \!\bigg[\! - \frac{b^{2}}{2 a} \!\bigg]  \nonumber
\\
& \, \times \bigg\{ \bigg[\! - \frac{b^{2}}{2 a} \!+\! 1 \!+\! |m| \!\bigg] \mathcal{I}_{m} \!\bigg[\! \frac{b^{2}}{2 a} \!\bigg] \!+\! \frac{b^{2}}{2 a} \mathcal{I}_{|m| + 1} \!\bigg[\! \frac{b^{2}}{2 a} \!\bigg] \bigg\} \, .
\label{formula_Grashteyn_II}
\end{align}
We do not detail here these calculations, and only give the final expressions of the drift and diffusion coefficients. Equations~\eqref{final_drift} and~\eqref{final_diff} become
\begin{align}
& A_{\bm{m}_{1}} \!(\!\bm{J}_{1}\!) \!=\! - G_{\bm{m}_{1}}^{(0)} \!(\!\bm{J}_{1}\!) \, G_{\bm{m}_{1}}^{(1)} \!(\!J_{\phi}^{1}\!) \big[ \alpha_{\bm{m}_{1}} \!(\!J_{\phi}^{1}\!) \!-\! \beta_{\bm{m}_{1}}^{r} \!(\!J_{\phi}^{1}\!) \!-\! \beta_{\bm{m}_{1}}^{z} \!(\!J_{\phi}^{1}\!) \big] , \nonumber
\\
& D_{\bm{m}_{1}} \!(\!\bm{J}_{1}\!) \!=\! G_{\bm{m}_{1}}^{(0)} \!(\!\bm{J}_{1}\!) \, G_{\bm{m}_{1}}^{(1)} \!(\!J_{\phi}^{1}\!) \, ,
\label{expression_A_D_integrated}
\end{align}
where we introduced the functions ${ G_{\bm{m}_{1}}^{(0)} \!(\bm{J}_{1}) }$ and ${ G_{\bm{m}_{1}}^{(1)} \!(J_{\phi}^{1}) }$ as
\begin{align}
& \! G_{\bm{m}_{1}}^{(0)} \!(\bm{J}_{1}) \!=\! \frac{8 \pi^{4} \mu}{(\bm{m}_{1} \!\cdot\! \bm{\Omega}_{1})'} \, F^{(0)} \!(J_{\phi}^{1}) \, C_{\bm{m}_{1}}^{\mathcal{D}} \!(\bm{J}_{1}) \, ,  \nonumber
\\
& \! G_{\bm{m}_{1}}^{(1)} \!(J_{\phi}^{1}) \!=\! \frac{\sigma_{r}^{2}}{\kappa_{1}} \mathcal{I}_{m_{1}^{r}} \!\big[ \chi_{r}^{\rm max} \big] \re^{- \chi_{r}^{\rm max}} \frac{\sigma_{z}^{2}}{\nu_{1}} \mathcal{I}_{m_{1}^{z}} \!\big[ \chi_{z}^{\rm max} \big] \re^{- \chi_{z}^{\rm max}} . 
\label{expression_G_integrated}
\end{align}
In equation~\eqref{expression_G_integrated}, we introduced the functions ${ F^{(0)} \!(J_{\phi}^{1}) }$ and ${ C_{\bm{m}_{1}}^{\mathcal{D}} \!(\bm{J}_{1}) }$, so that the quasi-isothermal DF from equation~\eqref{definition_DF_quasi_isothermal} and the susceptiblity coefficients from equation~\eqref{1/D_ASD} read
\begin{align}
& \!\!\!\! F (\bm{J}_{1}) \!=\! F^{(0)} \!(J_{\phi}^{1}) \, \exp \!\!\bigg[\! - \frac{\kappa_{1} J_{r}}{\sigma_{r}^{2}} \!\bigg] \, \exp \!\!\bigg[\! - \frac{\nu_{1} J_{z}}{\sigma_{z}^{2}} \!\bigg] \, ,  \nonumber
\\
& \!\!\!\! \frac{1}{| \mathcal{D}_{\bm{m}_{1} , \bm{m}_{1}}|^{2}} \!=\! C_{\bm{m}_{1}}^{\mathcal{D}} \!(\bm{J}_{1}) \mathcal{J}_{m_{1}^{r}}^{2} \!\!\bigg[\! \sqrt{\!\tfrac{2 J_{r}^{2}}{\kappa_{1}}} k_{r}^{\rm max} \!\bigg] \mathcal{J}_{m_{1}^{z}} \!\!\bigg[\! \sqrt{\tfrac{2 J_{z}^{2}}{\nu_{1}}} k_{z}^{\rm max} \!\bigg] .
\label{expression_F_1/D_integrated}
\end{align}
In equation~\eqref{expression_A_D_integrated}, we also introduced the coefficients ${ \alpha_{\bm{m}_{1}} \!(J_{\phi}^{1}) }$, ${ \beta_{\bm{m}_{1}}^{r} \!(J_{\phi}^{1}) }$ and ${ \beta_{\bm{m}_{1}}^{z} \!(J_{\phi}^{1}) }$ defined as
\begin{align}
& \alpha_{\bm{m}_{1}} \!(J_{\phi}^{1}) = m_{1}^{\phi} \frac{\partial \ln [F^{(0)}]}{\partial J_{\phi}^{1}} \!-\! m_{1}^{r} \frac{\kappa_{1}}{\sigma_{r}^{2}} \!-\! m_{1}^{z} \frac{\nu_{1}}{\sigma_{z}^{2}} \, , \nonumber
\\
& \beta_{\bm{m}_{1}}^{r} \!(J_{\phi}^{1}) = m_{1}^{\phi} \frac{\partial [\kappa_{1} / \sigma_{r}^{2}]}{\partial J_{\phi}^{1}} \frac{\sigma_{r}^{2}}{\kappa_{1}} \gamma_{m_{1}^{r}}^{r} \!(J_{\phi}^{1}) \, ,  \nonumber
\\
& \beta_{\bm{m}_{1}}^{z} \!(J_{\phi}^{1}) = m_{1}^{\phi} \frac{\partial [ \nu_{1} / \sigma_{z}^{2} ]}{\partial J_{\phi}^{1}} \frac{\sigma_{z}^{2}}{\nu_{1}} \gamma_{m_{1}^{z}}^{z} \!(J_{\phi}^{1}) \, ,
\label{definition_alpha_beta_integrated}
\end{align}
where the coefficient ${ \gamma_{m_{1}^{r}}^{r} \!(J_{\phi}^{1}) }$ is defined as
\begin{align}
\gamma_{m_{1}^{r}}^{r} (J_{\phi}^{1}) = \frac{1}{\mathcal{I}_{m_{1}^{r}} \!\big[ \chi_{r}^{\rm max} \big]} \bigg\{\! & \, (- \chi_{r}^{\rm max} \!+\! 1 \!+\! |m_{1}^{r}|) \, \mathcal{I}_{m_{1}^{r}} \!\big[ \chi_{r}^{\rm max} \big]  \nonumber
\\
& \, \!+\! \chi_{r}^{\rm max} \mathcal{I}_{|m_{1}^{r}| + 1} \!\big[ \chi_{r}^{\rm max} \big] \!\bigg\} \, .
\label{definition_gamma_integrated}
\end{align}
Equation~\eqref{definition_gamma_integrated} naturally extends to the definition of ${ \gamma_{m_{1}^{z}}^{z} \!(J_{\phi}^{1}) }$, thanks to the substitutions ${ m_{1}^{r} \!\to\! m_{1}^{z} }$ and ${ \chi_{r}^{\rm max} \!\to\! \chi_{z}^{\rm max} }$. The WKB approximation allowed us therefore to obtain in equation~\eqref{expression_A_D_integrated} explicit expressions for the drift and diffusion coefficients, where all quadratures have been computed.

Finally, let us note that if one assumes the system's DF to be at statistical equilibrium and to take the form of a Boltzmann DF, ${ F (\bm{J}) \!=\! C \, \re^{- \beta H (\bm{J})} }$, then the previous drift and diffusion coefficients are directly connected one to another. Indeed, for such a DF, one has ${ \partial F / \partial \bm{J} \!=\! - \beta \, F \, \bm{\Omega} (J_{\phi}) }$, where one notes that within the epicyclic approximation, the system's intrinsic frequencies $\bm{\Omega}$ only depend on the azimuthal action $J_{\phi}$. The drift coefficients from equation~\eqref{final_drift} may then be computed, and one gets ${ A_{\bm{m}_{1}} (\bm{J}_{1}) \!=\! \bm{m}_{1} \!\cdot\! \bm{\Omega} (J_{\phi}^{1}) \, \beta \, D_{\bm{m}_{1}} (\bm{J}_{1}) }$, which takes the form of a generalised Einstein relation for each resonance. This is a generic property of the Balescu-Lenard equation, which remains true beyond the present WKB approximation~\citep{Chavanis2012}.

The simple and tractable expressions of the drift and diffusion coefficients from equations~\eqref{final_drift} and~\eqref{final_diff} constitute one of the main results of this paper. Let us insist on the fact that the WKB formalism presented in this section is self-contained and that no ad hoc fittings were required. Finally, except for the explicit recovery of the amplification eigenvalues in equation~\eqref{lambda_even}, the previous calculations are not restricted to the quasi-isothermal DF from equation~\eqref{definition_DF_quasi_isothermal}. As a consequence, the collisional drift and diffusion coefficients from equations~\eqref{final_drift} and~\eqref{final_diff} are valid for any tepid disc's DF, provided that the epicyclic angle-action mapping from equation~\eqref{angles_mapping_epi} is applicable.

\section{Application: disc thickening}
\label{sec:application}

Let us now implement the previous thick WKB diffusion equations to get a better grasp of the various resonant processes at play during the secular evolution of a thick disc.
Let us already emphasise that describing self-consistently the secular evolution of a self-gravitating stellar disc is a very challenging task, which raises many difficulties. There exists no generic angle-action coordinates in the thickened geometry, nor appropriate basis elements, nor methods to compute the properties of the disc's collective effects. The previous WKB formalism allows for the simultaneous resolution of all these difficulties, at the cost of additional assumptions, e.g., epicyclic approximation, tightly wound perturbations, etc. The WKB framework appears therefore as a legitimate first step to investigate from first principles the complex dynamics of thickened discs. We present in section~\ref{sec:discmodel} the considered disc model. In section~\ref{sec:thickeningshotnoise} the formalism will first be applied to the formation of vertical ridges in action space found in the numerical experiments of~\cite{SolwaySellwood2012}. We will then consider in section~\ref{sec:timescales} the associated diffusion timescales and discuss the limitations of the WKB framework. In section~\ref{sec:thickeningbar}, we will focus on illustrating the thickening of galactic discs via resonant diffusion induced by central decaying bars, while in section~\ref{sec:GMC} we will consider the effect of the joint evolution of GMCs.

\subsection{The disc model}
\label{sec:discmodel}

In order to setup a typical thick disc, we follow the recent secular simulations of isolated thick discs presented in~\cite{SolwaySellwood2012}, hereafter So12 (specifically, the numerical parameters from the simulation named UCB, keeping only the most massive of its two components).
This simulation is especially relevant for the formalism presented here, since it models an unperturbed isolated stable and stationary thick disc, in which So12 observed the spontaneous appearance of transient spirals seeded by the disc's discreteness, and, only on secular timescales, the formation of a central bar.\footnote{Let us emphasise that this simulation UCB is significantly different from another simulation, M2, also presented in detail in~\cite{SolwaySellwood2012}. Indeed, M2 was tailored to support a ${ m \!=\! 2 }$ unstable spiral mode, in particular via a groove in the disc's DF. It contained a thin disc made of ${ N \!=\! 1.2 \!\times\! 10^{6} }$ particles, and was evolved up to ${ t \!\simeq\! 390 }$. On the other hand, the simulation UCB aimed at studying the effects of multiple transient spirals seeded by the finite number of particles in a quasi-stationary and stable disc. It contained only ${ N \!=\! 2 \!\times\! 10^{5} }$ particles and was evolved up to ${ t \!\simeq\! 3500 }$. Let us highlight the strong differences between the M2 and UCB simulations: unstable vs. stable, spiral mode vs. multiple transient spirals, large $N$ vs. small $N$, short integration time vs. long integration time, and collisionless dynamics vs. collisional dynamics.} The disc considered therein corresponds to a thickened Mestel disc.
\begin{figure}
\begin{center}
\epsfig{file=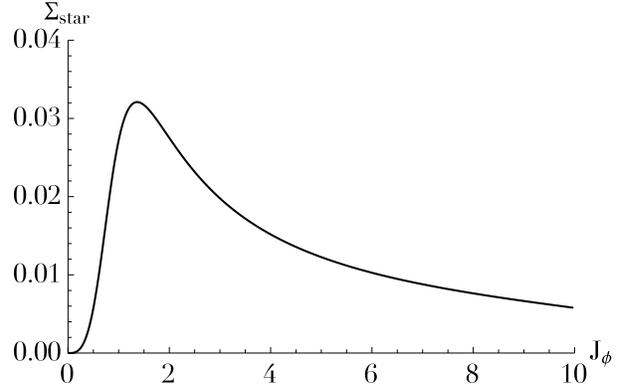,width=0.45\textwidth}
\caption{\small{Shape of the active surface density $\Sigma_{\rm star}$ from equation~\eqref{definition_Sigmastar}. Because of the tapering functions from equation~\eqref{definition_tapering}, the self-gravity of the disc is turned off in its inner and outer regions.
}}
\label{figSigmaStar}
\end{center}
\end{figure}
We start from an infinitely thin Mestel disc of surface density ${ \Sigma_{\rm M} (R) \!=\! V_{0}^{2} / (2 \pi G R)}$, where $V_{0}$ is a constant independent of radius. Assuming a vertical profile shape, one may thicken this surface density $\Sigma_{\rm M}$ to build up a density $\rho_{\rm M}$. Indeed, the ${3D}$ density ${ \rho_{\rm M} (R , z) }$ can be defined as
\begin{equation}
\rho_{\rm M} (R , z) = \Sigma_{\rm M} (R) \frac{1}{4 z_{0} (R)} \, \text{sech}^{2} \bigg[ \frac{z}{2 z_{0} (R)} \bigg] \, ,
\label{definition_rho_star}
\end{equation}
where a Spitzer vertical profile~\citep{Spitzer1942} was used, introducing $z_{0}$ the local thickness of the mean disc. Of course, note that the thickening was defined such that ${ \!\int \! \mathrm{d} z \, \rho_{\rm M} (R, z) \!=\! \Sigma_{\rm M} (R) }$. At this stage, we recall that one could have used alternative vertical profiles, e.g., exponential. Indeed, the results presented thereafter can straightforwardly be applied to different profiles, by adapting accordingly the relations between $h$, $z_{0}$ and ${ \sigma_{z} / \nu }$, obtained in equations~\eqref{definition_w} and~\eqref{link_sigmaz_z0_nu}. Once the total thickened density has been defined, one can then numerically determine the associated potential $\psi_{\rm M}$ via ${ \psi_{\rm M} (\bm{x}) \!=\! - \!\int\! \mathrm{d} \bm{x}_{1}  G \, \rho_{\rm M} (\bm{x}_{1})/|\bm{x} \!-\! \bm{x}_{1} |}$. Relying on the axisymmetry of the system, one obtains
\begin{align}
\psi_{\rm M} (R , z) =& \,  \!\! \int \!\! \mathrm{d} R_{1} \mathrm{d} z_{1} \, \frac{-4 G R_{1} \rho_{\rm M} (R_{1} , z_{1})}{\sqrt{(R \!-\! R_{1})^{2} \!+\! (z \!-\! z_{1})^{2}}} \nonumber
\\
& \, \times \, \mathcal{F}_{\rm ell} \bigg[ \frac{\pi}{2},  - \frac{4 R R_{1}}{(R \!-\! R_{1})^{2} \!+\! (z \!-\! z_{1})^{2}} \bigg] \, ,
\label{link_psi_rho}
\end{align}
where ${ \mathcal{F}_{\rm ell} [\phi,m] }$ is the elliptic integral of the first kind, defined as ${ \mathcal{F}_{\rm ell} [\phi , m] \!=\! \!\! \int_{0}^{\phi} \!\! \mathrm{d} \phi' \, [ 1 \!-\! m \sin^{2} \!(\phi') ]^{-1/2} }$.
Thanks to this numerical estimation of the thickened total potential $\psi_{\rm M}$ of the disc, one may then use equations~\eqref{definition_Rg},~\eqref{definition_Omega} and~\eqref{definitions_kappa_nu} to numerically determine the mapping ${ R_{\rg} \!\mapsto\! J_{\phi} }$ and the intrinsic frequencies $\Omega_{\phi}$, $\kappa$ and $\nu$. This completely characterises the epicyclic mapping to the angle-action coordinates presented in equation~\eqref{angles_mapping_epi}. For a sufficiently thin disc, one expects these mappings to be close to those obtained in the infinitely thin case, for which one immediately has
\begin{equation}
J_{\phi} = V_{0} \, R_{\rg}^{\rm thin} \;\;\; ; \;\;\; \Omega_{\phi}^{\rm thin} = \frac{V_{0}^{2}}{J_{\phi}} \;\;\; ; \;\;\; \kappa^{\rm thin} = \sqrt{2} \, \Omega_{\phi}^{\rm thin} \, .
\label{mapping_Mestel_thin}
\end{equation}
Given the thickened mean density profile $\rho_{\rm M}$ with its associated intrinsic frequencies, one may use the one-dimensional Jeans equation ~\citep[see, e.g., Eq. (4.271) in][]{BinneyTremaine2008} to constrain the value of the equilibrium vertical velocity dispersion $\sigma_{z}$. Indeed, one has
\begin{equation}
\frac{\partial (\rho_{\rm M} \, \sigma_{z}^{2})}{\partial z} = - \rho_{\rm M} \frac{\partial \psi_{\rm M}}{\partial z} \, ,
\label{vertical_Jeans_equation}
\end{equation}
where it is assumed that $\sigma_{z}$ is only a function of $R$.
Differentiating equation~\eqref{vertical_Jeans_equation} once w.r.t. $z$ and evaluating it at ${ z \!=\! 0 }$, one gets
\begin{equation}
\frac{\sigma_{z} (R)}{\nu (R)} = \sqrt{2} \, z_{0} (R) \, .
\label{link_sigmaz_z0_nu}
\end{equation}
Consequently, once the scale height $z_{0}$ of the disc and the intrinsic vertical frequency $\nu$ are numerically determined, the vertical velocity dispersion $\sigma_{z}$ within the disc follows immediately by equation~\eqref{link_sigmaz_z0_nu}.
One should note that the determination of the intrinsic frequencies required the use of the total potential of the system $\psi_{\rm M}$ from equation~\eqref{link_psi_rho}. However, our goal here is to model the secular evolution of the dynamically active component of the disc, i.e. the stars, whose density $\Sigma_{\rm star}$ is only one component of the total $\Sigma_{\rm M}$. Indeed, in order to build up a stable disc and deal with its central singularity and infinite extent, two tapering functions $T_{\rm inner}$ and $T_{\rm outer}$ must be introduced. They read
\begin{equation}
\begin{cases}
\displaystyle T_{\rm inner} (J_{\phi}) = \frac{J_{\phi}^{\nu_{\rm t}}}{(R_{\rm i} V_{0})^{\nu_{\rm t}} \!+\! J_{\phi}^{\nu_{\rm t}}} \, ,
\\
\displaystyle T_{\rm outer} (J_{\phi}) = \bigg[ 1 \!+\! \bigg[ \frac{J_{\phi}}{R_{\rm o} V_{0}} \bigg]^{\mu_{\rm t}} \bigg]^{-1} \, ,
\end{cases}
\label{definition_tapering}
\end{equation}
where $\nu_{\rm t}$ and $\mu_{\rm t}$ are two power indices controlling the sharpness of the two tapers, while $R_{\rm i}$ and $R_{\rm o}$ are two scale parameters. These two tapers mimic the presence of a bulge and the replacement of the outer disc by the dark halo. It is also assumed that only a fraction $\xi$ of the system is active (while the missing component will be a static contribution from the dark matter halo). As a consequence, the active surface density $\Sigma_{\rm star}$ of the disc may be written as
\begin{equation}
\Sigma_{\rm star} (J_{\phi}) = \xi \, \Sigma_{\rm M} (J_{\phi}) \, T_{\rm inner} (J_{\phi}) \, T_{\rm outer} (J_{\phi}) \, .
\label{definition_Sigmastar}
\end{equation}
The shape of the active surface density $\Sigma_{\rm star}$ is illustrated in figure~\ref{figSigmaStar}.
In order to follow the same setup as So12's UCB simulation, the numerical parameters are given the values
\begin{equation}
V_{0} = 1 \; ; \; G = 1 \; ; \; R_{\rm i} = 1 \; ; \; R_{\rm o} = 15 \; ; \; \nu_{\rm t} = 4 \; ; \; \mu_{\rm t} = 6 \, ,
\label{numerical_parameters}
\end{equation}
while the heat content of the disc is characterised by
\begin{equation}
\sigma_{r} = 0.227 \;\; ; \;\; \xi = 0.4 \, .
\label{numerical_parameters_II}
\end{equation}
It then only remains to define the height of the disc. So12 used a somewhat unusual vertical profile of constant vertical scale $z_{S}$, to define a thickened density $\rho_{\rm S}$ as
\begin{equation}
\rho_{\rm S} (z) = \frac{\Sigma_{\rm M}}{1.83 \, z_{\rm S}} \frac{1}{(\re^{|z/z_{\rm S}|/2} \!+\! 0.2 \, \re^{-5 |z/z_{\rm S}| /2})^{2}} \, .
\label{profile_So12}
\end{equation}
One can easily relate the Spitzer scale height $z_{0}$ from equation~\eqref{definition_rho_star} to the height $z_{\rm S}$ from equation~\eqref{profile_So12} by imposing the constraint ${ \rho_{\rm M} (z\!=\!0) \!=\! \rho_{\rm S} (z \!=\! 0)}$, which gives ${ z_{0} = 0.66 \, z_{\rm S} }$. As So12 used the choice ${ z_{\rm S} \!=\! 0.4 }$, we use here the value ${ z_{0} \!=\! 0.26 }$.

\begin{figure*}
\begin{center}
\epsfig{file=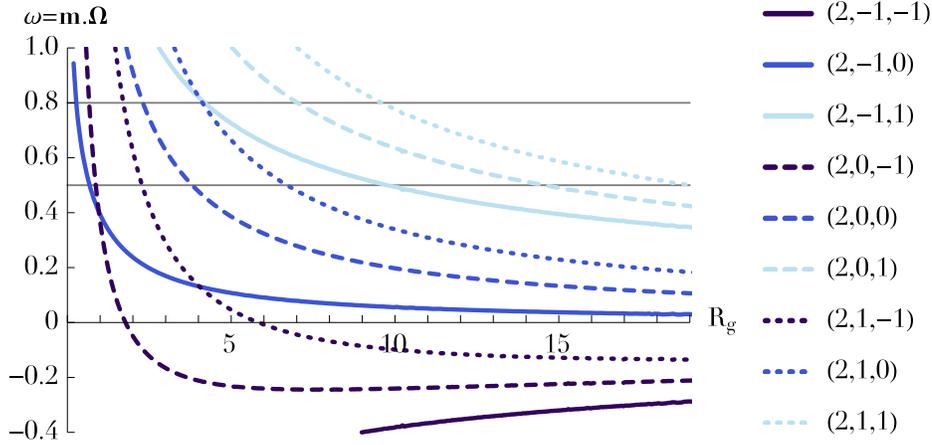,width=0.7\textwidth}
\caption{\small{Behaviour of the intrinsic frequency of resonance ${ \omega \!=\! \bm{m} \!\cdot\! \bm{\Omega} }$ as a function of the position within the disc and the resonant vector ${ \bm{m} \!=\! (m_{\phi} , m_{r} , m_{z}) }$. The grey lines correspond to the pattern frequency ${ m_{\rp} \Omega_{\rp} }$ introduced in the bar perturbations from equation~\eqref{assumption_C_bar} and considered in figure~\ref{fig_Contours_dFZdt_Bar}.
}}
\label{fig_Intrinsic_Frequencies}
\end{center}
\end{figure*}

Finally, it also important to note that So12's simulation was limited to the harmonic sector ${ 0 \!\leq\! m_{\phi} \!\leq\! 8 }$, except ${ m_{\phi} \!=\! 1 }$ (to avoid decentring). In our case, in order to clarify the dynamical mechanisms at play during the secular evolution, a more drastic limitation to the considered potential perturbations will be used and they will be restricted only to ${ m_{\phi} \!=\! 2 }$. In addition to this restriction, throughout the numerical calculations, the analysis will also be limited to only $9$ different resonances, i.e. 9 different vectors ${ \bm{m} \!=\! (m_{\phi} , m_{r} , m_{z}) }$. Indeed, we assume ${ m_{\phi} \!=\! 2 }$, $m_{r} \!\in\! \{ -1,0,1 \}$ and ${ m_{z} \!\in\! \{ -1,0,1 \} }$. Among these resonances, we define the corotation resonance (COR) as ${ \bm{m} \!=\! (2,0,0) }$, the radial (resp. vertical) inner Lindblad resonance (rILR) (resp. vILR) as ${ \bm{m} \!=\! (2,-1,0) }$ (resp. ${ \bm{m} \!=\! (2,0,-1) }$), and similarly the radial (resp. vertical) outer Lindblad resonance (rOLR) (resp. vOLR) as ${ \bm{m} \!=\! (2 , 1, 0) }$ (resp. ${ \bm{m} \!=\! (2,0,1) }$). Once the orbital frequencies $\bm{\Omega}$ and the considered resonance vectors $\bm{m}$ have been specified, one may study the behaviour of the resonance frequencies ${ \omega \!=\! \bm{m} \!\cdot\! \bm{\Omega} }$ as a function of the position within the disc. These frequencies, for which the amplification eigenvalues and the perturbation autocorrelation as in equation~\eqref{Dm_diagonalised_even} have to be evaluated, are illustrated in figure~\ref{fig_Intrinsic_Frequencies}.

When simulated on secular timescales, one observes sequences of transient spirals within the disc leading to an irreversible diffusion of the system's DF in action space (M. Solway, private communication). To probe such a secular thickening of the disc, one may consider the marginal distribution of vertical action $J_{z}$ as a function of the guiding radius $R_{\rg}$ within the disc. We define the function ${ F_{\rm Z} (R_{\rg} , J_{z} , t) }$ as
\begin{align}
F_{\rm Z} (R_{\rg} , J_{z} , t) & \, = \!\! \int \!\! \mathrm{d} \bm{\theta}' \mathrm{d} \bm{J}' \, \delta_{\rm D} (R_{\rg} \!-\! R_{\rg}') \, \delta_{\rm D} (J_{z} \!-\! J_{z}') \, F (\bm{J}' , t) \nonumber
\\
& \, = (2 \pi)^{3} \frac{\mathrm{d} J_{\phi}}{\mathrm{d} R_{\rg}} \! \int \!\! \mathrm{d} J_{r}' \, F (R_{\rg} , J_{r}' , J_{z} , t) \, .
\label{definition_FZ}
\end{align}
In equation~\eqref{definition_FZ}, starting from equation~\eqref{definition_Rg}, one can straightforwardly show that ${ \mathrm{d} J_{\phi} / \mathrm{d} R_{\rg} \!=\! (R_{\rg} \kappa^{2})/(2 \Omega_{\phi}) }$ (\,${ \!=\! V_{0} }$ for an infinitely thin Mestel disc, thanks to equation~\eqref{mapping_Mestel_thin}). 
The time variation of $F_{\rm Z}$ may generically be estimated via equations~\eqref{diffusion_equation_Ftot} and~\eqref{BL_div_Ftot} as
\begin{equation}
\frac{\partial F_{\rm Z}}{\partial t} = (2 \pi)^{3} \frac{\mathrm{d} J_{\phi}}{\mathrm{d} R_{\rg} } \! \int \!\! \mathrm{d} J_{r}' \, \text{div} (\bm{\mathcal{F}}_{\rm tot}) (R_{\rg} , J_{r}' , J_{z} , t) \, .
\label{FZ_DT}
\end{equation}
One can also rewrite equation~\eqref{FZ_DT} as the divergence of a flux ${ \bm{\mathcal{F}}_{Z} \!=\! (\mathcal{F}_{Z}^{\phi} , \mathcal{F}_{Z}^{z}) }$  defined in the ${ (J_{\phi} , J_{z})-}$plane, so as to have
\begin{equation}
\frac{\partial F_{\rm Z} (J_{\phi} , J_{z})}{\partial t} = \bigg(\! \frac{\partial }{\partial J_{\phi}} , \frac{\partial }{\partial J_{z}} \!\bigg) \!\cdot\! \bm{\mathcal{F}}_{\rm Z} = \frac{\partial \mathcal{F}_{\rm Z}^{\phi}}{\partial J_{\phi}} \!+\! \frac{\partial \mathcal{F}_{\rm Z}^{z} }{\partial J_{z}} \, ,
\label{FZ_div}
\end{equation}
where the flux components ${ (\mathcal{F}_{\rm Z}^{\phi} , \mathcal{F}_{\rm Z}^{z}) }$ are given by
\begin{equation}
\begin{cases}
\displaystyle \mathcal{F}_{\rm Z}^{\phi} = (2 \pi)^{3} \!\! \int \!\! \rd J_{r} ' \, \mathcal{F}_{\rm tot}^{\phi} (J_{\phi} , J_{r}' , J_{z}) \, ,
\\
\displaystyle \mathcal{F}_{\rm Z}^{z} = (2 \pi)^{3} \!\! \int \!\! \rd J_{r}' \, \mathcal{F}_{\rm tot}^{z} (J_{\phi} , J_{r}' , J_{z}) \, .
\end{cases}
\label{definition_FZ_flux}
\end{equation}
In equation~\eqref{definition_FZ_flux}, the total diffusion flux $\bm{\mathcal{F}}_{\rm tot}$ in the ${ (J_{\phi} , J_{r} , J_{z}) }$ space from equations~\eqref{diffusion_equation_Ftot} and~\eqref{BL_div_Ftot} was naturally written as ${ \bm{\mathcal{F}}_{\rm tot} \!=\! (\mathcal{F}_{\rm tot}^{\phi} , \mathcal{F}_{\rm tot}^{r} , \mathcal{F}_{\rm tot}^{z}) }$.

\begin{figure*}
\begin{center}
{\epsfig{file=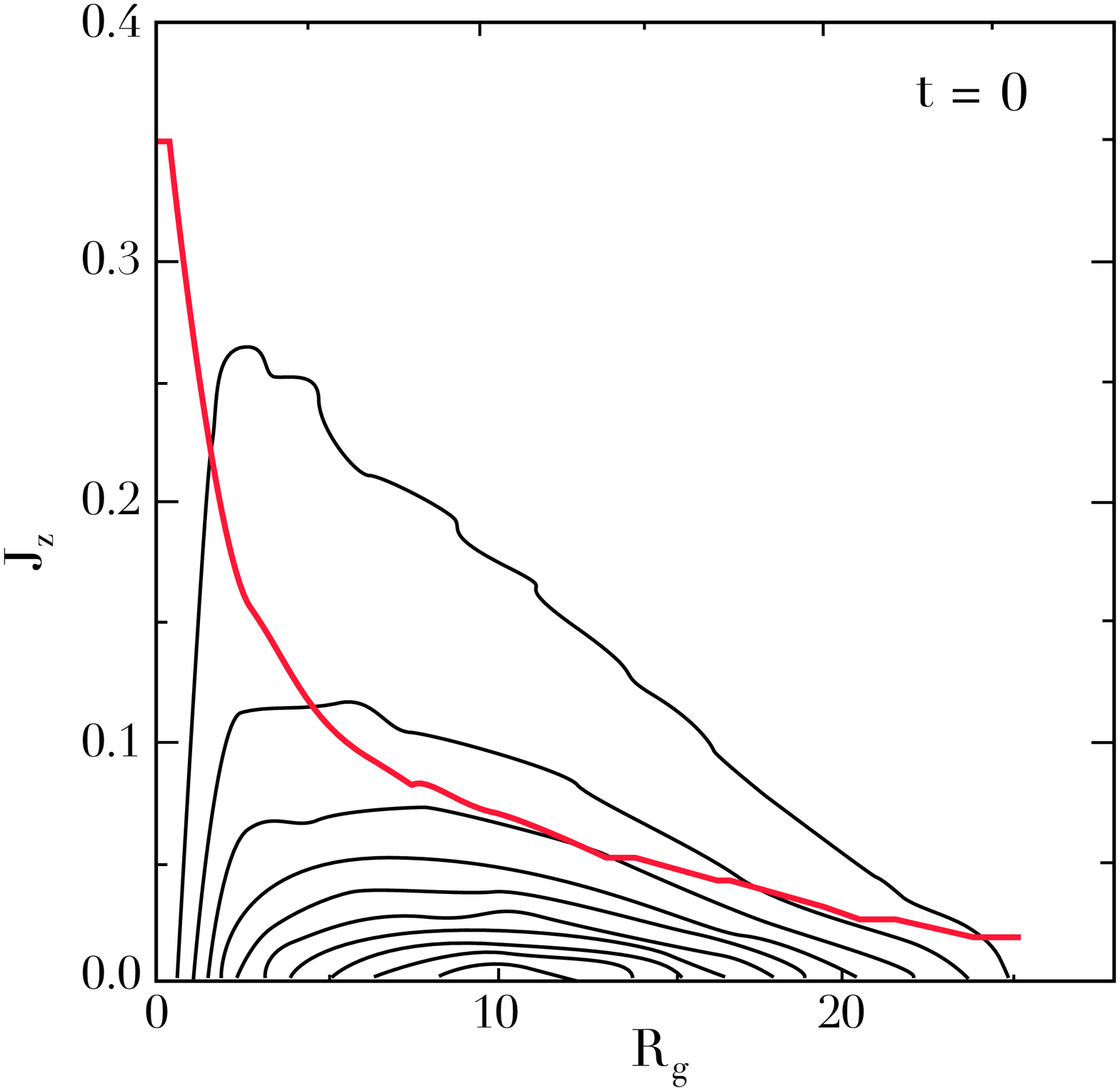,angle=-00,width=0.45\textwidth} }
{\epsfig{file=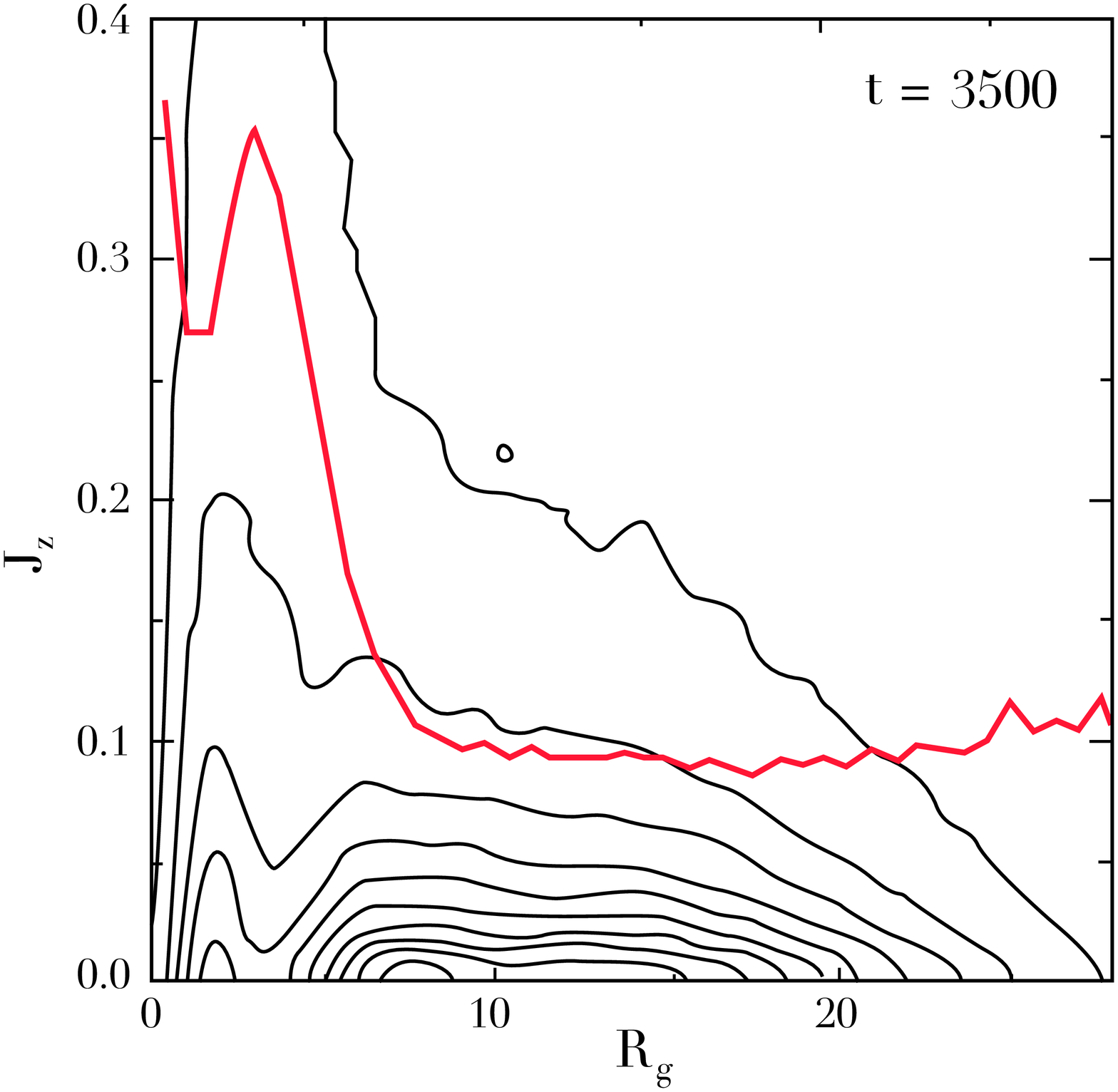,angle=-00,width=0.45\textwidth}}
\caption{\small{Results from the simulation UCB1 of So12. \textbf{Left panel}: Initial contours of the function ${ F_{\rm Z} (R_{\rg} , J_{z} , t) }$ from equation~\eqref{definition_FZ} for ${ t \!=\! 0 }$. This illustrates the distribution of vertical actions $J_{z}$ as a function of the guiding radius $R_{\rg}$ within the disc. Contours are spaced linearly between 95\% and 5\% of the function maximum. The red curve gives the mean value of $J_{z}$ for a given $R_{\rg}$. \textbf{Right panel}: Same as in the left panel but at a later stage of the evolution ${ t \!=\! 3500 }$. One can clearly note the formation on secular timescales of a narrow ridge of enhanced vertical actions $J_{z}$ in the inner regions of the disc.
}}
\label{fig_Solway}
\end{center}
\end{figure*}

The initial contours of $F_{\rm Z}$ are illustrated in the left panel of figure~\ref{fig_Solway},
while their long-term evolution is illustrated in the right panel of the same figure. When comparing the two panels of figure~\ref{fig_Solway}, one can clearly note the formation on secular timescales of a narrow ridge of enhanced vertical actions in the inner region of the disc, characterised by an increase of the mean value of the vertical action in these regions.\footnote{Figure~\ref{fig_Solway}, showing a vertical ridge, is a new figure, which was not presented nor discussed in~\cite{SolwaySellwood2012}. It was graciously provided to us by M. Solway. Although present in Solway's UCB simulations, it was never put forward nor discussed in previous papers.}
Let us note that the disc considered in So12 was purposely designed to be linearly stable, quasi-stationary, isolated and unperturbed. So12 could then explicitly check that this disc does not develop any spiral mode or bar instability for hundreds of dynamical times. The only source of fluctuations in the disc is due to weak transient spiral arms seeded by the disc's finite number of particles. 
In this context, the vertical ridge observed in figure~\ref{fig_Solway} has to be the signature of the spontaneous secular thickening of the disc sourced by its intrinsic shot noise amplified by self-gravity, since these are the only perturbations remaining in the system.\footnote{This conclusion is also reinforced by two additional tests presented in~\cite{Sellwood2012}, which investigated razor-thin analogs of So12's thickened simulations. The figure~2 of~\cite{Sellwood2012} shows that the larger the number of particles, the slower the evolution. The evolution is therefore induced by discreteness effects, as recovered quantitatively in~\cite{FouvryPichonMagorrianChavanis2015}. Moreover, figure~5 of~\cite{Sellwood2012} also shows that after redistributing randomly the azimuthal phases of the particles at some stage of the evolution, the ridge would still appear on the same timescale. The resonant ridge is therefore not a phase-dependent feature, and only depends on the system's mean orbital structure, i.e. its mean DF ${ F \!=\! F (\bm{J} , t) }$.}  
As already shown quantitatively in~\cite{FouvryPichonMagorrianChavanis2015} in the context of razor-thin discs, this corresponds to the exact dynamical regime of application of the Balescu-Lenard equation~\eqref{definition_BL}.

The aim of the upcoming sections is to discuss how the previous WKB limits of the collisionless and collisional secular diffusion equations provide a qualitative illustration of this ridge formation.
Secular evolution being by essence a slow process, we will restrict ourselves here to the estimation of the initial diffusion flux, $\bm{\mathcal{F}}_{\rm tot}$, at the time ${ t \!=\! 0 }$. In~\cite{FouvryPichonMagorrianChavanis2015} in the context of razor-thin discs, we already emphasised how the computation of the initial diffusion flux allows indeed for the recovery of the formation of resonant ridges in action space. Computing the evolution at later time, while theoretically interesting (and challenging), would not be astrophysically relevant in the present context, because it would describe an evolution on a timescale much larger than the age of the universe (see section~\ref{sec:GMC}).

\subsection{Shot noise driven resonant disc thickening}
\label{sec:thickeningshotnoise}

\begin{figure}
\begin{center}
\epsfig{file=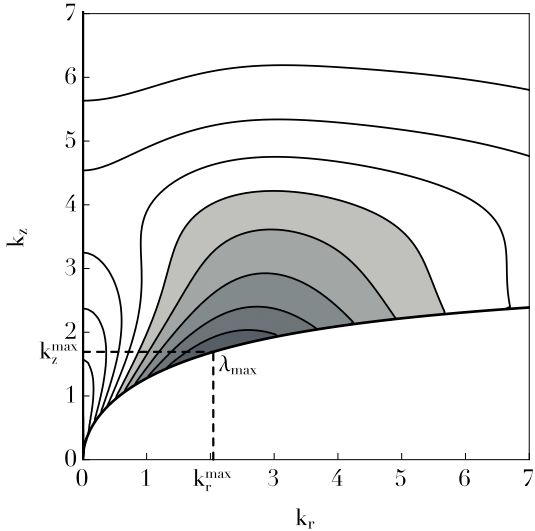,width=0.4\textwidth}
\caption{\small{Behaviour of the function ${ (k_{r} , k_{z} ) \!\mapsto\! \lambda (k_{r} , k_{z}) }$, as defined in equation~\eqref{lambda_even}, for ${ \bm{m} \!=\! \bm{m}_{\rm COR} }$ and ${ J_{\phi} \!=\! 1.5 }$. One should remember that the diffusion coefficients generically require to evaluate the amplification eigenvalues at the intrinsic frequency ${ \omega \!=\! \bm{m} \!\cdot\! \bm{\Omega} }$. Contours are spaced linearly between 90\% and 10\% of the function maximum $\lambda_{\rm max}$. The grey region corresponds to the domain $\mathcal{V}_{\rm max} \!=\! \{ (k_{r} , k_{z} ) \big| \lambda(k_{r} , k_{z}) \!\geq\! \lambda_{\rm max}/2 \}$, i.e. the region on which the integrations from equations~\eqref{Dm_diagonalised_even} and~\eqref{1/D_continuous} may be performed. One can finally note that here the maximum of amplification lies on the line ${ k_{z} \!=\! k_{z}^{1} (k_{r}) }$, i.e. along the line of the minimum quantised vertical frequency $k_{z}$.
}}
\label{figLambdakrkz}
\end{center}
\end{figure}

\begin{figure}
\begin{center}
\epsfig{file=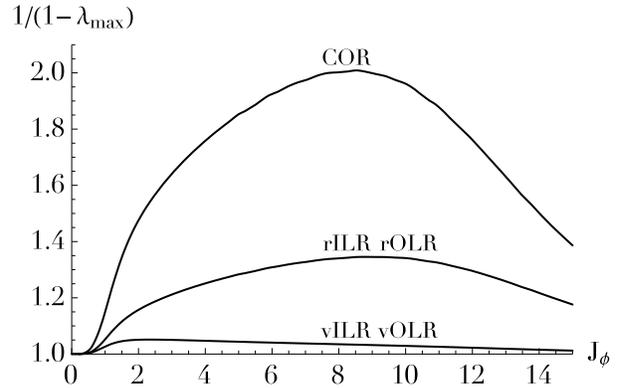,width=0.45\textwidth}
\caption{\small{Dependence of the amplification factor ${ 1 / (1 \!-\! \lambda_{\rm max} (\bm{m} , J_{\phi}) ) }$ as a function of the position $J_{\phi}$ within the disc, for various resonances $\bm{m}$. The amplification eigenvalues $\lambda$ are given by the simplified expression from equation~\eqref{lambda_simple}. The amplification associated with the COR is always larger than the ones associated with the other resonances. As expected, in the inner and outer regions of the disc, the strength of the amplification is turned off by the tapering functions from equation~\eqref{definition_tapering}.
}}
\label{figLambdamaxJphi}
\end{center}
\end{figure}

To compute the secular diffusion flux from equations~\eqref{diffusion_equation_Ftot} and~\eqref{BL_div_Ftot}, one first has to study the behaviour of the amplification eigenvalues ${ \lambda (k_{r} , k_{z}) }$ from equations~\eqref{lambda_even} and~\eqref{lambda_odd}, thanks to which the approximation of the small denominators may be performed. For a given resonance $\bm{m}$ and position $J_{\phi}$, the amplification function ${ (k_{r} , k_{z} ) \!\mapsto\! \lambda (k_{r} , k_{z}) }$ is illustrated in figure~\ref{figLambdakrkz}.
As presented in equations~\eqref{Dm_diagonalised_ASD_even} and~\eqref{1/D_ASD}, such a behaviour allows us to determine a region ${ \mathcal{V}_{\rm max} (\bm{m} , J_{\phi}) }$ over which the ${ (k_{r} , k_{z})-}$integrations from equations~\eqref{Dm_diagonalised_even} and~\eqref{1/D_continuous} may be performed. Figure~\ref{figLambdamaxJphi} illustrates the importance of the self-gravitating amplification by representing the behaviour of the function ${ J_{\phi} \!\mapsto\! 1/(1 \!-\! \lambda_{\rm max} (\bm{m} , J_{\phi}) ) }$ for various resonances.
After having estimated the system's amplification eigenvalues, one may in turn compute the induced collisionless diffusion (section~\ref{sec:verticalFP}) and the collisional one (section~\ref{sec:verticalBL}).

\subsubsection{Collisionless forced thickening}
\label{sec:verticalFP}

\begin{figure}
\begin{center}
{\epsfig{file=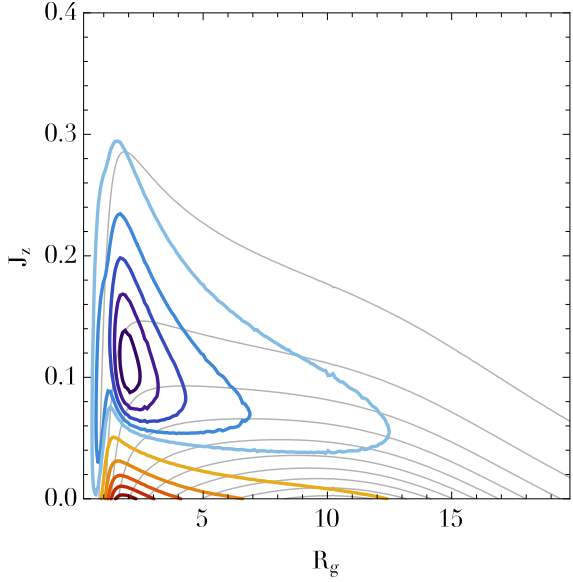,width=0.425\textwidth}}
\caption{\small{
Illustration of the initial contours of ${ \partial F_{\rm Z} / \partial t |_{t = 0} }$ predicted by the collisionless diffusion equation~\eqref{diffusion_equation}, when considering a secular forcing by shot noise as in equation~\eqref{assumption_C_shot_noise}. Red contours, for which ${ \partial F_{\rm Z} / \partial t |_{t = 0} \!<\! 0 }$, correspond to regions from which the orbits will be depleted and are spaced linearly between 90\% and 10\% of the function minimum, while blue contours, for which ${ \partial F_{\rm Z} / \partial t |_{t = 0} \!>\! 0 }$, correspond to regions where the number of orbits will increase during the diffusion and are spaced linearly between 90\% and 10\% of the function maximum. The background contours correspond to the initial contours of ${ F_{\rm Z} ( t \!=\! 0 ) }$, spaced linearly between 95\% and 5\% of the function maximum, and determined for the quasi-isothermal DF from equation~\eqref{definition_DF_quasi_isothermal}.
}}
\label{fig_Contours_dFZdt}
\end{center}
\end{figure}

In order to gain some qualitative insight on the formation of the vertical ridge observed in figure~\ref{fig_Solway}, one may first rely on the WKB limit of the collisionless diffusion formalism obtained in section~\ref{sec:FPWKB}. Because So12 considered an isolated disc, one has to assume some form for the perturbation power spectrum ${ \widehat{\mathcal{C}} [m_{\phi} , \omega , R_{\rg} , k_{r} , k_{z}] }$ that appears in equation~\eqref{Dm_diagonalised_even}. As in~\cite{FouvryPichon2015,FouvryBinneyPichon2015}, it will be assumed that the source of noise is given by shot noise, due to the finite number of stars in the disc. Such a type of perturbation may also mimic the perturbations induced by compact gas clouds within the disc. For such a Poisson shot noise, the perturbing potential varies radially like ${ \psi^{\rm e} \!\propto\! \sqrt{\Sigma_{\rm star}} }$. For simplicity, the dependence of $\widehat{\mathcal{C}}$ with $\omega$, $k_{r}$, $k_{z}$ is neglected. Moreover, as detailed below equation~\eqref{profile_So12}, as perturbations were restricted to the sole harmonic sector ${ m_{\phi} \!=\! 2 }$, the same restriction applies to $\widehat{\mathcal{C}}$. As a consequence, let us assume for our illustration purposes that, up to a normalisation, the autocorrelation of the external perturbations takes the simple form
\begin{equation}
\widehat{\mathcal{C}} [m_{\phi} , \omega , R_{\rg} , k_{r} , k_{z}] = \delta_{m_{\phi}}^{2} \, \Sigma_{\rm star} (R_{\rg}) \, .
\label{assumption_C_shot_noise}
\end{equation}
One should note that shot noise is not per se an external perturbation. To account in a more rigourous way for such intrinsic finite${-N}$ effects, one should rely on the inhomogeneous Balescu-Lenard equation, as will be presented in section~\ref{sec:verticalBL}. One should finally note that the noise assumption from equation~\eqref{assumption_C_shot_noise} is rather crude, since we only included a dependence w.r.t. $R_{\rg}$. Here the lack of dependence w.r.t. $\omega$ implies that at a given location in the disc, all resonances undergo the same perturbations, even if they are not associated with the same resonant frequencies ${ \bm{m} \!\cdot\! \bm{\Omega} }$.
Thanks to the estimation of the disc's amplification eigenvalues, and the perturbation power spectrum from equation~\eqref{assumption_C_shot_noise}, one may compute the WKB collisionless diffusion flux $\bm{\mathcal{F}}_{\rm tot}$ from equation~\eqref{definition_Ftot}, and subsequently its divergence ${ \text{div} (\bm{\mathcal{F}}_{\rm tot}) }$. One can then estimate the initial time variation of the function $F_{\rm Z}$ from equation~\eqref{FZ_DT}. The initial contours of ${ \partial F_{\rm Z} / \partial t |_{t = 0} }$ are illustrated in figure~\ref{fig_Contours_dFZdt}.
In this figure, one recovers qualitatively the formation of a resonant ridge of increased vertical actions in the inner region of the disc, as was observed in figure~\ref{fig_Solway}. This illustrates qualitatively how the Poisson shot noise induced by the finite number of particles -- as approximated by equation~\eqref{assumption_C_shot_noise} -- may lead to a secular thickening of the disc.

\subsubsection{Collisional thickening}
\label{sec:verticalBL}

\begin{figure}
\begin{center}
\epsfig{file=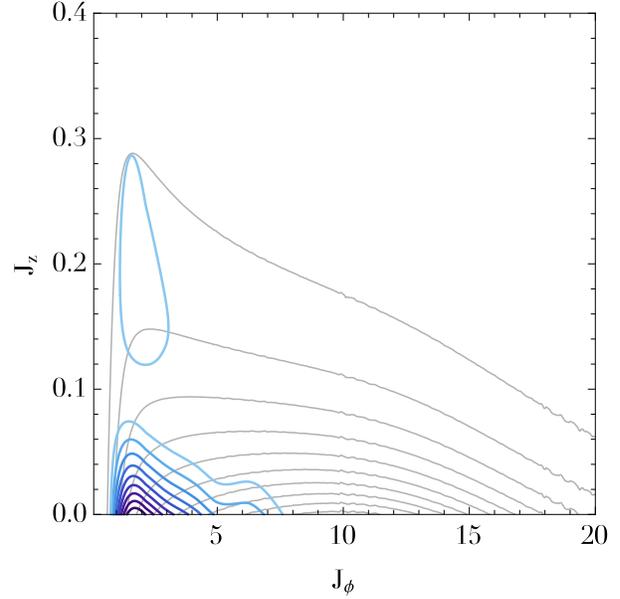,angle=-00,width=0.45\textwidth}
\caption{\small{
Illustration of the norm of the collisional diffusion flux ${ | N \bm{\mathcal{F}}_{\rm Z} | (t \!=\! 0) }$ in the ${ (J_{\phi} , J_{z})-}$plane predicted by the Balescu-Lenard equation~\eqref{definition_BL}. The contours are spaced linearly between 90\% and 10\% of the maximum norm. The background contours correspond to the initial contours of ${ F_{\rm Z} ( t \!=\! 0 ) }$, spaced linearly between 95\% and 5\% of the function maximum, and determined for the initial quasi-isothermal DF from equation~\eqref{definition_DF_quasi_isothermal}. One can clearly note the presence of an enhanced diffusion flux in the inner region of the disc, compatible with the localised increase of the vertical actions observed in figure~\ref{fig_Solway}.
}}
\label{fig_Contours_Norm_Jz}
\end{center}
\end{figure}

The previous section relied on the WKB collisionless diffusion equation~\eqref{diffusion_equation}. In order to better account for the intrinsic Poisson shot noise, one may now proceed to the same estimations, while relying on the WKB Balescu-Lenard equation~\eqref{definition_BL}. Thanks to the previous estimations of the amplification eigenvalues, one may straightforwardly compute the collisional susceptibility coefficients from equation~\eqref{1/D_ASD}. This allows us to determine the drift and diffusion coefficients from equations~\eqref{final_drift} and~\eqref{final_diff}, and consequently the total collisional diffusion flux introduced in equation~\eqref{definition_F_tot}. Because the mass of the particles is given by ${ \mu \!=\! M_{\rm tot} / N }$, we will rather consider the quantity ${ N \bm{\mathcal{F}}_{\rm tot} }$, which is independent of $N$. Following equation~\eqref{FZ_div}, one can then compute the diffusion flux ${ N \bm{\mathcal{F}}_{\rm Z} }$ in the ${ (J_{\phi} , J_{z})-}$plane. The initial contours of the norm ${ | N \bm{\mathcal{F}}_{\rm Z} | (t \!=\! 0) }$ are illustrated in figure~\ref{fig_Contours_Norm_Jz}.
In this figure, one can note how the diffusion flux ${ N \bm{\mathcal{F}}_{\rm Z} }$ is localised in the inner region of the disc. Both figures~\ref{fig_Contours_dFZdt} and~\ref{fig_Contours_Norm_Jz} are in qualitative agreement and predict a localised increase in the vertical actions as observed in direct numerical simulations. The crude approximation of the Poisson shot noise from equation~\eqref{assumption_C_shot_noise} also allows us to qualitatively recover with the collisionless secular diffusion formalism, the results obtained here thanks to the collisional formalism, within which the spectral properties of the Poisson shot noise are self-consistently accounted for.

\subsubsection{Vertical kinetic heating}
\label{sec:verticalsigmaz}

\begin{figure}
\begin{center}
\begin{tabular}{@{}cc@{}}
{\epsfig{file=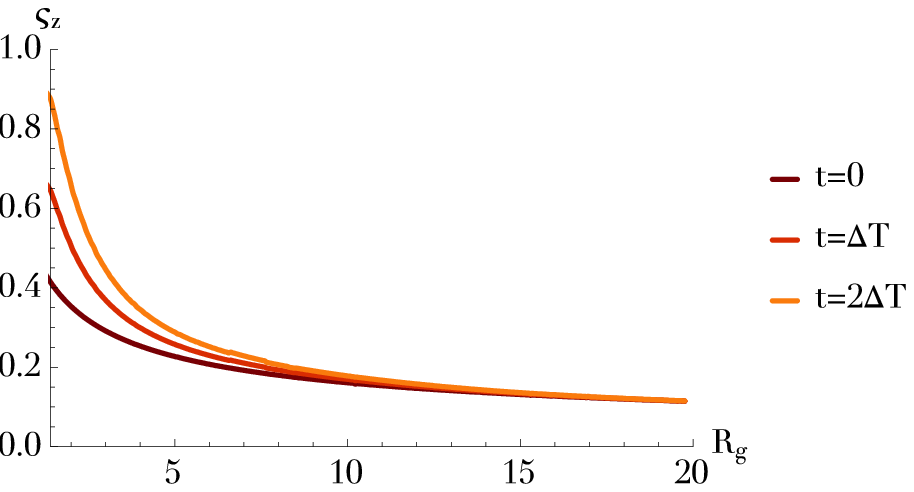,width=0.45\textwidth}} \\
{\epsfig{file=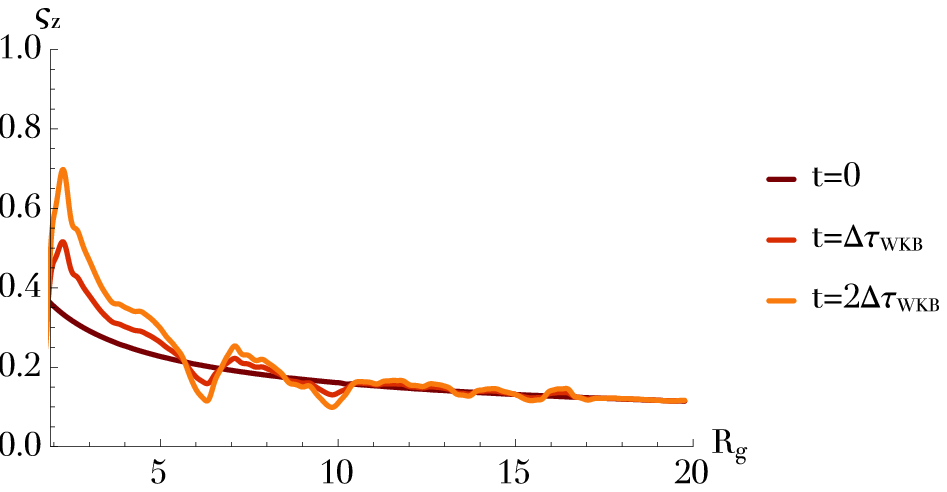,width=0.45\textwidth} }
\end{tabular}
\caption{\small{\textbf{Top panel}: Illustration of the expected increase in the vertical velocity dispersion $\varsigma_{z} (R_{\rg} , t)$ as a function of the position within the disc, at various stages of the diffusion, relying on the collisionless WKB diffusion from equation~\eqref{diffusion_equation}. For ${ t \!=\! 0 }$, one has ${ \varsigma_{z} (R_{\rg} , t \!=\! 0) \!=\! \sigma_{z} (R_{\rg}) }$, while for larger values of $t$ (here ${ \Delta T }$ is an arbitrary timestep), we used the estimation ${ \varsigma_{z} (R_{\rg} , t) \!\simeq\! \sigma_{z} (R_{\rg}) \!+\! t \, \partial \varsigma_{z} / \partial t |_{t = 0} }$, and equation~\eqref{derivative_sigma_z}. \textbf{Bottom panel}: Same as the top panel, for the collisional WKB limit of the Balescu-Lenard equation~\eqref{definition_BL}. Here ${ \Delta \tau_{\rm WKB} }$ is a timestep introduced in section~\eqref{sec:timescales}.
}}
\label{fig_sigma_z}
\end{center}
\end{figure}

In order to better assess the properties of the diffusion induced by finite${-N}$ effects, let us now consider the induced increase in the vertical velocity dispersion. Indeed, disc thickening can observationally best be probed by determining the evolution of the vertical velocity dispersion ${ \varsigma_{z}^{2} (R_{\rg} , t) \!=\! \langle v_{z}^{2} \rangle (R_{\rg} , t ) }$, defined as
\begin{equation}
\varsigma_{z}^{2} (R_{\rg} , t) = \frac{\displaystyle\int \!\! \mathrm{d} \bm{\theta}' \mathrm{d} \bm{J}' \, \delta_{\rm D} (R_{\rg} \!-\! R_{\rg}') \, F (\bm{J}' , t) \, (v_{z}')^{2}}{\displaystyle\int \!\! \mathrm{d} \bm{\theta}' \mathrm{d} \bm{J}' \, \delta_{\rm D} (R_{\rg} \!-\! R_{\rg}') \, F (\bm{J}' , t)} \, .
\label{definition_sigma_z}
\end{equation}
Thanks to the epicyclic approximation from equations~\eqref{actions_Jr_Jz} and~\eqref{angles_mapping_epi}, one immediately has ${ v_{z}^{2} \!=\! 2 J_{z} \nu \sin^{2} (\theta_{z}) }$. In equation~\eqref{definition_sigma_z}, one can perform the integrations over ${ \bm{\theta}' }$ and ${ J_{\phi}' }$ to obtain
\begin{equation}
\varsigma_{z}^{2} (R_{\rg} , t) = \nu (R_{\rg}) \, \frac{\displaystyle\int \!\! \mathrm{d} J_{r}' \mathrm{d} J_{z}' \, F (R_{\rg} , J_{r}' , J_{z}' , t) \, J_{z}'}{\displaystyle\int \!\! \mathrm{d} J_{r}^{'} \mathrm{d} J_{z}' \, F (R_{\rg} , J_{r}' , J_{z}' , t) } \, .
\label{calculation_sigma_z}
\end{equation}
Because for ${ t \!=\! 0 }$, ${ F(\bm{J}' , t) }$ is given by the quasi-isothermal DF from equation~\eqref{definition_DF_quasi_isothermal}, one immediately recovers ${ \varsigma_{z}^{2} (R_{\rg} , t \!=\! 0) \!=\! \sigma_{z}^{2} (R_{\rg}) }$. One can also compute the initial time derivative of $\varsigma_{z}^{2}$. After some simple algebra, it reads
\begin{equation}
\frac{\partial \varsigma_{z}^{2}}{\partial t} \bigg|_{t = 0} \!\!\!\!\! = \nu \, \frac{\displaystyle \int \!\! \mathrm{d} J_{r}' \mathrm{d} J_{z}' \, J_{z}' \frac{\partial F}{\partial t} \!\bigg|_{t = 0} \!\! \!-\!  \frac{\sigma_{z}^{2}}{\nu} \!\! \int \!\! \mathrm{d} J_{r}' \mathrm{d} J_{z}' \, \frac{\partial F}{\partial t} \!\bigg|_{t = 0 }}{\displaystyle \!\! \int \!\! \mathrm{d} J_{r}' \mathrm{d} J_{z}' \, F (t \!=\! 0)}  \, ,
\label{derivative_sigma_z}
\end{equation}
where ${ \partial F / \partial t \!=\! \text{div} (\bm{\mathcal{F}}_{\rm tot}) }$ is given by the diffusion equations~\eqref{diffusion_equation_Ftot} and~\eqref{BL_div_Ftot}. Using the fact that ${ \partial \varsigma_{z}^{2} / \partial t \!=\! 2 \varsigma_{z} \partial \varsigma_{z} / \partial t }$, one can anticipate a secular increase in the vertical velocity dispersion $\varsigma_{z}$ under the effect of the Poisson shot noise perturbations. This is illustrated in figure~\ref{fig_sigma_z}, where we represent ${ \varsigma_{z} (R_{\rg} , t ) \!\simeq\! \sigma_{z} (R_{\rg} ) \!+\! t \, \partial \varsigma_{z} / \partial t |_{t = 0} }$, as predicted by both collisionless and collisional formalisms.
Consistently with figures~\ref{fig_Contours_dFZdt} and~\ref{fig_Contours_Norm_Jz}, the WKB formalisms predict that the most significant increase in the vertical velocity dispersion occurs in the inner region of the disc, as already observed in figure~\ref{fig_Solway}. This illustrates qualitatively how the discrete Poisson shot noise may lead on secular timescales to a thickening of the disc. Finally, recall that a strength of the Balescu-Lenard formalism is that it is self-contained and does not involve any ad hoc fittings of the system's perturbations. Thanks to the calculation of the induced collisional increase in $\varsigma_{z}$ presented in the bottom panel of figure~\ref{fig_sigma_z}, one may now study the typical timescale of collisional diffusion predicted by the thick WKB Balescu-Lenard equation and compare it to the one observed in So12's simulation. This is the purpose of the next section.

\subsection{Diffusion timescale}
\label{sec:timescales}

Thanks to the previous estimates of the collisional diffusion flux ${ N \bm{\mathcal{F}}_{\rm Z} }$, one may now compare the diffusion timescale of appearance of the finite${-N}$ effects predicted by the Balescu-Lenard equation with So12's numerical measurements. Indeed, one can note that the Balescu-Lenard equation~\eqref{definition_BL} depends on the number $N$ of particles through the mass of the individual particles ${ \mu \!=\! M_{\rm tot} / N }$. Equation~\eqref{definition_BL} may therefore be rewritten as
\begin{equation}
\frac{\partial F}{\partial t} = \frac{1}{N} C_{\rm BL} [F] \, ,
\label{rewrite_BL}
\end{equation} 
where ${ C_{\rm BL} [F] \!=\! N \text{div} (\bm{\mathcal{F}}_{\rm tot}) }$ is the ${ N-}$independent Balescu-Lenard collisional operator, i.e. the r.h.s. of equation~\eqref{definition_BL} multiplied by ${ N \!=\! M_{\rm tot}/ \mu }$. Equation~\eqref{rewrite_BL} illustrates the fact that the larger the number of particles, the slower the secular evolution. Introducing the rescaled time ${ \tau \!=\! t/N }$, one may rewrite equation~\eqref{rewrite_BL} as
\begin{equation}
\frac{\partial F}{\partial \tau} = C_{\rm BL} [F] \, ,
\label{rewrite_BL_tau}
\end{equation}
so as to write the Balescu-Lenard equation without any explicit appearance of $N$. This allows us to compare the time during which So12's simulation was performed with the collisional timescale of evolution predicted by the Balescu-Lenard formalism. The right panel of figure~\ref{fig_Solway} was observed in So12 with ${ N \!=\! 2 \!\times\! 10^{5} }$ particles, after a time ${ \Delta t_{\rm So12} \!=\! 3500 }$. As a consequence, So12 observed the resonant ridge after a rescaled time ${ \Delta \tau_{\rm So12} \!=\! \Delta t_{\rm So12} / N \!\simeq\! 2 \!\times\!10^{-2} }$. In figure~\ref{fig_Solway}, looking at the evolution of the mean value of $J_{z}$, one can note that during the rescaled time ${ \Delta \tau_{\rm So12} }$, the mean vertical action in the inner region of the disc was approximately doubled. One can then compare this time with the typical time predicted by the thick WKB Balescu-Lenard formalism to lead to a similar increase of the mean vertical action.
Thanks to equations~\eqref{actions_Jr_Jz} and~\eqref{angles_mapping_epi}, one has ${ v_{z}^{2} \!=\! 2 \nu J_{z} \sin^{2} (\theta_{z}) }$, so that ${ \varsigma_{z}^{2} \!=\! \nu \, \big< J_{z} \big> }$. As a consequence, doubling the mean vertical action ${ \big< J_{z} \big> }$ only requires to multiply the vertical velocity dispersion $\varsigma_{z}$ by $\sqrt{2}$.
Thanks to figure~\ref{fig_sigma_z}, one can note that such an increase in $\varsigma_{z}$ is reached after a rescaled time ${ \Delta \tau_{\rm WKB} \!=\! 10^{3} }$. Comparing the numerically measured time ${ \Delta \tau_{\rm So12} }$ and the thick WKB Balescu-Lenard prediction ${ \Delta \tau_{\rm WKB} }$, one obtains
\begin{equation}
\frac{\Delta \tau_{\rm So12}}{\Delta \tau_{\rm WKB}} \sim 2 \!\times\! 10^{-5} \, .
\label{comparison_delta}
\end{equation}
The disagreement between the measured and the predicted timescales is even larger than what was obtained in FPC15 in the razor-thin case (${\sim\! 10^{-3}}$) for the ${J_{r}-}$diffusion. The timescale discrepancy observed in FPC15, while using the razor-thin WKB limit of the Balescu-Lenard formalism -- which was solved in~\cite{FouvryPichonMagorrianChavanis2015} by resorting to a global evaluation of the Balescu-Lenard diffusion flux -- was interpreted to be due to the incompleteness of the WKB basis. Indeed, by restricting ourselves only to tightly wound perturbartions, this WKB limit was not able to capture the swing amplification mechanism~\citep{GoldreichLyndenBell1965a,JulianToomre1966,Toomre1981} which describes the strong amplification of unwinding perturbations.
The thickened WKB formalism presented in section~\ref{sec:WKBlimit} suffers from the same flaw, and this is illustrated in the timescale mismatch from equation~\eqref{comparison_delta}, that can be directly attributed to the neglect of some components of the self-gravitating amplification in the qualitative illustrations obtained via the WKB frameworks.

\subsection{Thickening induced by bars}
\label{sec:thickeningbar}

\begin{figure*}
\centering
\begin{tabular}{@{}cc@{}}
{\epsfig{file=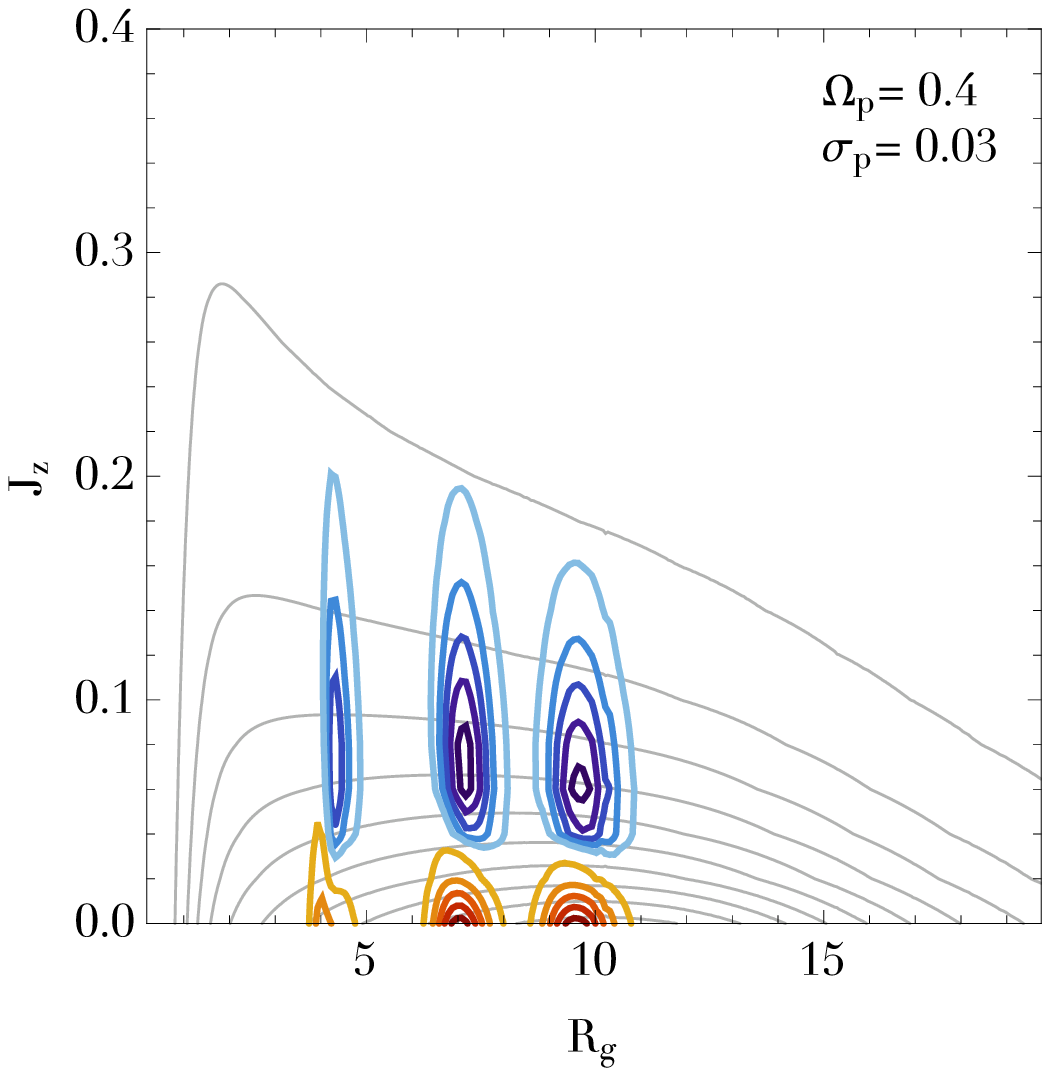,width=0.425\textwidth}}
&
{\epsfig{file=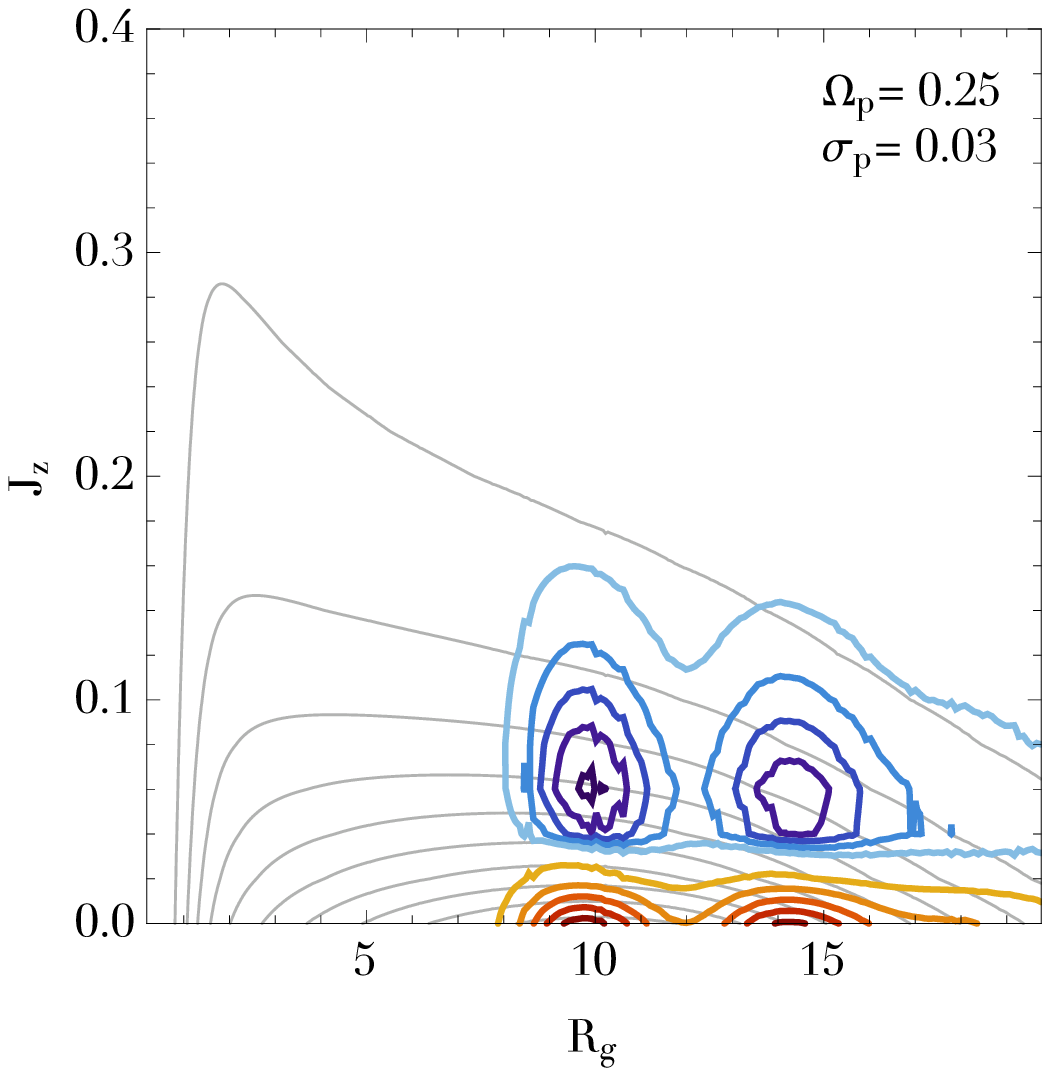,width=0.425\textwidth}}
\\
{\epsfig{file=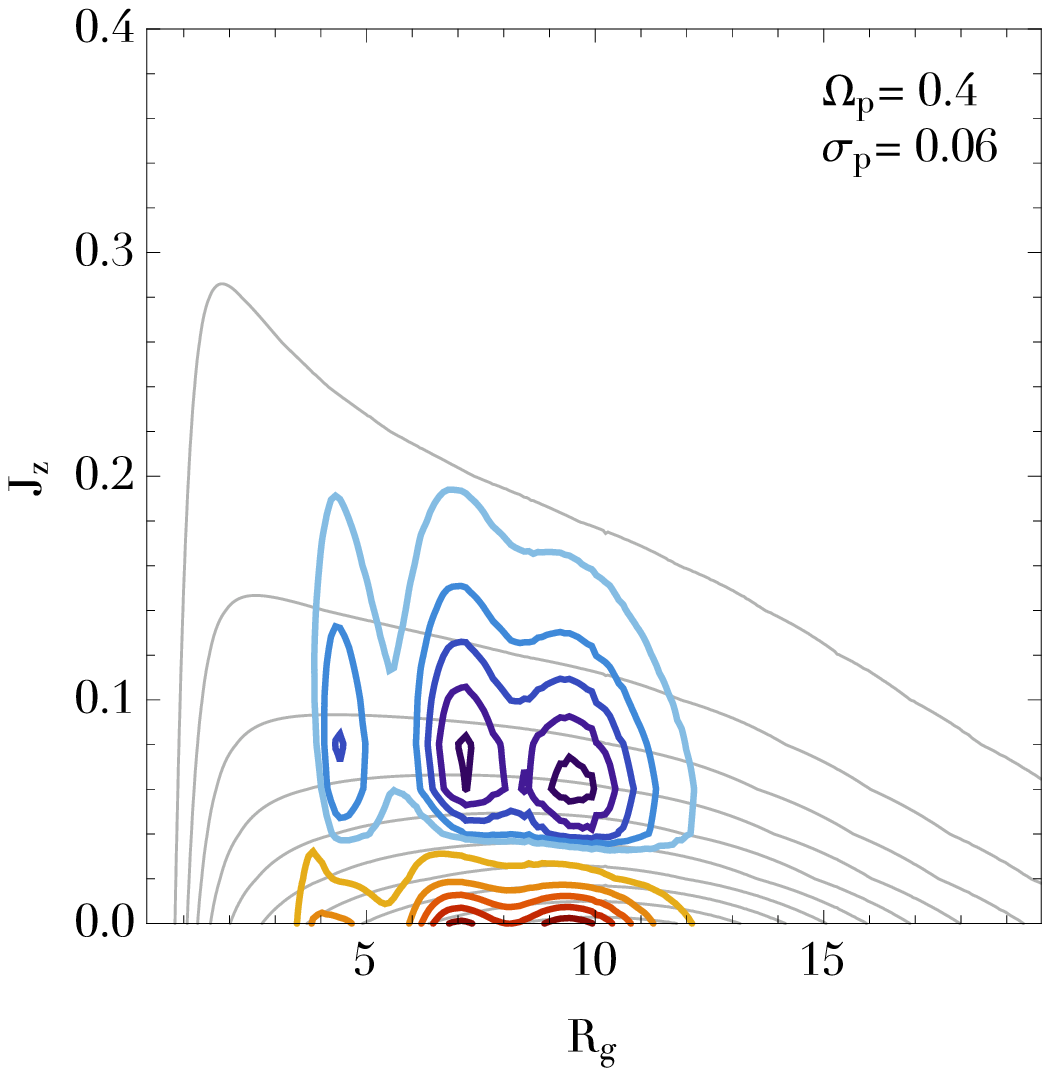,width=0.425\textwidth}}
&
{\epsfig{file=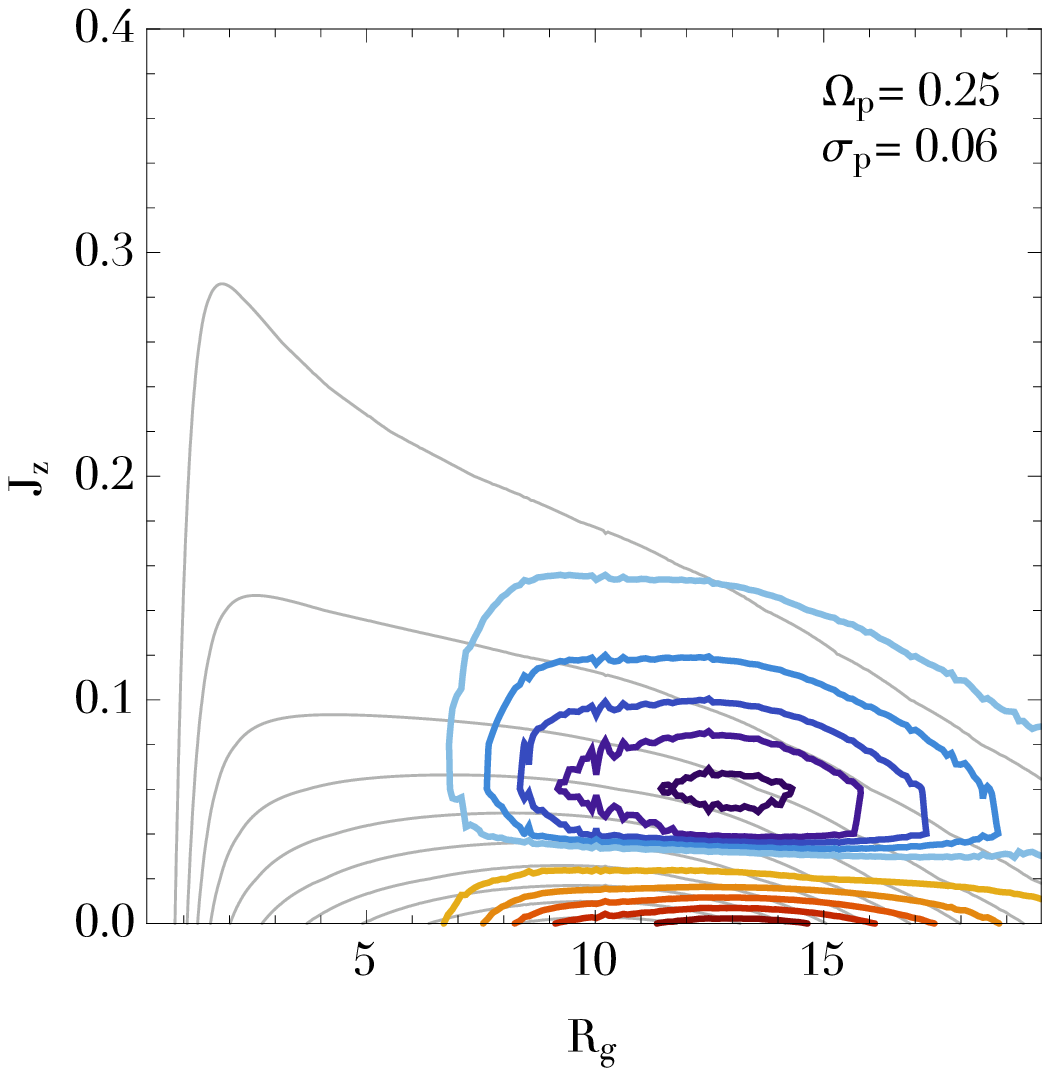,width=0.425\textwidth}}
\end{tabular}
\caption{\small{Illustration of the initial contours of ${ \partial F_{\rm Z} / \partial t |_{t = 0} }$ using the same conventions as in figure~\ref{fig_Contours_dFZdt}, when considering a secular collisionless forcing by a series of bar as in equation~\eqref{assumption_C_bar}, for different precession rates ${ \Omega_{\rp} }$ and temporal decays ${ \sigma_{\rp} }$. The diffusion in the inner regions has been turned off by considering a perturbation amplitude ${ A_{\rb} (R_{\rg}) \!=\! H [R_{\rg} \!-\! R_{\rm cut}] }$, with ${ R_{\rm cut} \!=\! 2.5 }$. The position of the various resonance radii can be determined thanks to the behaviours of the intrinsic frequencies ${ \omega \!=\! \bm{m} \!\cdot\! \bm{\Omega} }$ from figure~\ref{fig_Intrinsic_Frequencies}. \textbf{Top-left panel}: ${ \Omega_{\rp} \!=\! 0.4 }$ and ${ \sigma_{\rp} \!=\! 0.03 }$, i.e. long-lived fast bars. \textbf{Top-right panel}: ${ \Omega_{\rp} \!=\! 0.25 }$ and ${ \sigma_{\rp} \!=\! 0.03 }$, i.e. long-lived slow bars. \textbf{Bottom-left panel}: ${ \Omega_{\rp} \!=\! 0.4 }$ and ${ \sigma_{\rp} \!=\! 0.06 }$, i.e. short-lived fast bars. \textbf{Bottom-right panel}: ${ \Omega_{\rp} \!=\! 0.25 }$ and ${ \sigma_{\rp} \!=\! 0.06 }$, i.e. short-lived slow bars.
}}
\label{fig_Contours_dFZdt_Bar}
\end{figure*}

In order to investigate another mechanism of secular thickening, one may modify the perturbations sourcing the WKB collisionless diffusion coefficients from equation~\eqref{Dm_diagonalised_ASD_even}. Instead of considering the effect of shot noise as in equation~\eqref{assumption_C_shot_noise}, we may now study the secular effect of a stochastic series of central bars on the galactic disc thickness. Let us then assume that the autocorrelation of the external perturbations takes the simple form
\begin{equation}
\widehat{\mathcal{C}} [m_{\phi} , \!\omega , \!R_{\rg} , \!k_{r} , \!k_{z}\!] \!=\! \delta_{m_{\rp}}^{m_{\phi}} A_{\rb} (R_{\rg}) \exp \!\!\bigg[\! - \frac{(\!\omega \!-\! m_{\rp} \Omega_{\rp} \!)^{2}}{2 \sigma_{\rp}^{2}} \!\bigg] \, ,
\label{assumption_C_bar}
\end{equation}
where ${ m_{\rp} \!=\! 2 }$ is the pattern number of the bar, ${ \Omega_{\rp} }$ is its typical pattern speed, and ${ \sigma_{\rp} \!\sim\! 1/T_{\rb} \!\sim\! (1/\Omega_{\rp}) (\partial \Omega_{\rp} / \partial t)}$, with $T_{\rb}$ the typical bar's lifetime, describes the typical decay time of the bar frequency. The slower $\Omega_{\rp}$ evolves, the narrower the frequency window from equation~\eqref{assumption_C_bar} will be, and therefore the smaller $\sigma_{\rp}$. Finally, in equation~\eqref{assumption_C_bar}, ${ A_{\rb} (R_{\rg}) }$ is an amplitude factor depending on the position within the disc, describing the radial profile and extension of the bar. One should note that equation~\eqref{assumption_C_bar} is a rather crude assumption, since for simplicity, we neglect here any dependence w.r.t. $k_{r}$ and $k_{z}$ (which in turn implies that the perturbation is radially and vertically decorrelated). We consider the same thickened Mestel disc as in section~\ref{sec:discmodel}, perturbed by various series of bars characterised by ${ \Omega_{\rp} \!\in\! \{ 0.4 , 0.25 \} }$ and ${ \sigma_{\rp} \!\in\! \{ 0.03 , 0.06 \} }$. Finally, in order to focus on the intermediate regions of the disc, i.e. belonging to neither the bulge nor the bar, we assume that ${ A_{\rb} (R_{\rg}) \!=\! H [ R_{\rg} \!-\! R_{\rm cut} ] }$, where ${ H[x] }$ is an Heaviside function such that ${ H [x] \!=\! 1}$ for ${ x \!\geq\! 0 }$ and $0$ otherwise, and ${ R_{\rm cut} \!=\! 2.5 }$ is a truncation radius, below which the bar is present. The initial contours of ${ \partial F_{\rm Z} / \partial t |_{t = 0} }$, for these various choices of bar perturbations, are illustrated in figure~\ref{fig_Contours_dFZdt_Bar}.
From the various panels of figure~\ref{fig_Contours_dFZdt_Bar}, one should first note how the frequency selection present in the noise assumption from equation~\eqref{assumption_C_bar} tends to localise as expected the resonant ridge of enhanced thickness. This figure also emphasises how the dynamical properties of the bars may change the orbital signature of diffusion. Indeed, by comparing the left-hand panels with the right-hand ones, one recovers that the slower the bar, the further out the ridge of diffusion, i.e. as $\Omega_{\rp}$ decreases, the ridges move outwards. Similarly, by comparing the top panels with the bottom ones, one observes that the more long-lived the bars, the narrower the diffusion features, i.e. as $\sigma_{\rp}$ decreases, the ridges get sharper and do not overlay anymore. Finally, the position of the various ridges observed in figure~\ref{fig_Contours_dFZdt_Bar} can be straightforwardly interpreted thanks to figure~\ref{fig_Intrinsic_Frequencies}, which illustrates the behaviour of the resonance frequencies ${ \omega \!=\! \bm{m} \!\cdot\! \bm{\Omega} }$ as a function of the position in the disc. This allows us to determine the dominant resonance associated with each of the ridges observed in figure~\ref{fig_Contours_dFZdt_Bar}. Because shot noise perturbations as in equation~\eqref{assumption_C_shot_noise} and perturbations associated with bars as in equation~\eqref{assumption_C_bar} do not have the same spectral structure, the diffusion features observed in figures~\ref{fig_Contours_dFZdt} and~\ref{fig_Contours_dFZdt_Bar} are significantly different. The perturbations' spectral characteristics (equations~\eqref{assumption_C_shot_noise} or \eqref{assumption_C_bar}) shape the diffusion coefficients from equation~\eqref{Dm_general}.

The process of secular thickening induced by a bar-like perturbation should have a clear chemo-dynamical signature in the radial and vertical distribution of stars of a given age and vertical dispersion.
Indeed, gas inflow will (re)-generate a cold component of stars within a razor-thin disc throughout a Hubble time.
Conversely, potential fluctuations near the disc will trigger radial and vertical migrations in regions which resonate with the perturbations. Hence, depending on the spectral properties of the perturbations, the rate of star formation, the gas infall within the disc, and the underlying orbital structure, the distribution of stellar ages, metallicities and vertical velocities should reflect the net effect of all these processes.

\subsection{GMCs triggered thickening}
\label{sec:GMC}

In a realistic galactic disc, we do not expect the self-induced diffusion of stars alone to drive the disc's thickening within a Hubble time. However, the predicted collisional timescale of diffusion from section~\ref{sec:timescales} should be updated when accounting for the joint evolution of the galaxy's GMCs. So12 gives a possible scaling to physical units as
\begin{equation}
R_{\rm i} = 0.75 \, \text{kpc} \;\;\; ; \;\;\; \tau_{0} = \frac{R_{\rm i}}{V_{0}} = 3.0 \, \text{Myr} \, .
\label{units_So12}
\end{equation}
A typical Milky Way like galaxy is such that ${ N_{\rm MW} \!\sim\! 10^{11} }$. As a consequence, the rescaled time ${ \Delta \tau_{\rm So12} \!\simeq\! 2 \!\times\! 10^{-2} }$ becomes for such a system
\begin{equation}
\Delta t_{\rm MW} \simeq 6 \!\times\! 10^{6} \, \text{Gyr} \simeq 6 \!\times\! 10^{5} \, t_{\rm Hub.} \, ,
\label{time_MW}
\end{equation}
where we introduced the Hubble time as ${ t_{\rm Hub.} \!\simeq\! 10 \, \text{Gyr} }$. This shows that the mechanism of self-induced thickening of stellar disc investigated in So12 is too slow to be relevant per se for a Milky Way like galaxy.
However, it has been suggested~\citep[e.g.,][]{SpitzerSchwarzschild1953,Wielen1977,Lacey1984,BinneyLacey1988,Jenkins1992,IdaKokubo1993,ShiidsukaIda1999,HanninenFlynn2002,AumerBinney2016} that the joint evolution of a stellar disc and a population of forming and dissolving GMCs could be responsible for such thickening through local deflections.
As already emphasised in~\cite{Heyvaerts2010,Chavanis2012}, the Balescu-Lenard formalism may describe simultaneously multiple populations of various masses, while accounting by construction for transient spiral structures and non-local resonant encounters between dressed orbits. Let us emphasise that the resonant diffusion captured by the Balescu-Lenard equation is a different mechanism from the close encounters associated with the scattering mechanism from~\cite{SpitzerSchwarzschild1953}.\footnote{See~\cite{Chavanis2013} for a detailed discussion of the links between the self-consistent Balescu-Lenard equation and other kinetic theories based on two-body encounters.}
We now briefly discuss how the joint evolution of a population of stars and GMCs could lead to a global thickening of the disc on a much shorter timescale.

One can write the Balescu-Lenard equation for a system with multiple components (corresponding to say, stars and GMCs, of different mass). The different components will be indexed by the letters ${ \rm ``a" }$ and ${ \rm ``b" }$. The particles of the component ${ \rm ``a" }$ have a mass $\mu_{\rm a}$ and follow the DF $F^{\rm a}$. Each DF $F^{\rm a}$ is normalised such that ${ \int \!\! \rd \bm{x} \rd \bm{v} F^{\rm a} \!=\! M_{\rm tot}^{\rm a} }$, where $M_{\rm tot}^{\rm a}$ is the total active mass of the component ${ \rm ``a" }$. The evolution of each DF is given by
\begin{align}
& \, \frac{\partial F^{\rm a}}{\partial t} \!=\! \pi (\!2 \pi\!)^{d} \frac{\partial }{\partial \bm{J}_{1}} \!\cdot\! \bigg[ \sum\limits_{\bm{m}_{1} , \bm{m}_{2}} \!\!\!\!\bm{m}_{1}  \!\!\! \int \!\!\! \rd \bm{J}_{2} \frac{\delta_{\rm D} (\bm{m}_{1} \!\cdot\! \bm{\Omega}_{1} \!-\! \bm{m}_{2} \!\cdot\! \bm{\Omega}_{2}) }{| \mathcal{D}_{\bm{m}_{1} , \bm{m}_{2}} \!(\!\bm{J}_{1} \!, \!\bm{J}_{2} \!, \!\bm{m}_{1} \!\cdot\! \bm{\Omega}_{1}) |^{2}} \nonumber
\\
& \times \,  \sum\limits_{\rm b} \! \bigg\{ \mu_{\rm b} \, F^{\rm b} \!(\bm{J}_{2}\!) \, \bm{m}_{1} \!\cdot\! \frac{\partial F^{\rm a}}{\partial \bm{J}_{1}} \!-\! \mu_{\rm a} \, F^{\rm a} \!(\bm{J}_{1}\!) \, \bm{m}_{2} \!\cdot\! \frac{\partial F^{\rm b}}{\partial \bm{J}_{2}} \bigg\}  \bigg]  \, .
\label{LB_multi}
\end{align}
In the multi-component case, the susceptibility coefficients are still given by equation~\eqref{definition_1/D}.
However, now the response matrix encompasses all the active components of the system, so that
\begin{equation}
\widehat{\mathbf{M}}_{pq} (\omega) \!=\! (\!2 \pi\!)^{d} \! \sum\limits_{\bm{m}} \!\! \int \!\! \rd \bm{J} \, \frac{\bm{m} \!\cdot\! \partial (\sum_{\rm b} \! F^{\rm b}\!) / \partial \bm{J}}{\omega \!-\! \bm{m} \!\cdot\! \bm{\Omega}} \big[ \psi_{\bm{m}}^{(p)} (\bm{J}) \big]^{*} \psi_{\bm{m}}^{(q)} (\bm{J}) \, .
\label{Fourier_M_multi}
\end{equation}
Introducing drift and diffusion coefficients, equation~\eqref{LB_multi} may be rewritten under the form
\begin{equation}
\frac{\partial F^{\rm a}}{\partial t} \!\!=\!\! \sum\limits_{\bm{m}_{1}} \!\! \frac{\partial }{\partial \bm{J}_{1}} \!\cdot\!  \bigg[ \bm{m}_{1} \!\! \sum\limits_{\rm b} \!\!\bigg\{\! \mu_{\rm a} A_{\bm{m}_{1}}^{\rm b} \!\!(\!\bm{J}_{1}\!) F^{\rm a} \!(\!\bm{J}_{1}\!) \!+\! \mu_{\rm b} D_{\bm{m}_{1}}^{\rm b} \!\!(\!\bm{J}_{1}\!) \bm{m}_{1} \!\cdot\! \frac{\partial F^{\rm a}}{\partial \bm{J}_{1}} \!\!\bigg\} \!\!\bigg] \, ,
\label{LB_multi_short}
\end{equation}
where the drift and diffusion coefficients ${ A_{\bm{m}_{1}}^{\rm b} (\bm{J}_{1}) }$ and ${ D_{\bm{m}_{1}}^{\rm b} (\bm{J}_{1}) }$ both depend on the location $\bm{J}_{1}$ in action-space, the considered resonance $\bm{m}_{1}$ and the component ${ \rm ``b" }$ which is used as the underlying DF to estimate them. Indeed, the drift coefficients are generically given by
\begin{equation}
A_{\bm{m}_{1}}^{\rm b} \!\!(\!\bm{J}_{1}\!) \!=\! - \pi (\!2 \pi\!)^{d} \!\! \sum\limits_{\bm{m}_{2}} \!\! \int \!\!\! \rd \bm{J}_{2} \frac{\delta_{\rm D} (\bm{m}_{1} \!\cdot\! \bm{\Omega}_{1} \!-\! \bm{m}_{2} \!\cdot\! \bm{\Omega}_{2} )}{| \mathcal{D}_{\bm{m}_{1} , \bm{m}_{2}} \!(\!\bm{J}_{1} \!, \!\bm{J}_{2} \! , \!\bm{m}_{1} \!\!\cdot\! \bm{\Omega}_{1} \!) |^{2}} \, \bm{m}_{2} \!\cdot\! \frac{\partial F^{\rm b}}{\partial \bm{J}_{2}} \, ,
\label{drift_multi}
\end{equation}
while the diffusion coefficients read
\begin{equation}
D_{\bm{m}_{1}}^{\rm b} \!(\bm{J}_{1}\!) \!=\! \pi (\!2 \pi\!)^{d} \!\! \sum\limits_{\bm{m}_{2}} \!\! \int \!\!\! \rd \bm{J}_{2} \frac{\delta_{\rm D} (\bm{m}_{1} \!\cdot\! \bm{\Omega}_{1} \!-\! \bm{m}_{2} \!\cdot\! \bm{\Omega}_{2} )}{| \mathcal{D}_{\bm{m}_{1} , \bm{m}_{2}} \!(\bm{J}_{1} , \!\bm{J}_{2} , \!\bm{m}_{1} \!\!\cdot\! \bm{\Omega}_{1} \!) |^{2}} \, F^{\rm b} (\bm{J}_{2}) \, .
\label{diff_multi}
\end{equation}
One should pay attention to the fact that the drift and diffusion coefficients from equations~\eqref{drift_multi} and~\eqref{diff_multi} do not have the same dimensions as the mono-component ones introduced in equations~\eqref{initial_drift} and~\eqref{initial_diff}.
One can finally rewrite equation~\eqref{LB_multi_short} as
\begin{equation}
\frac{\partial F^{\rm a}}{\partial t} \!=\! \! \sum\limits_{\bm{m}_{1}} \!\! \frac{\partial }{\partial \bm{J}_{1}} \!\cdot\!  \bigg[ \bm{m}_{1} \!\bigg\{\! \mu_{\rm a} \, A_{\bm{m}_{1}}^{\rm tot} \!(\!\bm{J}_{1}\!) \, F^{\rm a} (\bm{J}_{1}\!) \!+\! D_{\bm{m}_{1}}^{\rm tot} \!(\!\bm{J}_{1}\!) \, \bm{m}_{1} \!\cdot\! \frac{\partial F^{\rm a}}{\partial \bm{J}_{1}} \!\bigg\}\! \bigg] \, ,
\label{LB_short_2}
\end{equation}
where we defined the total drift and diffusion coefficients $A_{\bm{m}_{1}}^{\rm tot}$ and $ D_{\bm{m}_{1}}^{\rm tot} $ as
\begin{equation}
A_{\bm{m}_{1}}^{\rm tot} (\bm{J}_{1}) = \sum\limits_{\rm b} A_{\bm{m}_{1}}^{\rm b} (\bm{J}_{1}) \;\;\;  ; \;\;\; D_{\bm{m}_{1}}^{\rm tot} (\bm{J}_{1}) = \sum\limits_{\rm b} \mu_{\rm b} \, D_{\bm{m}_{1}}^{\rm b} (\bm{J}_{1}) \, .
\label{A_D_tot_multi}
\end{equation}
In equation~\eqref{LB_short_2}, the drift coefficients are multiplied by the mass $\mu_{\rm a}$ of the considered component. This essentially captures the known process of segregation, when a spectrum of masses is involved. This can be seen for instance by seeking asymptotic stationary solutions to equation~\eqref{LB_short_2} by nulling the curly brace on the r.h.s., leading to the multi-component Boltzmann distribution.

Let us now emphasise some properties of the multi-component Balescu-Lenard equation~\eqref{LB_short_2} when considering the joint evolution of stars and GMCs in a stellar disc. Let us assume that the disc contains a total mass $M_{\rm tot}^{\star}$ of $N_{\star}$ stars of individual mass $\mu_{\star}$, described by the DF $F^{\star}$. In addition, the system contains a total mass $M_{\rm tot}^{\rG}$ of $N_{\rG}$ GMCs of individual mass $\mu_{\rG}$ described by the DF $F^{\rG}$. For simplicity, it will also be assumed that the stars and the GMCs are distributed according to a similar distribution (keeping in mind that in reality the GMCs are typically colder). Therefore, because of their normalisation, one has the relation
\begin{equation}
F^{\rG} = \frac{M_{\rm tot}^{\rG}}{M_{\rm tot}^{\star}} \, F^{\star} \, .
\label{relation_DF_GMC}
\end{equation}
One may then estimate the total drift and diffusion coefficients from equation~\eqref{A_D_tot_multi} which take the form
\begin{equation}
A_{\bm{m}_{1}}^{\rm tot} =  ( 1 \!+\! \alpha_{A} ) \, A_{\bm{m}_{1}}^{\star} \;\;\; ; \;\;\; D_{\bm{m}_{1}}^{\rm tot} = ( 1 \!+\! \alpha_{D} ) \, \mu_{\star} \, D_{\bm{m}_{1}}^{\star} \, ,
\label{A_D_tot_GMC}
\end{equation}
where we introduced the dimensionless quantities $\alpha_{A}$ and $\alpha_{D}$ as
\begin{equation}
\alpha_{A} \!=\! \frac{M_{\rm tot}^{\rG}}{M_{\rm tot}^{\star}} \!=\!  \frac{\mu_{\rG}}{\mu_{\star}} \,  \frac{N_{\rG}}{N_{\star}} \; ; \; \alpha_{D} \!=\! \frac{\mu_{\rG} M_{\rm tot}^{\rG}}{\mu_{\star} M_{\rm tot}^{\star}} \!=\! \left(\! \frac{\mu_{\rG}}{\mu_{\star}} \!\right)^{2}  \frac{N_{\rG}}{N_{\star}} \, .
\label{definition_g_BL}
\end{equation}
Thanks to equation~\eqref{LB_short_2}, the evolution equation for the stars' distribution becomes
\begin{equation}
\frac{\partial F^{\star}}{\partial t} \!\! = \!\! \sum_{\bm{m}_{1}} \!\frac{\partial }{\partial \bm{J}_{1}} \!\cdot\! \bigg[ \bm{m}_{1}  \mu_{\star} \!\bigg\{\! (\!1 \!+\! \alpha_{A}\!) A_{\bm{m}_{1}}^{\star} F^{\star} \!\!+\! (\!1 \!+\! \alpha_{D}\!) D_{\bm{m}_{1}}^{\star} \bm{m}_{1} \!\cdot\! \frac{\partial F^{\star}}{\partial \bm{J}_{1}} \!\bigg\} \!\bigg] \, ,
\label{LB_star_multi}
\end{equation}
where the dependences w.r.t. $\bm{J}_{1}$ have not been written out to simplify the notations. In equation~\eqref{LB_star_multi}, the case without GMCs can be recovered by assuming ${ \alpha_{A} \!=\! \alpha_{D} \!=\! 0 }$.~\cite{Murray2011} gives the typical current properties of the Milky Way's GMCs\footnote{A more involved modelling would also account for the expected secular variability of these populations, due to the exponential decay in the system's star formation throughout cosmic times and the rapid disappearance of GMCs.} as
\begin{equation}
\mu_{\rG} \!\simeq\! 10^{5} M_{\odot} \;\;\; ; \;\;\; N_{\rG} \!\simeq\! 10^{4} \;\;\; ; \;\;\; M_{\rm tot}^{\rG} \!\simeq\! 10^{9} M_{\odot} \, .
\label{values_Murray}
\end{equation}
As a consequence, for a Milky Way like galaxy, with ${ N_{\star} \!\simeq\! 10^{11} }$ and ${ \mu_{\star} \!\simeq\! 1 M_{\odot} }$, one obtains
\begin{equation}
\frac{\mu_{\rG}}{\mu_{\star}} \!\!\sim\!\! \frac{10^{5}}{1} \!\!\sim\!\! 10^{5}  ; \frac{N_{\rG}}{N_{\star}} \!\!\sim\!\! \frac{10^{4}}{10^{11}} \!\!\sim\!\! 10^{-7} \Longrightarrow \alpha_{A} \!\sim\! 10^{-2} \, ; \, \alpha_{D} \!\sim\! 10^{3} \, .
\label{order_MW_multi}
\end{equation}
Using the fact that ${ \alpha_{A} \!\ll\! 1 }$ and ${ \alpha_{D} \gg 1 }$, equation~\eqref{LB_star_multi} becomes
\begin{equation}
\frac{\partial F^{\star}}{\partial t} \!\! = \!\! \sum_{\bm{m}_{1}} \!\frac{\partial }{\partial \bm{J}_{1}} \!\cdot\! \bigg[ \bm{m}_{1} \, \mu_{\star} \bigg\{ A_{\bm{m}_{1}}^{\star} F^{\star} \!+ \alpha_{D} \, D_{\bm{m}_{1}}^{\star} \bm{m}_{1} \!\cdot\! \frac{\partial F^{\star}}{\partial \bm{J}_{1}} \!\bigg\} \!\bigg] .
\label{LB_star_multi_2}
\end{equation}
The presence of the GMCs therefore tends to boost the diffusion coefficients both in absolute terms and w.r.t. the drift ones. Since ${ \alpha_{D} \!\gg\! 1 }$, the GMCs will act as a catalyst and will significantly hasten the diffusion of the stars and therefore the thickening of the disc.
The multi-component Balescu-Lenard formalism captures the secular effect of multiple resonant deflections of stars by GMCs: the lighter stellar population will drift towards the high altitude ``atmosphere'' (larger ${J}_z$), while the GMCs sink in. 
If this selective boost of the diffusion component w.r.t. the drift is directly translated into the diffusion timescale of secular thickening of the disc, one obtains
\begin{equation}
\Delta t_{\rG+\star} = \frac{\Delta t_{\star}}{\alpha_{D}} \, ,
\label{boost_Delta_tau}
\end{equation}
where ${ \Delta t_{\star} }$ corresponds to the timescale of the disc's spontaneous thickening when only stars are considered, while ${ \Delta t_{\rG+\star} }$ corresponds to the case where the joint evolution of the GMCs is taken into account. The presence of the GMCs, which are less numerous but more massive than the stars, can therefore significantly alter how stars diffuse compared to the case where they diffuse alone. Let us emphasise that these considerations are generic and independent from the thickened WKB approximation presented in the previous sections.
When applied to equation~\eqref{time_MW}, the timescale boost from equation~\eqref{boost_Delta_tau} immediately translates to 
\begin{equation}
\Delta t_{\mathrm{MW}+\rG} \simeq 6 \!\times\! 10^{2} \,  t _{\rm Hub.} \, ,
\label{time_MW_G}
\end{equation}
where ${ \Delta t_{\mathrm{MW}+\rG} }$ corresponds to the timescale of thickening of a Milky Way like galaxy when the joint evolution of the GMCs is also taken into account. Equation~\eqref{time_MW_G} emphasises how the presence of GMCs tends to significantly hasten the secular thickening of stellar discs induced by discrete resonant encounters. However, despite this diffusion boost, the secular broadening mechanism described previously still appears as too slow compared to the typical lifetime of a Milky Way like galaxy.
The previous analysis therefore tends to show that the self-induced collisional mechanism of secular thickening sourced by finite${-N}$ fluctuations, captured by the Balescu-Lenard equation~\eqref{definition_BL}, and numerically studied in So12, even when accounting for the diffusion acceleration due to the presence of the more massive and less numerous GMCs, is not sufficiently rapid to lead to a significant secular thickening of a Milky Way like stellar disc on a Hubble time.~\cite{AumerBinney2016} reached a similar conclusion on the efficiency of the GMCs heating to thicken stellar discs when studying the quiescent growth of isolated galactics discs in numerical simulations. One could finally perform the same calculations to determine the typical timescale of appearance of the radial ridge observed in~\cite{Sellwood2012}. There, the radial ridge in the ${ (J_{\phi} , J_{r})-}$plane appears after a time ${ \Delta t_{\rm S12}^{\rm radial} \!=\! 1500 }$ for ${ N \!=\! 5\!\times\! 10^{7} }$ particles. The associated rescaled time of diffusion is then given by ${ \Delta \tau^{\rm radial} \!=\! 3 \!\times\! 10^{-5} }$. Relying on the physical units from equation~\eqref{units_So12}, for a Milky Way-like galaxy, the radial ridge would appear after a time ${ \Delta t_{\rm MW}^{\rm radial} \!=\! 10^{3} t_{\rm Hub.} }$, when only the stars are considered. We showed in equation~\eqref{boost_Delta_tau} that the simultaneous presence of the GMCs would hasten the system's diffusion and would therefore lead to an appearance of the radial ridge on a timescale of the order ${ \Delta t_{\rm MW+G}^{\rm radial}\!\simeq\! \Delta t_{\rm MW}^{\rm radial} / (10^{3}) \!\simeq\! t_{\rm Hub.} }$. As a consequence, while we showed in equation~\eqref{time_MW_G} that the presence of the GMCs would still not allow for the appearance of a vertical ridge on the typical lifetime of a Milky Way like galaxy, such a self-induced diffusion mechanism would be fast enough to induce a radial ridge in the galaxy's DF. This could lead for example to a signature in the Milky Way's DF, soon probed by the GAIA spacecraft.

\section{Conclusion}
\label{sec:conclusion}

The thickening of thin and thick galactic discs is the topic of very active research \citep[e.g.,][]{Minchev2015,GrandSpringel2016}.
In this context, two equations describing the orbital diffusion of a self-gravitating system were investigated: the collisionless evolution induced by external stochastic perturbations or the spontaneous collisional evolution described by the inhomogeneous Balescu-Lenard equation.
These diffusion equations were applied to a thickened tepid galactic disc.
Relying on the epicyclic approximation, their thick WKB limits were found while assuming that only radially tightly wound transient spirals are sustained by the disc. An ad hoc uniform cavity was assumed in particular in order to solve Poisson's equation in a closed form.
This yielded equation~\eqref{Dm_diagonalised_even}, a simple double quadrature for the collisionless diffusion coefficients, and equations~\eqref{final_drift} and~\eqref{final_diff} for the collisional drift and diffusion coefficients (and equations~\eqref{drift_multi} and~\eqref{diff_multi} for the multi-component counterparts), providing a straightforward understanding of the positions of maximum orbital diffusion within the disc.
 A scale-height dependent thick disc Toomre parameter was also derived correspondingly.

When applied to a shot noise perturbed tepid Toomre-stable tapered thick disc, these formalisms predict the formation of vertical ridges of resonant orbits towards larger vertical actions, in qualitative agreement with the vertical ridges identified numerically by~\cite{SolwaySellwood2012} via direct ${N-}$body simulations.
This extends the findings of~\cite{BinneyLacey1988} to the self-gravitating case, as in the present work we treat in a coherent manner the dressing of the perturbations, the associated spiral response and the induced thickening.
Potential fluctuations within the disc statistically induce a vertical bending of a subset of resonant orbits, triggering the corresponding increase in vertical velocity dispersion.
Such a process provides a possible mechanism allowing for galactic discs to thicken on secular timescales, either perturbed by their own Poisson shot noise or, e.g., by a set of dynamically dragged bars, or catalised by the joint evolution of GMCs.
In the case of decaying bars, we have shown that, as expected, the diffusion is strongest at resonances and tightest when the rate of change of the pattern is slowest.
When considering the collisional effects of GMCs, we showed that such a mechanism is not sufficiently fast to lead to a significant secular thickening of a Milky Way like galaxy on a Hubble time (see~\cite{Donghia2013} and references therein for the effects of GMCs on spiral activity).
Determining which of these processes are the dominant ones depends on the relative amplitude of the various external and internal potential fluctuations sourcing the diffusion coefficients. The amplitude of the former will have to be quantified on simulations. Both should have a clear signature in vertical metallicity gradients to be quantified by GAIA, consistent with radial churning~\citep{SellwoodBinney2002} and migration.

It should be emphasised that various approximations were made in order to reach these conclusions: 
\begin{itemize}
\item we relied on the epicyclic approximation and the plane parallel Schwarzschild approximation to build an integrable model for a tepid thickened disc.
\item we approximated the edge of the disc with a sharp edge to solve Poisson's equation vertically.
\item we relied on the WKB approximation to describe the radial component of spiral waves.
\item when computing the susceptibility of the disc, we neglected the relative importance of vertical action gradients of the DF compared to radial ones.
\item we also assumed when computing the susceptibility of the disc that the orbits are closed on resonance.
\item when considering the dressed collisionless diffusion, we assumed some partially ad hoc external source of perturbations to describe shot noise or sequences of slowing down bars.
\end{itemize}

One should keep in mind that the WKB approximation significantly underestimates the amplitude of the resonant ridges (but less so for thin rather than razor-thin disc, given the increased $Q$ number), as it cannot account for swing amplification~\citep{GoldreichLyndenBell1965a,JulianToomre1966,Toomre1981}, the strong self-gravitating amplification of unwinding perturbations.

Beyond the scope of this paper, it would be worthwhile to implement anharmonic corrections in the vertical oscillation to better account for the stiffness of the vertical potential.
As emphasised here, one should eventually not restrict one's description to WKB waves as they do not capture swing amplification which boosts the amplitude of the diffusion coefficients, and narrows the ridge.
One would then solve the exact field equations without assuming separability and deal with a full response matrix while considering both secular processes (dressed collisionless Fokker-Planck and Balescu-Lenard) simultaneously.
While it was clearly already a numerical challenge in the 2D case presented in~\cite{FouvryPichonMagorrianChavanis2015}, its implementation in 3D is all the more difficult that we do not have angle-action coordinates for thick discs beyond the epicyclic approximation. One would have to resort to constructions such as the torus machine to first build perturbatively a mapping of action space from an integrable model to the non integrable one via fits of generating functions~\citep{KaasalainenBinney1994a,KaasalainenBinney1994b}. Should chaos around regular islands become important,
one could resort to the dual stochastic Langevin formulation~\citep[see][]{FouvryPichonMagorrian2017} and account for the corresponding induced chaotic diffusion.
Finally, evolving forward in time a diffusion equation such as the Balescu-Lenard equation still remains a challenging numerical problem, in particular because of the self-consistency requirement. Indeed, as the diffusion occurs, i.e. as the system's orbital structure gets distorted, the system's drift and diffusion coefficients have to be recomputed in order to account for the new system's DF. One possibility to integrate in time such an equation is to resort to its associated stochastic Langevin rewriting~\citep{FouvryPichonMagorrian2017}, which describes the stochastic dynamics of one test star instead of the diffusion of the system's whole DF.
The choice of bar-like correlation in equation~\eqref{assumption_C_bar} would also need to be revisited in view of statistical measurements of bar formation and dissolution in cosmological simulations. More generally, it would be useful to quantify the statistics of cosmic noise at the disc length scale, extending the work of~\cite{Aubert2007}, which focused on the virial radius.
Such formalisms could also give some insight on the thickening of debris protoplanetary or galacto-centric discs in the quasi-Keplerian regime~\citep{FouvryPichonMagorrian2017}.


\subsection*{Acknowledgements}
{\small
We thank the Institute of Astronomy, Cambridge, for hospitality while this investigation was initiated.
Special thanks to James Binney for a careful reading of the manuscript at various stages of its completion. 
We are grateful to D. Lynden-Bell, J.~Magorrian, S.~Prunet, G.~Ogilvie and J.~Papaloizou for fruitful discussions, and to M.~Solway for agreeing to reproduce figure~\ref{fig_Solway}.
JBF, CP and PHC also thank the CNRS {\tt Inphyniti} program for funding.
CP thanks Clare and Churchill college, Cambridge, 
the French Oversea's program in London, and the community of
${\text{\url{http://mathematica.stackexchange.com}}}$ for their help.
Many thanks to Eric Pharabod for some figures, and St\'ephane Rouberol for customing the Horizon cluster for our purposes.
Support for Program number HST-HF2-51374 was provided by NASA through a grant from the Space Telescope Science Institute, which is operated by the Association of Universities for Research in Astronomy, Incorporated, under NASA contract NAS5-26555.
This work is partially supported by the Spin(e) grants ANR-13-BS05-0005 of the French \textit{Agence Nationale de la Recherche}, and by the LABEX Institut Lagrange de Paris (under reference ANR-10-LABX-63) which is funded by ANR-11-IDEX-0004-02.
The Horizon cluster is hosted by the Institut d'Astrophysique de Paris.\par
}

\bibliographystyle{mn2e}
\bibliography{references}

\begin{thebibliography}{91}
\expandafter\ifx\csname natexlab\endcsname\relax\def\natexlab#1{#1}\fi

\bibitem[{{Abadi} {et~al}\mbox{.}(2003){Abadi}, {Navarro}, {Steinmetz}, \&
  {Eke}}]{AbadiNavarro2003}
{Abadi} M.~G., {Navarro} J.~F., {Steinmetz} M., {Eke} V.~R., 2003, \apj, 597,
  21

\bibitem[{{Aubert} \& {Pichon}(2007)}]{Aubert2007}
{Aubert} D., {Pichon} C., 2007, \mnras, 374, 877

\bibitem[{{Aumer} {et~al}\mbox{.}(2016){Aumer}, {Binney}, \&
  {Sch{\"o}nrich}}]{AumerBinney2016}
{Aumer} M., {Binney} J., {Sch{\"o}nrich} R., 2016, \mnras, 459, 3326

\bibitem[{Balescu(1960)}]{Balescu1960}
Balescu R., 1960, Physics of Fluids, 3, 52

\bibitem[{{Barbanis} \& {Woltjer}(1967)}]{BarbanisWoltjer1967}
{Barbanis} B., {Woltjer} L., 1967, \apj, 150, 461

\bibitem[{{Binney} \& {Lacey}(1988)}]{BinneyLacey1988}
{Binney} J., {Lacey} C., 1988, \mnras, 230, 597

\bibitem[{{Binney} \& {McMillan}(2011)}]{BinneyMcMillan2011}
{Binney} J., {McMillan} P., 2011, \mnras, 413, 1889

\bibitem[{Binney \& Tremaine(2008)}]{BinneyTremaine2008}
Binney J., Tremaine S., 2008, Galactic Dynamics (Second Edition). Princeton
  University Press

\bibitem[{{Bird} {et~al}\mbox{.}(2012){Bird}, {Kazantzidis}, \&
  {Weinberg}}]{Bird2012}
{Bird} J.~C., {Kazantzidis} S., {Weinberg} D.~H., 2012, \mnras, 420, 913

\bibitem[{Born(1960)}]{Born1960}
Born M., 1960, The Mechanics of the Atom. F. Ungar Pub. Co.

\bibitem[{{Bournaud} {et~al}\mbox{.}(2009){Bournaud}, {Elmegreen}, \&
  {Martig}}]{Bournaud2009}
{Bournaud} F., {Elmegreen} B.~G., {Martig} M., 2009, \apjl, 707, L1

\bibitem[{{Bovy} {et~al}\mbox{.}(2012){Bovy}, {Rix}, {Liu}, {Hogg}, {Beers}, \&
  {Lee}}]{Bovy2012}
{Bovy} J., {Rix} H.-W., {Liu} C., {Hogg} D.~W., {Beers} T.~C., {Lee} Y.~S.,
  2012, \apj, 753, 148

\bibitem[{{Brook} {et~al}\mbox{.}(2004){Brook}, {Kawata}, {Gibson}, \&
  {Freeman}}]{Brook2004}
{Brook} C.~B., {Kawata} D., {Gibson} B.~K., {Freeman} K.~C., 2004, \apj, 612,
  894

\bibitem[{{Carlberg} \& {Sellwood}(1985)}]{CarlbergSellwood1985}
{Carlberg} R.~G., {Sellwood} J.~A., 1985, \apj, 292, 79

\bibitem[{{Chandrasekhar}(1942)}]{Chandrasekhar1942}
{Chandrasekhar} S., 1942, Principles of Stellar Dynamics. University of Chicago
  Press

\bibitem[{{Chavanis}(2012)}]{Chavanis2012}
{Chavanis} P.-H., 2012, Physica A, 391, 3680

\bibitem[{{Chavanis}(2013)}]{Chavanis2013}
{Chavanis} P.-H., 2013, \aap, 556, A93

\bibitem[{Courant \& Hilbert(1953)}]{CourantHilbert1953}
Courant R., Hilbert D., 1953, Methods of Mathematical Physics, Vol. 1. New
  York: Interscience

\bibitem[{Daubechies(1990)}]{Daubechies1990}
Daubechies I., 1990, Information Theory, IEEE Transactions on, 36, 961

\bibitem[{{Di Matteo} {et~al}\mbox{.}(2011){Di Matteo}, {Lehnert}, {Qu}, \&
  {van Driel}}]{DiMatteo2011}
{Di Matteo} P., {Lehnert} M.~D., {Qu} Y., {van Driel} W., 2011, \aap, 525, L3

\bibitem[{{D'Onghia} {et~al}\mbox{.}(2013){D'Onghia}, {Vogelsberger}, \&
  {Hernquist}}]{Donghia2013}
{D'Onghia} E., {Vogelsberger} M., {Hernquist} L., 2013, \apj, 766, 34

\bibitem[{{Fouvry} {et~al}\mbox{.}(2015{\natexlab{a}}){Fouvry}, {Binney}, \&
  {Pichon}}]{FouvryBinneyPichon2015}
{Fouvry} J.-B., {Binney} J., {Pichon} C., 2015{\natexlab{a}}, \apj, 806, 117

\bibitem[{{Fouvry} \& {Pichon}(2015)}]{FouvryPichon2015}
{Fouvry} J.-B., {Pichon} C., 2015, \mnras, 449, 1982

\bibitem[{{Fouvry} {et~al}\mbox{.}(2015{\natexlab{b}}){Fouvry}, {Pichon}, \&
  {Chavanis}}]{FouvryPichonChavanis2015}
{Fouvry} J.-B., {Pichon} C., {Chavanis} P.-H., 2015{\natexlab{b}}, \aap, 581,
  A139

\bibitem[{{Fouvry} {et~al}\mbox{.}(2017){Fouvry}, {Pichon}, \&
  {Magorrian}}]{FouvryPichonMagorrian2017}
{Fouvry} J.-B., {Pichon} C., {Magorrian} J., 2017, \aap, 598, A71

\bibitem[{{Fouvry} {et~al}\mbox{.}(2015{\natexlab{c}}){Fouvry}, {Pichon},
  {Magorrian}, \& {Chavanis}}]{FouvryPichonMagorrianChavanis2015}
{Fouvry} J.-B., {Pichon} C., {Magorrian} J., {Chavanis} P.-H.,
  2015{\natexlab{c}}, \aap, 584, A129

\bibitem[{{Fouvry} {et~al}\mbox{.}(2015{\natexlab{d}}){Fouvry}, {Pichon}, \&
  {Prunet}}]{FouvryPichonPrunet2015}
{Fouvry} J.-B., {Pichon} C., {Prunet} S., 2015{\natexlab{d}}, \mnras, 449, 1967

\bibitem[{{Freeman}(1987)}]{Freeman1987}
{Freeman} K.~C., 1987, \araa, 25, 603

\bibitem[{Gabor(1946)}]{Gabor1946}
Gabor D., 1946, Electrical Engineers, 93, 429

\bibitem[{{Gilmore} \& {Reid}(1983)}]{GilmoreReid1983}
{Gilmore} G., {Reid} N., 1983, \mnras, 202, 1025

\bibitem[{{Goldreich} \& {Lynden-Bell}(1965)}]{GoldreichLyndenBell1965a}
{Goldreich} P., {Lynden-Bell} D., 1965, \mnras, 130, 125

\bibitem[{{Goldstein}(1950)}]{Goldstein1950}
{Goldstein} H., 1950, {Classical mechanics}. Addison-Wesley

\bibitem[{Gradshteyn \& Ryzhik(2007)}]{Gradshteyn2007}
Gradshteyn I.~S., Ryzhik I.~M., 2007, Table of integrals, series, and products.
  Elsevier Academic Press

\bibitem[{{Grand} {et~al}\mbox{.}(2016){Grand}, {Springel}, {G{\'o}mez},
  {Marinacci}, {Pakmor}, {Campbell}, \& {Jenkins}}]{GrandSpringel2016}
{Grand} R.~J.~J., {Springel} V., {G{\'o}mez} F.~A., {Marinacci} F., {Pakmor}
  R., {Campbell} D.~J.~R., {Jenkins} A., 2016, \mnras, 459, 199

\bibitem[{{Griv} \& {Gedalin}(2012)}]{GrivGedalin2012}
{Griv} E., {Gedalin} M., 2012, \mnras, 422, 600

\bibitem[{{H{\"a}nninen} \& {Flynn}(2002)}]{HanninenFlynn2002}
{H{\"a}nninen} J., {Flynn} C., 2002, \mnras, 337, 731

\bibitem[{{Haywood}(2008)}]{Haywood2008}
{Haywood} M., 2008, \mnras, 388, 1175

\bibitem[{{Heyvaerts}(2010)}]{Heyvaerts2010}
{Heyvaerts} J., 2010, \mnras, 407, 355

\bibitem[{{Ida} {et~al}\mbox{.}(1993){Ida}, {Kokubo}, \&
  {Makino}}]{IdaKokubo1993}
{Ida} S., {Kokubo} E., {Makino} J., 1993, \mnras, 263, 875

\bibitem[{{Ivezi{\'c}} {et~al}\mbox{.}(2008){Ivezi{\'c}}, {Sesar}, {Juri{\'c}},
  {Bond}, {Dalcanton}, {Rockosi}, {Yanny}, {Newberg}, {Beers}, {Allende
  Prieto}, {Wilhelm}, {Lee}, {Sivarani}, {Norris}, {Bailer-Jones}, {Re
  Fiorentin}, {Schlegel}, {Uomoto}, {Lupton}, {Knapp}, {Gunn}, {Covey},
  {Smith}, {Miknaitis}, {Doi}, {Tanaka}, {Fukugita}, {Kent}, {Finkbeiner},
  {Munn}, {Pier}, {Quinn}, {Hawley}, {Anderson}, {Kiuchi}, {Chen}, {Bushong},
  {Sohi}, {Haggard}, {Kimball}, {Barentine}, {Brewington}, {Harvanek},
  {Kleinman}, {Krzesinski}, {Long}, {Nitta}, {Snedden}, {Lee}, {Harris},
  {Brinkmann}, {Schneider}, \& {York}}]{Izevic2008}
{Ivezi{\'c}} {\v Z}. {et~al.}, 2008, \apj, 684, 287

\bibitem[{{Jeans}(1929)}]{Jeans1929}
{Jeans} J., 1929, Astronomy and Cosmogony. Cambridge University Press

\bibitem[{{Jenkins}(1992)}]{Jenkins1992}
{Jenkins} A., 1992, \mnras, 257, 620

\bibitem[{{Julian} \& {Toomre}(1966)}]{JulianToomre1966}
{Julian} W.~H., {Toomre} A., 1966, \apj, 146, 810

\bibitem[{{Juri{\'c}} {et~al}\mbox{.}(2008){Juri{\'c}}, {Ivezi{\'c}}, {Brooks},
  {Lupton}, {Schlegel}, {Finkbeiner}, {Padmanabhan}, {Bond}, {Sesar},
  {Rockosi}, {Knapp}, {Gunn}, {Sumi}, {Schneider}, {Barentine}, {Brewington},
  {Brinkmann}, {Fukugita}, {Harvanek}, {Kleinman}, {Krzesinski}, {Long},
  {Neilsen}, {Nitta}, {Snedden}, \& {York}}]{Juric2008}
{Juri{\'c}} M. {et~al.}, 2008, \apj, 673, 864

\bibitem[{{Kaasalainen} \&
  {Binney}(1994{\natexlab{a}})}]{KaasalainenBinney1994a}
{Kaasalainen} M., {Binney} J., 1994{\natexlab{a}}, Phys. Rev. Lett., 73, 2377

\bibitem[{{Kaasalainen} \&
  {Binney}(1994{\natexlab{b}})}]{KaasalainenBinney1994b}
{Kaasalainen} M., {Binney} J., 1994{\natexlab{b}}, \mnras, 268, 1033

\bibitem[{{Kalnajs}(1965)}]{Kalnajs1965}
{Kalnajs} A.~J., 1965, Ph.D. thesis. Harvard University

\bibitem[{{Kalnajs}(1976)}]{Kalnajs1976II}
{Kalnajs} A.~J., 1976, \apj, 205, 745

\bibitem[{{Lacey}(1984)}]{Lacey1984}
{Lacey} C.~G., 1984, \mnras, 208, 687

\bibitem[{Lenard(1960)}]{Lenard1960}
Lenard A., 1960, Annals of Physics, 10, 390

\bibitem[{{Lin} \& {Shu}(1966)}]{LinShu1966}
{Lin} C.~C., {Shu} F.~H., 1966, Proc. Natl. Acad. Sci. USA, 55, 229

\bibitem[{{Liouville}(1837)}]{Liouville1837}
{Liouville} J., 1837, J. Math. Pures Appl., 1, 16

\bibitem[{{Loebman} {et~al}\mbox{.}(2011){Loebman}, {Ro{\v s}kar},
  {Debattista}, {Ivezi{\'c}}, {Quinn}, \& {Wadsley}}]{LoebmanRoskar2011}
{Loebman} S.~R., {Ro{\v s}kar} R., {Debattista} V.~P., {Ivezi{\'c}} {\v Z}.,
  {Quinn} T.~R., {Wadsley} J., 2011, \apj, 737, 8

\bibitem[{{Lynden-Bell} \& {Kalnajs}(1972)}]{LyndenBell1972}
{Lynden-Bell} D., {Kalnajs} A.~J., 1972, \mnras, 157, 1

\bibitem[{{Meza} {et~al}\mbox{.}(2005){Meza}, {Navarro}, {Abadi}, \&
  {Steinmetz}}]{Meza2005}
{Meza} A., {Navarro} J.~F., {Abadi} M.~G., {Steinmetz} M., 2005, \mnras, 359,
  93

\bibitem[{{Minchev} {et~al}\mbox{.}(2013){Minchev}, {Chiappini}, \&
  {Martig}}]{Minchev2013}
{Minchev} I., {Chiappini} C., {Martig} M., 2013, \aap, 558, A9

\bibitem[{{Minchev} {et~al}\mbox{.}(2014){Minchev}, {Chiappini}, \&
  {Martig}}]{Minchev2014}
{Minchev} I., {Chiappini} C., {Martig} M., 2014, \aap, 572, A92

\bibitem[{{Minchev} \& {Famaey}(2010)}]{MinchevFamaey2010}
{Minchev} I., {Famaey} B., 2010, \apj, 722, 112

\bibitem[{{Minchev} {et~al}\mbox{.}(2012){Minchev}, {Famaey}, {Quillen},
  {Dehnen}, {Martig}, \& {Siebert}}]{MinchevFamaey2012}
{Minchev} I., {Famaey} B., {Quillen} A.~C., {Dehnen} W., {Martig} M., {Siebert}
  A., 2012, \aap, 548, A127

\bibitem[{{Minchev} {et~al}\mbox{.}(2015){Minchev}, {Martig}, {Streich},
  {Scannapieco}, {de Jong}, \& {Steinmetz}}]{Minchev2015}
{Minchev} I., {Martig} M., {Streich} D., {Scannapieco} C., {de Jong} R.~S.,
  {Steinmetz} M., 2015, \apjl, 804, L9

\bibitem[{{Minchev} \& {Quillen}(2006)}]{MinchevQuillen2006}
{Minchev} I., {Quillen} A.~C., 2006, \mnras, 368, 623

\bibitem[{{Monari} {et~al}\mbox{.}(2016){Monari}, {Famaey}, \&
  {Siebert}}]{Monari2016}
{Monari} G., {Famaey} B., {Siebert} A., 2016, \mnras, 457, 2569

\bibitem[{{Murray}(2011)}]{Murray2011}
{Murray} N., 2011, \apj, 729, 133

\bibitem[{{Noguchi}(1998)}]{Noguchi1998}
{Noguchi} M., 1998, \nat, 392, 253

\bibitem[{O'Leary \& Stewart(1990)}]{OLearyStewart1990}
O'Leary D., Stewart G., 1990, J. Comput. Phys., 90, 497

\bibitem[{Palmer(1994)}]{Palmer1994}
Palmer P., 1994, Stability of Collisionless Stellar Systems. Springer
  Netherlands

\bibitem[{{Palmer} {et~al}\mbox{.}(1989){Palmer}, {Papaloizou}, \&
  {Allen}}]{Palmer1989}
{Palmer} P.~L., {Papaloizou} J., {Allen} A.~J., 1989, \mnras, 238, 1281

\bibitem[{{Pichon} \& {Aubert}(2006)}]{PichonAubert2006}
{Pichon} C., {Aubert} D., 2006, \mnras, 368, 1657

\bibitem[{{Polyachenko} \& {Shukhman}(1982)}]{PolyachenkoShukman1982}
{Polyachenko} V.~L., {Shukhman} I.~G., 1982, \sovast, 26, 140

\bibitem[{{Purcell} {et~al}\mbox{.}(2011){Purcell}, {Bullock}, {Tollerud},
  {Rocha}, \& {Chakrabarti}}]{Purcell2011}
{Purcell} C.~W., {Bullock} J.~S., {Tollerud} E.~J., {Rocha} M., {Chakrabarti}
  S., 2011, \nat, 477, 301

\bibitem[{{Quillen} {et~al}\mbox{.}(2009){Quillen}, {Minchev},
  {Bland-Hawthorn}, \& {Haywood}}]{Quillen2009}
{Quillen} A.~C., {Minchev} I., {Bland-Hawthorn} J., {Haywood} M., 2009, \mnras,
  397, 1599

\bibitem[{{Quinn} {et~al}\mbox{.}(1993){Quinn}, {Hernquist}, \&
  {Fullagar}}]{Quinn1993}
{Quinn} P.~J., {Hernquist} L., {Fullagar} D.~P., 1993, \apj, 403, 74

\bibitem[{{Romeo}(1992)}]{Romeo1992}
{Romeo} A.~B., 1992, \mnras, 256, 307

\bibitem[{{Sch{\"o}nrich} \&
  {Binney}(2009{\natexlab{a}})}]{SchonrichBinney2009a}
{Sch{\"o}nrich} R., {Binney} J., 2009{\natexlab{a}}, \mnras, 396, 203

\bibitem[{{Sch{\"o}nrich} \&
  {Binney}(2009{\natexlab{b}})}]{SchonrichBinney2009b}
{Sch{\"o}nrich} R., {Binney} J., 2009{\natexlab{b}}, \mnras, 399, 1145

\bibitem[{{Sellwood}(2012)}]{Sellwood2012}
{Sellwood} J.~A., 2012, \apj, 751, 44

\bibitem[{{Sellwood} \& {Binney}(2002)}]{SellwoodBinney2002}
{Sellwood} J.~A., {Binney} J.~J., 2002, \mnras, 336, 785

\bibitem[{{Sellwood} \& {Carlberg}(1984)}]{SellwoodCarlberg1984}
{Sellwood} J.~A., {Carlberg} R.~G., 1984, \apj, 282, 61

\bibitem[{{Shiidsuka} \& {Ida}(1999)}]{ShiidsukaIda1999}
{Shiidsuka} K., {Ida} S., 1999, \mnras, 307, 737

\bibitem[{{Solway} {et~al}\mbox{.}(2012){Solway}, {Sellwood}, \&
  {Sch{\"o}nrich}}]{SolwaySellwood2012}
{Solway} M., {Sellwood} J.~A., {Sch{\"o}nrich} R., 2012, \mnras, 422, 1363

\bibitem[{{Spitzer}(1942)}]{Spitzer1942}
{Spitzer}, Jr. L., 1942, \apj, 95, 329

\bibitem[{{Spitzer} \& {Schwarzschild}(1953)}]{SpitzerSchwarzschild1953}
{Spitzer}, Jr. L., {Schwarzschild} M., 1953, \apj, 118, 106

\bibitem[{{Toomre}(1964)}]{Toomre1964}
{Toomre} A., 1964, \apj, 139, 1217

\bibitem[{{Toomre}(1981)}]{Toomre1981}
{Toomre} A., 1981, in Structure and Evolution of Normal Galaxies, pp. 111--136

\bibitem[{{Toth} \& {Ostriker}(1992)}]{Toth1992}
{Toth} G., {Ostriker} J.~P., 1992, \apj, 389, 5

\bibitem[{{Vandervoort}(1970)}]{Vandervoort1970}
{Vandervoort} P.~O., 1970, \apj, 161, 87

\bibitem[{{Villalobos} \& {Helmi}(2008)}]{Villalobos2008}
{Villalobos} {\'A}., {Helmi} A., 2008, \mnras, 391, 1806

\bibitem[{{Weinberg}(2001)}]{Weinberg2001a}
{Weinberg} M.~D., 2001, \mnras, 328, 311

\bibitem[{{Weinberg}(2015)}]{Weinberg2015I}
{Weinberg} M.~D., 2015, ArXiv e-prints

\bibitem[{{Wielen}(1977)}]{Wielen1977}
{Wielen} R., 1977, \aap, 60, 263

\bibitem[{{Yoachim} \& {Dalcanton}(2006)}]{YoachimDalcanton2006}
{Yoachim} P., {Dalcanton} J.~J., 2006, \aj, 131, 226

\end{thebibliography}

\appendix

\section{Antisymmetric basis }
\label{sec:appendixantisymbasis}

 Section~\ref{sec:thickbasiselements} was restricted to symmetric basis elements. A very similar construction can also be made for antisymmetric basis elements. Assuming ${ \psi_{z} (-z) \!=\! - \psi_{z} (z) }$, the ansatz from equation~\eqref{ansatz_psi_z} leads to ${ D \!=\! -A }$ and ${ C \!=\! - B }$, so that the system from equation~\eqref{continuity_conditions_even} becomes
\begin{equation}
\begin{cases}
\displaystyle A \re^{- k_{r} h} = 2 \ri B \sin (k_{z} h) \, ,
\\
\displaystyle k_{r} A \re^{- k_{r} h} = - 2 \ri k_{z} B \cos (k_{z} h) \, .
\end{cases}
\label{continuity_conditions_odd}
\end{equation}
Similarly to equation~\eqref{quantisation_even}, it imposes the quantisation relation
\begin{equation}
\tan (k_{z} h) = - \frac{k_{z}}{k_{r}} \, .
\label{quantisation_odd}
\end{equation}
It then leads to the same typical step distance ${ \Delta k_{z} }$ as in equation~\eqref{step_distance_kz}.
Similarly to equations~\eqref{full_psi_p_even} and~\eqref{full_rho_p_even}, the full expressions of the antisymmetric potential and density basis elements can straightforwardly be obtained as
\begin{align}
\psi^{[k_{\phi} , k_{r} , R_{0} , n]} & \, (R , \phi , z) = \mathcal{A} \, \psi_{r}^{[k_{\phi} , k_{r} , R_{0}]} (R , \phi) \nonumber
\\
& \times \,
\begin{cases}
\begin{aligned}
\displaystyle & \! \sin (k_{z}^{n} z) & \!\!\!\! \text{if}& \; |z| \leq h \, ,
\\
\displaystyle & \! \re^{k_{r} h} \! \sin (k_{z}^{n} h) \, \re^{- k_{r} |z|} & \!\!\!\! \text{if}& \; z \geq h \, ,
\\
\displaystyle & \! \!-\! \re^{k_{r} h} \! \sin (k_{z}^{n} h) \, \re^{- k_{r} |z|} & \!\!\!\! \text{if}& \; z \leq h \, .
\end{aligned}
\end{cases}
\label{full_psi_p_odd}
\end{align}
and
\begin{align}
\rho^{[k_{\phi} , k_{r} , R_{0} , n]} (R , \phi , z) = & \, - \frac{k_{r}^{2} \!+\! (k_{z}^{n})^{2}}{4 \pi G} \nonumber
\\
& \times \psi^{[k_{\phi} , k_{r} , R_{0} , n]} (R , \phi , z) \, \Theta \!\bigg[\! \frac{z}{h} \!\bigg] \, .
\label{full_rho_p_odd}
\end{align}
As for the symmetric case, the relative orthogonality of the antisymmetric elements is immediately satisfied. In addition, for a given set of indices ${ [ k_{\phi} , k_{r} , R_{0} ] }$, the symmetric elements are naturally orthogonal w.r.t. the antisymmetric ones. As a consequence, the thick WKB basis, when extended with the antisymmetric basis elements, still constitutes a biorthogonal basis. In analogy with equation~\eqref{amplitude_basis_even}, the amplitude of the antisymmetric basis elements is given by
\begin{equation}
\mathcal{A} = \sqrt{\frac{G}{R_{0} h (k_{r}^{2} \!+\! (k_{z}^{n})^{2})}} \, \beta_{n} \, ,
\label{amplitude_basis_odd}
\end{equation}
where similarly to equation~\eqref{alpha_even}, $\beta_{n}$ is a numerical prefactor given by
\begin{equation}
\beta_{n} = \sqrt{\frac{2}{1 \!-\! \sin (2 k_{z}^{n} h) / (2 k_{z}^{n} h )}} \, .
\label{beta_odd}
\end{equation}
Note that in the antisymmetric case, the quantisation relation~\eqref{quantisation_odd} imposes ${ k_{z}^{1} \!>\! \pi /(2h) }$ (see figure~\ref{fig_quantisation}), so that in this domain ${ 1.3 \!\lesssim\! \beta_{n} \!\lesssim\! 1.5 }$. Following equation~\eqref{psi_m_even}, the Fourier transformed antisymmetric basis elements read
\begin{align}
\psi_{\bm{m}}^{[k_{\phi} , k_{r} , R_{0} , n]} (\bm{J}) = & \, \delta_{m_{\phi}}^{k_{\phi}} \, \delta_{m_{z}}^{\rm odd} \, \mathcal{A} \, \re^{\ri k_{r} R_{\rg}} \, \ri^{m_{z} - 1 - m_{r}} \, \mathcal{B}_{R_{0}} (R_{\rg}) \nonumber
\\
& \, \times \, \mathcal{J}_{m_{r}} \!\bigg[\! \sqrt{\tfrac{2 J_{r}}{\kappa}} k_{r} \!\bigg] \, \mathcal{J}_{m_{z}} \!\bigg[\! \sqrt{\tfrac{2 J_{z}}{\nu}} k_{z}^{n} \!\bigg] \, . 
\label{psi_m_odd}
\end{align}

\section{A diagonal response matrix?}
\label{sec:appendixMdiagonal}

In this Appendix, let us detail why it may be assumed as in equation~\eqref{Matrix_diagonal} that the system's response matrix is diagonal. First of all, because the symmetric (resp. antisymmetric) Fourier transformed basis elements from equation~\eqref{psi_m_even} (resp. equation~\eqref{psi_m_odd}) involve a $\delta_{m_{z}}^{\rm even}$ (resp. $\delta_{m_{z}}^{\rm odd}$), one may immediately conclude that the response matrix coefficients from equation~\eqref{Fourier_M} are equal to zero as soon as the two considered basis elements do not have the same symmetry. As a consequence, the symmetric and antisymmetric cases may be treated separately.

\begin{figure}
\begin{center}
\begin{tabular}{@{}c@{}}
\epsfig{file=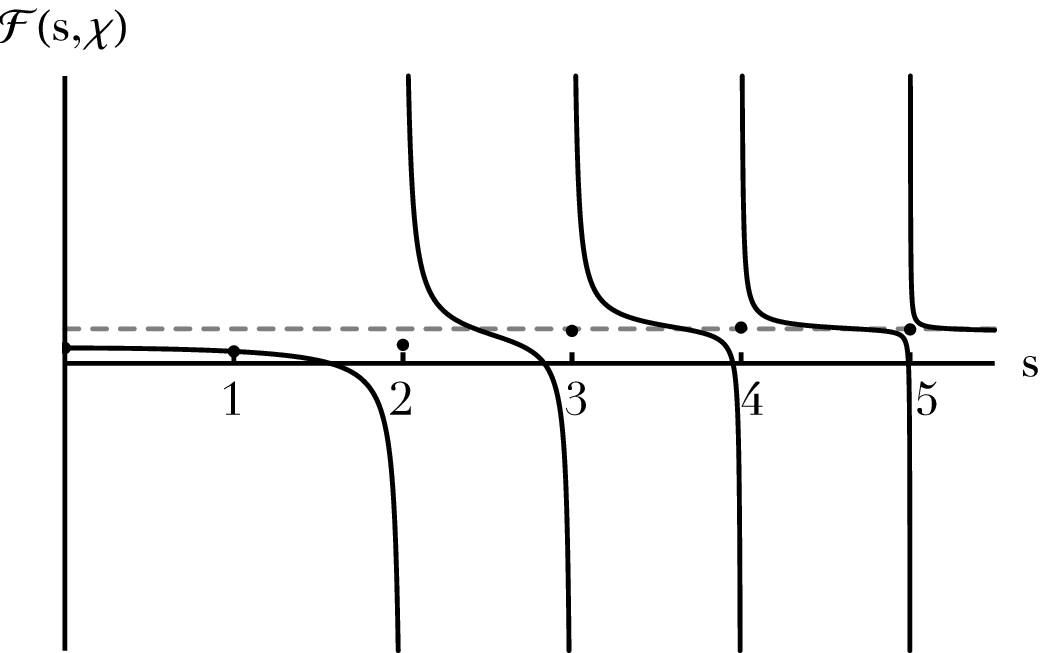,width=0.45\textwidth}
\\
\epsfig{file=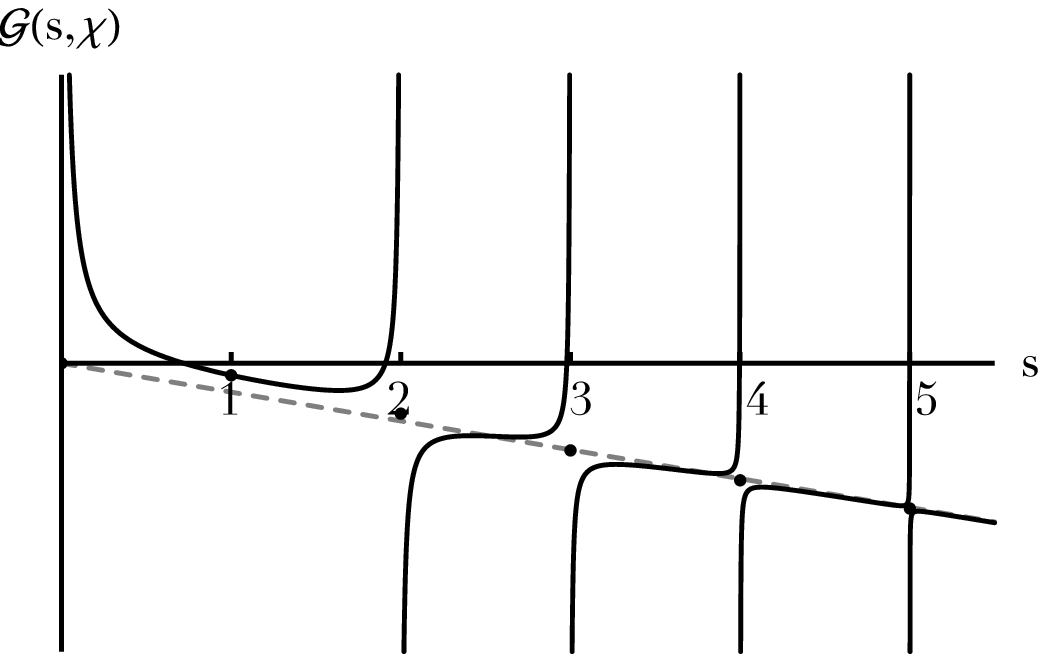,width=0.45\textwidth}
\end{tabular}
\caption{\small{Illustration of the behaviours of the functions ${ s \!\mapsto\! \mathcal{F} (s , \chi) }$ and ${ s \!\mapsto\! \mathcal{G} (s , \chi) }$ (black curves), for a given value ${ \chi \!=\! 1 }$, along with their approximations from equation~\eqref{asymptotics_F_G} (grey lines). One should note the diverging branches of these functions in the neighbourhood of integers. However, these functions are well defined when evaluated for integer values of $s$, provided that one considers ${ \lim_{\eta \to 0} \text{Re} [ \mathcal{F} (n \!+\! \ri \eta , \chi) ] }$ (similarly for $\mathcal{G}$), as illustrated with the black dots.
}}
\label{fig_FandG}
\end{center}
\end{figure}

\begin{figure}
\begin{center}
\epsfig{file=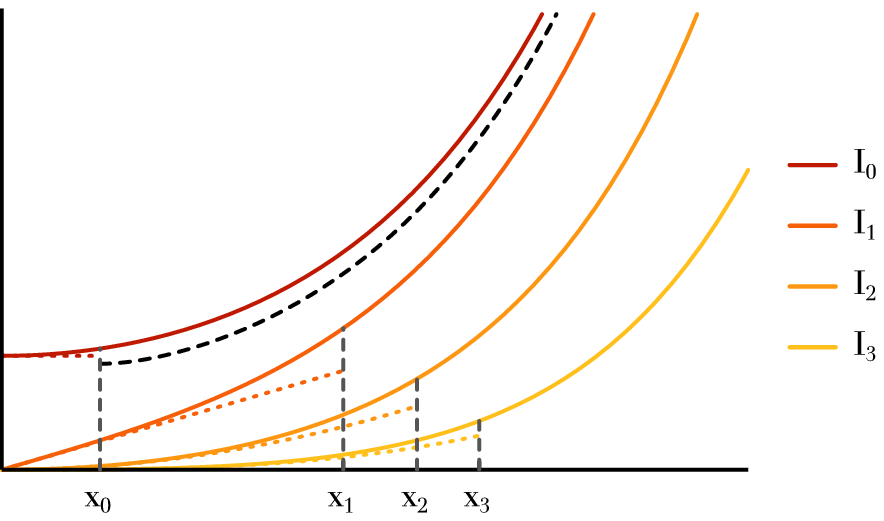,width=0.45\textwidth}
\caption{\small{Illustration of the asymptotics of the modified Bessel functions of the first kind from equation~\eqref{asymptotics_Bessel}. The full lines are the four first Bessel functions, alongside with their polynomial approximations in zero (dashed curves). The black dashed curve is their exponential approximation.
}}
\label{fig_AsymptoticsBessel}
\end{center}
\end{figure}

The basis elements from equation~\eqref{definition_psi_p} depend on four indices ${ [ k_{\phi} , k_{r} , R_{0} , n] }$. As was obtained in FPP15 by relying on the tight-winding approximation, the response matrix can be considered as diagonal w.r.t. the indices ${ [ k_{\phi} , k_{r} , R_{0} ] }$. Therefore, for a given set ${ [k_{\phi} , k_{r} , R_{0}] }$, it remains to check whether or not the response matrix is diagonal w.r.t. the $k_{z}^{n}$ index.
It is straightforward to generalise the expression~\eqref{lambda_even} of the symmetric diagonal coefficients to the non-diagonal ones as
\begin{align}
\widehat{\mathbf{M}}_{pq} = & \, \frac{2 \pi G \Sigma \alpha_{p} \alpha_{q}}{h \kappa^{2} \sqrt{(1 \!+\! (k_{z}^{p}/k_{r})^{2}) (1 \!+\! (k_{z}^{q} / k_{r})^{2})}} \nonumber
\\
& \, \times \sum_{\ell_{z} \rm even} \!\! \exp \!\bigg[\! - \frac{(k_{z}^{p})^{2} \!+\! (k_{z}^{q})^{2}}{2 \nu^{2} / \sigma_{z}^{2}} \!\bigg] \, \mathcal{I}_{\ell_{z}} \!\bigg[\! \frac{k_{z}^{p} k_{z}^{q}}{\nu^{2} / \sigma_{z}^{2}} \!\bigg] \nonumber
\\
& \, \times \frac{1}{(1 \!-\! s_{\ell_{z}}^{2})} \bigg\{\! \mathcal{F} (s_{\ell_{z}} , \chi_{r}) \!-\! \ell_{z} \frac{\nu}{\sigma_{z}^{2}} \frac{\sigma_{r}^{2}}{\kappa} \mathcal{G} (s_{\ell_{z}} , \chi_{r} ) \!\bigg\} \, . 
\label{off_diagonal_M}
\end{align}
As in equation~\eqref{lambda_odd}, the expression of the antisymmetric non-diagonal matrix coefficients can straightforwardly be obtained from equation~\eqref{off_diagonal_M} by making the substitution ${ \alpha \!\to\! \beta }$ and restricting the sum on $\ell_{z}$ only to odd values. Thanks to its symmetry, showing that the response matrix may be assumed as diagonal amounts to proving that for ${ p \!\neq\! q }$, one has ${ \widehat{\mathbf{M}}_{pq} \!\ll\! \widehat{\mathbf{M}}_{pp} }$. In order to perform such a comparison, one has to focus on the quantities which depend on $k_{z}^{p}$ and $k_{z}^{q}$ in equation~\eqref{off_diagonal_M}. Let us therefore define the dimensionless quantity $K_{pq}^{(\ell_{z})}$ as
\begin{align}
K_{pq}^{(\ell_{z})} = & \, \frac{1}{\sqrt{(1 \!+\! (k_{z}^{p} / k_{r})^{2}) (1 \!+\! (k_{z}^{q} / k_{r})^{2})}} \nonumber
\\
& \, \times \exp \!\! \bigg[\! - \frac{(k_{z}^{p})^{2} \!+\! (k_{z}^{q})^{2}}{2 \nu^{2} / \sigma_{z}^{2}} \!\bigg] \, \mathcal{I}_{\ell_{z}} \!\bigg[\! \frac{k_{z}^{p} k_{z}^{q}}{\nu^{2} / \sigma_{z}^{2}} \!\bigg] \, .
\label{definition_K_M}
\end{align}
Equation~\eqref{definition_K_M} does account for the prefactors $\alpha_{p}$ and $\alpha_{q}$ as they are always of order unity. While present in equation~\eqref{off_diagonal_M}, note that the definition of $K_{pq}^{(\ell_{z})}$ from equation~\eqref{definition_K_M} does not involve the terms ${ \mathcal{F} (s_{\ell_{z}} , \chi_{r}) }$, ${ \mathcal{G} (s_{\ell_{z}} , \chi_{r}) }$ and ${ 1 / (1 \!-\! s_{\ell_{z}}^{2}) }$ since they do not depend on the choice of $k_{z}^{p}$ and $k_{z}^{q}$. 
As illustrated in figure~\ref{fig_FandG}, the functions ${ s_{\ell_{z}} \!\mapsto\! \mathcal{F}(s_{\ell_{z}}, \chi_{r}) }$ and
${ s_{\ell_{z}} \!\mapsto\! \mathcal{G}(s_{\ell_{z}}, \chi_{r}) }$ are ill-defined when $s_{\ell_{z}}$ is an integer.
To regularise these values, a small imaginary part must be added to $s_{\ell_{z}}$. However, while regularising the values of these functions for exact integers, this procedure does not prevent the divergences of $\mathcal{F}$ and $\mathcal{G}$ in the neighbourhood of integers. To avoid these diverging behaviours, the functions $\mathcal{F}$ and $\mathcal{G}$ are approximated by smooth functions as
\begin{equation}
\mathcal{F}(s_{\ell_{z}}, \chi_{r}) \simeq f_{r} \;\;\; ; \;\;\; \mathcal{G}(s_{\ell_{z}},\chi_{r}) \simeq - g_{r} \, s_{\ell_{z}} \, ,
\label{asymptotics_F_G}
\end{equation}
where $f_{r}$ and $g_{r}$ do not depend on $s_{\ell_{z}}$. This is illustrated in figure~\ref{fig_FandG}.
When evaluating the response matrix to compute the collisionless diffusion coefficients from equation~\eqref{Dm_general} or the collisional dressed susceptibility coefficients from equation~\eqref{definition_1/D}, one has to consider $\omega$ to be at
resonance so that ${ \omega \!=\! \bm{m} \!\cdot\! \bm{\Omega} }$. In that situation,
\begin{equation}
s_{\ell_{z}} = m_{r} \!+\! (m_{z} \!-\! \ell_{z}) \frac{\nu}{\kappa} \, .
\label{s_resonance}
\end{equation}
Therefore, the value of $s_{\ell_{z}}$ is either an integer (for $\ell_{z} \!=\! m_{z}$) or far from one,
provided that ${ \nu / \kappa }$ is of high rational order. This justifies the
approximations from equation~\eqref{asymptotics_F_G}. One may then cut the sum on $\ell_{z}$ defining $\widehat{\mathbf{M}}_{pq}$ in equation~\eqref{off_diagonal_M} according to the resulting powers of $\ell_{z}$. In order to prove that for ${ p \!\neq\! q }$,
one has ${ \widehat{\mathbf{M}}_{pq} \!\ll\! \widehat{\mathbf{M}}_{pp} }$, one is left to prove that 
\begin{equation}
S_{\gamma} (p,q) = \sum_{\ell_{z} } \frac{\ell_{z}^{\gamma} K_{pq}^{(\ell_{z})}}{1 \!-\! s_{\ell_{z}}^{2}} \!\ll\! S_{\gamma} (p,p) \, ,
\label{definition_Sgamma}
\end{equation}
where the power index $\gamma$ is such that ${ \gamma \!\in\! \{ 0, 1, 2 \} }$. To further dedimensionalise the problem, let us define the typical dynamical height of the disc ${ d \!=\! \sigma_{z} / \nu }$ and introduce the dimensionless quantities $\ell_{p}$, $\ell_{q}$ and $\ell_{r}$ as
\begin{equation}
\ell_{p} = k_{z}^{p} d \;\;\; ; \;\;\; \ell_{q} = k_{z}^{q} d \;\;\; ; \;\;\; \ell_{r} = k_{r} d \, ,
\label{definition_ell_M}
\end{equation}
so that the expression~\eqref{definition_K_M} of $K_{pq}^{(\ell_{z})}$ may be rewritten as
\begin{equation}
K_{pq}^{(\ell_{z})} = \frac{ \mathcal{I}_{\ell_{z}} [ \ell_{p} \ell_{q} ] \,\, \re^{ - (\ell_{p}^{2} + \ell_{q}^{2})/2}}{\sqrt{(1 \!+\! (\ell_{p} / \ell_{r})^{2}) (1 \!+\! (\ell_{q} / \ell_{r})^{2})}} \, .
\label{definition_K_M_dedimensions}
\end{equation}
As illustrated in figure~\ref{fig_quantisation}, the quantisation of the vertical frequencies implies that the fundamental symmetric mode plays a different role than all the other quantised frequencies (both symmetric and antisymmetric), since it is the only frequency inferior to ${ \pi/(2h) }$. In order to emphasise this specific role, let us renumber, in this Appendix only, the indices $p$, such that ${ p \!=\! 0 }$ corresponds to the fundamental symmetric mode, while ${ p \!\geq\! 1 }$ corresponds to the other quantised frequencies superior to ${ \pi / (2h) }$. With such a choice, the numbering of the antisymmetric basis elements only starts at ${ p \!=\! 1 }$. As shown in figure~\ref{fig_quantisation}, one has the inequalities
\begin{equation}
0 \!<\! k_{z}^{0} \!<\! \frac{\pi}{2 h} \; ; \; \frac{(p \!-\! \tfrac{1}{2}) \pi}{h} \!<\! k_{z}^{p} \!<\! \frac{(p \!+\! \tfrac{1}{2}) \pi}{h} \;\; \text{(for ${ p \!\geq\! 1 }$)}\, .
\label{inequality_kz_M}
\end{equation}
Finally note that in the infinitely thin limit, equation~\eqref{kz1_even} has the asymptotic behaviour ${ k_{z}^{0} \!\sim\! \sqrt{k_{r} / h} }$. Given the relation~\eqref{link_sigmaz_z0_nu} between the sharp cavity of height $h$ and the physical scale $d$ of the disc, the relation ${ h \!=\! 2 d }$ holds, so that equation~\eqref{inequality_kz_M} may be rewritten as
\begin{align}
0 \!<\! \ell_{0} \!<\! \frac{\pi}{2 \sqrt{2}} \; ; \; \frac{(p \!-\! \tfrac{1}{2}) \pi}{\sqrt{2}} \!<\! \ell_{p} \!<\! \frac{(p \!+\! \tfrac{1}{2}) \pi}{\sqrt{2}} \;\; \text{(for ${ p \!\geq\! 1 }$)} \, .
\label{inequality_ell_M}
\end{align}
Similarly, ${ \ell_{r} \!=\! (k_{r} h )/\sqrt{2} }$. Notice that expression~\eqref{definition_K_M_dedimensions} of $K_{pq}^{(n)}$ involves a modified Bessel function ${ \mathcal{I}_{n} (\ell_{p} \ell_{q}) }$, that needs to be approximated carefully. Indeed, equivalents in $0$ and ${ + \infty }$ of $\mathcal{I}_{n}$ are respectively given by
\begin{equation}
\mathcal{I}_{n} (x) \underset{0}{\sim} \frac{1}{n!} \left( \frac{x}{2} \right)^{n} \;\;\; ; \;\;\; \mathcal{I}_{n} (x) \underset{+\infty}{\sim} \frac{\re^{x}}{\sqrt{2 \pi x}} \, .
\label{asymptotics_Bessel}
\end{equation}
As illustrated in figure~\ref{fig_AsymptoticsBessel}, one must determine which approximation (polynomial or exponential) is relevant for a given
value of $n$ and $x$.
Therefore, for each ${ n \!\geq\! 0 }$, let us introduce
$x_{n}$, such that for ${ x \!\leq\! x_{n} }$ (resp. ${ x \!\geq\! x_{n} }$),
one uses the asymptotic development from equation~\eqref{asymptotics_Bessel}
in $0$ (resp. ${ + \infty }$). 
In the expression~\eqref{definition_K_M_dedimensions} of the matrix coefficients, notice that the Bessel function is only
evaluated in $\ell_{p} \ell_{q}$, with $p$ and $q$ two integers. For $p$ and $q$ given, 
there exists an integer $n_{pq}$ such that:
\begin{align}
\forall \ell_{z} < n_{pq}, & \quad \mathcal{I}_{\ell_{z}}(\ell_{p} \ell_{q} ) \simeq \frac{\re^{\ell_{p} \ell_{q}}}{\sqrt{2 \pi \ell_{p} \ell_{q}}} \, ,
\nonumber
\\
\forall \ell_{z} \geq n_{pq}, & \quad \mathcal{I}_{\ell_{z}} (\ell_{p} \ell_{q}) \simeq \frac{1}{\ell_{z}!} \left( \frac{\ell_{p} \ell_{q}}{2} \right)^{\ell_{z}} \, .
\end{align}
Notice in figure~\ref{fig_AsymptoticsBessel}, that except for ${ \ell_{z} \!=\! 0 }$, the exponential approximation is significantly bigger than the actual value of ${ \mathcal{I}_{\ell_{z}} }$. This does not impact the following calculation, since when proving that ${ \widehat{\mathbf{M}}_{pq} \!\ll\! \widehat{\mathbf{M}}_{pp} }$ the exponential approximation will only be applied on $\widehat{\mathbf{M}}_{pq}$ alone, or on $\widehat{\mathbf{M}}_{pq}$ and $\widehat{\mathbf{M}}_{pp}$ simultaneously with similar errors, so that the comparison between the approximated values will also hold for the exact values.
A naive approach to compare ${ S_{\gamma} (p,q) }$ and ${ S_{\gamma} (p,p) }$ as required by equation~\eqref{definition_Sgamma}, would be to compare them term by term, that is to prove that ${ K_{pq}^{(\ell_{z})} \!\ll\! K_{pp}^{(\ell_{z})} }$ for all $\ell_{z}$.
This is unfortunately not sufficient, and one must be more cautious, and cut the sum on $\ell_{z}$ in ${ S_{\gamma} (p,q) }$ from equation~\eqref{asymptotics_Bessel}, between three different contributions, for which one can directly show:
\begin{itemize}
\item For the first terms (with ${ |\ell_{z}| \!<\! n_{pp} }$ and ${ |\ell_{z}| \!<\! n_{pq} }$):
\begin{equation}
K_{pq}^{(\ell_{z})} \!\ll\! K_{pp}^{(1)} \, .
\nonumber
\end{equation}
\item For the intermediate terms (with ${ n_{pp} \!\leq\! |\ell_{z}| \!<\! n_{pq} }$):
\begin{equation}
\sum_{n_{pp} \leq |\ell_{z}| < n_{pq}} \!\!\!\!\!\! \frac{\ell_{z}^{\gamma}
K_{pq}^{(\ell_{z})}}{1 \!-\! s_{\ell_{z}}^{2}} \!\ll\! K_{pp}^{(1)} \, .
\nonumber
\end{equation}
\item For the last terms (with ${ |\ell_{z}| \!\geq\! n_{pq} }$):
\begin{equation}
\sum_{|\ell_{z}| \geq n_{pq}} \frac{\ell_{z}^{\gamma}
K_{pq}^{(\ell_{z})}}{1 \!-\! s_{\ell_{z}}^{2}} \!\ll\! K_{pp}^{(1)} \, .
\nonumber
\end{equation}
\end{itemize}
This last relation holds when ${ k_{r} h \!\gtrsim\! 0.03 }$, but gets violated in the limit
of a razor-thin disc, when ${ q \!=\! 0} $.
These comparisons are easy to obtain, and only require to use the appropriate approximations
of the Bessel functions from equation~\eqref{asymptotics_Bessel} for the two elements which are compared, and rely on the step distances between two consecutive basis elements from equation~\eqref{inequality_ell_M}. These inequalities show that, when ${ k_{r} h \!\gtrsim\! 0.03 }$, for all $p$ and $q$, one has
${ \widehat{\mathbf{M}}_{pq} \!\ll\! \widehat{\mathbf{M}}_{pp} }$. The same result also holds for ${ k_{r} h \!\lesssim\! 0.03 }$, but only when ${ q \!\neq\! 0 }$. We therefore reached the following conclusions:
\begin{itemize}
\item The antisymmetric response matrix can always be assumed to be diagonal.
\item For ${ k_{r} h \!\gtrsim 0.03 }$, the symmetric response matrix can be assumed to be diagonal.
\item For ${ k_{r} h \!\lesssim\! 0.03 }$, i.e. in the limit of a razor-thin disc, the symmetric response matrix takes the form of an arrowhead matrix.
\end{itemize}

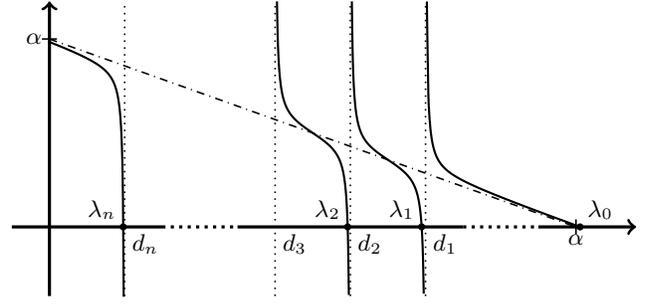
\begin{figure}
\begin{center}
\begin{tikzpicture}
\pgfmathsetmacro{\dn}{1};
\pgfmathsetmacro{\ddd}{3};
\pgfmathsetmacro{\dd}{4};
\pgfmathsetmacro{\d}{5};
\pgfmathsetmacro{\aaaa}{7}; \pgfmathsetmacro{\va}{2.5};
\pgfmathsetmacro{\xmin}{-0.5}; \pgfmathsetmacro{\xmax}{7.8};
\pgfmathsetmacro{\ymin}{-1+0.08}; \pgfmathsetmacro{\ymax}{3};
\pgfmathsetmacro{\dotmin}{1.5} ; \pgfmathsetmacro{\dotmax}{2.5};
\pgfmathsetmacro{\ddotmin}{5.5}; \pgfmathsetmacro{\ddotmax}{6.5};
\pgfmathsetmacro{\lt}{0.1};
\draw [very thick] (\xmin , 0) -- (\dotmin , 0) ; \draw [very thick , dotted] (\dotmin , 0) -- (\dotmax , 0) ; \draw [very thick] (\dotmax, 0) -- (\ddotmin , 0) ;
\draw [very thick , dotted] (\ddotmin, 0) -- (\ddotmax, 0) ; \draw [-> , very thick] (\ddotmax , 0) -- (\xmax , 0) ;
\draw [-> , very thick] (0 , \ymin) -- (0 , \ymax);
\draw (\dn,0) node[font = \normalsize, below right] {$d_{n}$};
\draw (\ddd,0) node[font = \normalsize, below right] {$d_{3}$};
\draw (\dd,0) node[font = \normalsize, below right] {$d_{2}$};
\draw (\d,0) node[font = \normalsize, below right ] {$d_{1}$};
\draw (\aaaa,0) node[font = \normalsize, below] {$\alpha$}; \draw [semithick] (\aaaa , 0 - \lt) -- (\aaaa , 0 + \lt) ;
\draw (0,\va) node[font = \normalsize , left] {$\alpha$}; \draw [semithick] (0 - \lt , \va) -- (0 + \lt , \va) ;
\draw [dotted, semithick] (\dn , \ymin) -- (\dn , \ymax);
\draw [dotted, semithick] (\ddd , \ymin) -- (\ddd , \ymax);
\draw [dotted, semithick] (\dd , \ymin) -- (\dd , \ymax);
\draw [dotted, semithick] (\d , \ymin) -- (\d , \ymax);
\draw [dashdotted, semithick] (0, \va) -- (\aaaa , 0) ;
\pgfmathsetmacro{\expchar}{0.04};
\pgfmathsetmacro{\dminzero}{5.01745} ;
\pgfmathsetmacro{\dmaxzero}{7.05451} ;
\pgfmathsetmacro{\lzero}{7.05451} ;
\pgfmathsetmacro{\dminone}{4.02024} ;
\pgfmathsetmacro{\dmaxone}{4.97732} ;
\pgfmathsetmacro{\lone}{4.94838} ;
\pgfmathsetmacro{\dmintwo}{3.02467} ;
\pgfmathsetmacro{\dmaxtwo}{3.98112} ;
\pgfmathsetmacro{\ltwo}{3.96446} ;
\pgfmathsetmacro{\dminn}{0} ;
\pgfmathsetmacro{\dmaxn}{0.987291} ;
\pgfmathsetmacro{\ln}{0.981391} ;
\draw [thick , smooth] plot[domain= \dminzero : \dmaxzero , variable = \x , samples = \nbpFLambda] (\x , { \va - \x * (\va/\aaaa) + \expchar/(\x - \d) }) ;
\draw [thick , smooth] plot[domain= \dminone : \dmaxone , variable = \x , samples = \nbpFLambda] (\x , { \va - \x * (\va/\aaaa) + \expchar/(\x - \dd) + \expchar/(\x - \d) }) ;
\draw [thick , smooth] plot[domain= \dmintwo : \dmaxtwo , variable = \x , samples = \nbpFLambda] (\x , { \va - \x * (\va/\aaaa) + \expchar/(\x - \ddd) + \expchar/(\x - \dd) }) ;
\draw [thick , smooth] plot[domain= \dminn : \dmaxn , variable = \x , samples = \nbpFLambda] (\x , { \va - \x * (\va/\aaaa) + \expchar/(\x - \dn) }) ;
\draw (\lzero , 0) node[font = \normalsize , above right] {$\lambda_{0}$}; \draw [fill] (\lzero , 0) circle [radius = 0.04];
\draw (\lone , 0) node[font = \normalsize , above left] {$\lambda_{1}$}; \draw [fill] (\lone , 0) circle [radius = 0.04];
\draw (\ltwo , 0) node[font = \normalsize , above left] {$\lambda_{2}$}; \draw [fill] (\ltwo , 0) circle [radius = 0.04];
\draw (\ln , 0) node[font = \normalsize , above left] {$\lambda_{n}$}; \draw [fill] (\ln , 0) circle [radius = 0.04];
\end{tikzpicture}
\caption{\small{Illustration of the behaviour of the function ${ \lambda \!\mapsto\! f_{\widehat{\mathbf{M}}} (\lambda) }$, thanks to which the eigenvalues of the arrowhead response matrix from equation~\eqref{arrowheadmatrix} may be determined.
}}
\label{fig_shape_flambda}
\end{center}
\end{figure}

Finally, let us now justify why for a sufficiently thin disc, for which the symmetric response matrix takes the form of an arrowhead matrix, the diagonal response matrix case is recovered. In this limit, the symmetric response matrix takes the form 
\begin{equation}
\widehat{\mathbf{M}} = 
\left( \begin{array}{cccc}
\alpha & z_{1} & \cdots & z_{n}
\\
z_{1} & d_{1} & & 
\\
\vdots & & \ddots & 
\\
z_{n} & & & d_{n}
\end{array}
\right) \, .
\label{arrowheadmatrix}
\end{equation}
Assuming that ${ \forall i \, , z_{i} \!\neq\! 0 }$ and ${ \forall i \!\neq\! j \, , d_{i} \!\neq\! d_{j} }$, it can be shown~\citep{OLearyStewart1990} that the eigenvalues ${ (\lambda_{i})_{0 \leq i \leq n} }$ of the arrowhead matrix from equation~\eqref{arrowheadmatrix} are the ${ (n \!+\! 1) }$ solutions of the equation
\begin{equation}
f_{\widehat{\mathbf{M}}} (\lambda) = \alpha \!-\! \lambda - \! \sum_{i = 1}^{n} \! \frac{z_{i}^{2}}{d_{i} \!-\! \lambda} = 0 \, .
\label{Mdiag_polynom}
\end{equation}
Provided that the $d_{i}$ are in descending order, the eigenvalues $\lambda_{i}$ of $\widehat{\mathbf{M}}$ are interlaced so that
\begin{equation}
\lambda_{0} > d_{1} > \lambda_{1} > ... > d_{n} > \lambda_{n} \, .
\label{interlaced_eigenvalues}
\end{equation}
Finally, the eigenvectors $\bm{x}_{i}$ associated with the eigenvalue $\lambda_{i}$ are proportional to
\begin{equation}
\bm{x}_{i} = \bigg( 1 \, ; \, \frac{z_{1}}{\lambda_{i} \!-\! d_{1}} \, ; \, ... \, ; \, \frac{z_{j}}{\lambda_{i} \!-\! d_{j}} \, ; \, ... \, ; \, \frac{z_{n}}{\lambda_{i} \!-\! d_{n}} \bigg) \, .
\label{expression_eigenvector}
\end{equation}
In our case, the comparison relations ${ \alpha \!\gg\! z_{i} }$ and ${ z_{i} \!\gg\! d_{i} }$ also hold. An illustration in this regime of the behaviour of the function ${ \lambda \!\mapsto\! f_{\widehat{\mathbf{M}}} (\lambda) }$ from equation~\eqref{Mdiag_polynom} is shown in figure~\ref{fig_shape_flambda}.
To justify why the arrowhead response matrix from equation~\eqref{arrowheadmatrix} may be considered as diagonal, one has to justify that despite the first line and column, the matrix eigenvalues remain close to the matrix diagonal coefficients, so as to have
\begin{equation}
\lambda_{0} \simeq \alpha \;\;\; \text{and} \;\;\; \lambda_{i} \simeq d_{i} \;\; \text{(for ${ i \!\geq\! 1 }$)} \, .
\label{approximation_lambda_M}
\end{equation}
Similarly, it must also be ensured that the associated eigenvectors $\bm{x}_{i}$ remain close the natural basis elements so that
\begin{equation}
\bm{x}_{i} \simeq (0 \, ; \, ... \, ; 1 \, ; \, 0 \, ; \, ....) \, , 
\label{approximation_xi_M}
\end{equation}
where the only non-zero index is at the $i^{\rm th}$ position.
As illustrated in figure~\ref{fig_shape_flambda}, to determine the eigenvalues $\lambda_{i}$, the equation ${ f_{\widehat{\mathbf{M}}} (\lambda_{i}) \!=\! 0 }$ introduced in equation~\eqref{Mdiag_polynom} must be solved. 
This may be rewritten as
\begin{equation}
1 \!-\! \frac{\lambda_{i}}{\alpha} \!-\! \sum_{i = 1}^{n} \frac{(z_{i} / \alpha)^{2}}{(d_{i}/\alpha) \!-\! (\lambda_{i} / \alpha)} = 0 \, .
\label{rewriting_Mdiag_polynom}
\end{equation} 
Since one has ${ (z_{i} / \alpha) \!\ll\! 1 }$, in order for equation~\eqref{rewriting_Mdiag_polynom} to be fulfilled, one must necessarily have ${ \lambda_{i} / \alpha \!\simeq\! 1}$ or ${ ((d_{i}/\alpha) \!-\! (\lambda_{i} / \alpha)) \!\ll\! 1 }$. It follows immediately that ${ \lambda_{0} \!\simeq\! \alpha }$ and ${ \lambda_{i} \!\simeq\! d_{i} }$. As a consequence, equation~\eqref{approximation_lambda_M} holds: the matrix eigenvalues $\lambda_{i}$ remain close to the matrix diagonal coefficients ${ (\alpha , d_{1} , ... , d_{n}) }$. The eigenvectors $\bm{x}_{i}$ introduced in equation~\eqref{expression_eigenvector}, can be rewritten as
\begin{equation}
\bm{x}_{i} \!=\! \!\bigg(\!\! 1 ; \!\frac{(\!z_{1} \!/\! \alpha\!)^{2}}{(\!\lambda_{i}\!/\!\alpha\!) \!\!-\!\! (\!d_{1} \!/\! \alpha\!)} \frac{1}{(\!z_{1} \!/\! \alpha\!)} ; \!... ; \!\frac{(\!z_{j} \!/\! \alpha\!)^{2}}{(\!\lambda_{i} \!/\! \alpha\!) \!\!-\!\! (\!d_{j} \!/\! \alpha\!)} \frac{1}{(\!z_{j} \!/\! \alpha\!)} ; \!... \!\!\bigg) .
\label{rewriting_eigenvector}
\end{equation}
In equation~\eqref{rewriting_eigenvector}, if one considers the case ${ i \!=\! 0 }$, thanks to equation~\eqref{approximation_lambda_M}, one has ${ \lambda_{0} \!\simeq\! \alpha }$, so that using the fact that ${ d_{j} \!\ll\! \alpha }$, the generic term from equation~\eqref{rewriting_eigenvector} takes the form
\begin{equation}
\frac{(z_{j}/\alpha)^{2}}{(\lambda_{0} / \alpha) \!-\! (d_{j} / \alpha)} \frac{1}{(z_{j} / \alpha)} \simeq \frac{(z_{j} / \alpha)}{1} \ll 1 \, ,
\label{rewriting_eigenvector_I}
\end{equation}
where we used the fact ${ z_{j} \!\ll\! \alpha }$. As a consequence, for ${ i \!=\! 0 }$, in equation~\eqref{rewriting_eigenvector}, all the terms except the first one are negligible in front of $1$, so that one has ${ \bm{x}_{0} \!\simeq\! (1 ; 0;...;0) }$. In equation~\eqref{rewriting_eigenvector}, if one considers the case ${ i \!\neq\! 0 }$, one has to study the $i^{\rm th}$ term of equation~\eqref{rewriting_eigenvector} which takes the form
\begin{equation}
\frac{(z_{i}/\alpha)^{2}}{(\lambda_{i} / \alpha) \!-\! (d_{i} / \alpha)} \frac{1}{ (z_{i} / \alpha) } \simeq \frac{1}{(z_{i} / \alpha)} \gg 1 \, .
\label{rewriting_eigenvector_II}
\end{equation}
In this calculation, equation~\eqref{rewriting_Mdiag_polynom} was used to show that for ${ i \!\neq\! 0 }$, one has ${ \tfrac{(z_{i}/\alpha)^{2}}{(d_{i} / \alpha) \!-\! (\lambda_{i} / \alpha)} \!\simeq\! 1 }$. As a consequence, for ${ i \!\neq\! 0 }$, the eigenvector $\bm{x}_{i}$ is dominated by its $i^{\rm th}$ coefficient and can therefore be assumed to be proportional to ${ (0 ; ..;1;0;...) }$, where the non-zero index is as the $i^{\rm th}$ position. Consequently, we may assume that the response matrix eigenvectors remain close to the natural basis elements. As a conclusion, even in the limit of a razor-thin disc, the arrowhead symmetric response matrix from equation~\eqref{arrowheadmatrix} may still be assumed to be diagonal. This justifies the generic use of the diagonal amplification eigenvalues in equation~\eqref{Matrix_diagonal}, when computing the diffusion coefficients.

\section{Antisymmetric collisionless diffusion}
\label{sec:appendixantisymDm}

In this Appendix, let us show how one may obtain the counterparts of equation~\eqref{Dm_autocorrelation_even} for the antisymmetric components. Here, the main differences are that the quantised $k_{z}$ frequencies are given by equation~\eqref{quantisation_odd} (with the same step distance as in equation~\eqref{step_distance_kz}), the $m_{z}$ considered will necessarily be odd, and the Fourier transformed basis elements from equation~\eqref{psi_m_odd} must be considered, which involve a different normalisation constant $\beta_{n}$. In analogy with equation~\eqref{Dm_autocorrelation_even}, the antisymmetric diffusion coefficients are given by
\begin{align}
D_{\bm{m}}^{\rm anti} (\bm{J}) & \, =  \delta_{m_{z}}^{\rm odd} \frac{1}{(2 \pi)^{2}} \nonumber
\\
& \times \, \!\! \int \!\! \mathrm{d} k_{r}^{p} \mathrm{d} k_{z}^{p} \, \mathcal{J}_{m_{r}} \!\bigg[\! \sqrt{\tfrac{2 J_{r}}{\kappa}} k_{r}^{p} \!\bigg] \mathcal{J}_{m_{z}} \!\bigg[\! \sqrt{\tfrac{2 J_{z}}{\nu}} k_{z}^{p} \!\bigg] \frac{\beta_{p}^{2}}{1 \!-\! \lambda_{p}} \nonumber
\\
& \times \, \int \!\! \mathrm{d} k_{r}^{q} \mathrm{d} k_{z}^{q} \, \mathcal{J}_{m_{r}} \!\bigg[\! \sqrt{\tfrac{2 J_{r}}{\kappa}} k_{r}^{q} \!\bigg] \mathcal{J}_{m_{z}} \!\bigg[\! \sqrt{\tfrac{2 J_{z}}{\nu}} k_{z}^{q} \!\bigg] \frac{\beta_{q}^{2}}{1 \!-\! \lambda_{q}} \nonumber
\\
& \times \, \widehat{C}_{\psi^{\rm e}} [m_{\phi} , \bm{m} \!\cdot\! \bm{\Omega} , R_{\rg} , k_{r}^{p} , k_{r}^{q} , k_{z}^{p} , k_{z}^{q} ] \, ,
\label{Dm_autocorrelation_odd}
\end{align}
where one should pay attention to the fact that the pertubation autocorrelation $\widehat{C}_{\psi^{\rm e}}$ introduced in equation~\eqref{definition_general_autocorrelation} for the symmetric case has to be computed slighlty differently for the antisymmetric case. Indeed, because the antisymmetric basis elements from equation~\eqref{full_psi_p_odd} possess an odd vertical dependence, the even-restricted vertical Fourier transform from equation~\eqref{definition_FT} should be replaced by an odd-restricted vertical Fourier transform defined as
\begin{equation}
f_{k_{z}} = \!\! \int_{-h}^{+ h} \!\!\!\!\! \mathrm{d} z \, \sin (k_{z} z) \, f[z] \, .
\label{definition_odd_vertical}
\end{equation}
In equation~\eqref{Dm_autocorrelation_odd}, notice that the integrations on $k_{z}^{p}$ and $k_{z}^{q}$ should only be made for ${ k_{z} \!\geq\! k_{z,a}^{1} }$, i.e. for $k_{z}$ larger than the fundamental antisymmetric mode $k_{z,a}^{1}$ as illustrated in figure~\ref{fig_quantisation}.
Using the antisymmetric diagonalisation of the autocorrelation obtained in equation~\eqref{approximation_Ga} and following equation~\eqref{Dm_diagonalised_even} for the symmetric diffusion coefficients, equation~\eqref{Dm_autocorrelation_odd} may be simplified as
\begin{align}
D_{\bm{m}}^{\rm anti} (\bm{J}) = & \, \delta_{m_{z}}^{\rm odd} \frac{\pi}{(2 \pi)^{2}} \!\! \int \!\! \mathrm{d} k_{r}^{p} \mathrm{d} k_{z}^{p} \, \mathcal{J}_{m_{r}}^{2} \!\!\bigg[\! \sqrt{\!\tfrac{2 J_{r}}{\kappa}} k_{r}^{p} \!\bigg] \mathcal{J}_{m_{z}}^{2} \!\!\bigg[\! \sqrt{\!\tfrac{2 J_{z}}{\nu}} k_{z}^{p} \!\bigg] \nonumber
\\
& \times \, \bigg[\! \frac{\beta_{p}^{2}}{1 \!-\! \lambda_{p}} \!\bigg]^{2} \widehat{\mathcal{C}} [m_{\phi} , \bm{m} \!\cdot\! \bm{\Omega} , R_{\rg} , k_{r}^{p} , k_{z}^{p}] \, .
\label{Dm_diagonalised_odd}
\end{align}
In equation~\eqref{Dm_diagonalised_odd}, despite the fact that one is considering antisymmetric diffusion coefficients, it is important to note that here $\widehat{\mathcal{C}}$ contains an even-restricted vertical Fourier transform, as detailed in equation~\eqref{approximation_Ga}. Such a property underlines how the symmetric and antisymmetric diffusion coefficients are indeed similar. Proceeding as in equation~\eqref{Dm_diagonalised_ASD_even}, the approximation of the small denominators simplifies equation~\eqref{Dm_diagonalised_odd} as
\begin{align}
D_{\bm{m}}^{\rm anti} (\bm{J}) = & \, \delta_{m_{z}}^{\rm odd} \frac{\pi |\mathcal{V}_{\rm max}|}{(2 \pi)^{2}} \mathcal{J}_{m_{r}}^{2} \!\!\bigg[\! \sqrt{\!\tfrac{2 J_{r}}{\kappa}} k_{r}^{\rm max} \!\bigg] \mathcal{J}_{m_{z}}^{2} \!\!\bigg[\! \sqrt{\!\tfrac{2 J_{z}}{\nu}} k_{z}^{\rm max} \!\bigg] \nonumber
\\
& \, \!\!\!\!\! \times \bigg[\! \frac{\beta_{\rm max}^{2}}{ 1 \!-\!  \lambda_{\rm max}} \!\bigg]^{2}\,  \widehat{\mathcal{C}} [m_{\phi} , \bm{m} \!\cdot\! \bm{\Omega} , R_{\rg} , k_{r}^{\rm max} , k_{z}^{\rm max} ] \, .
\label{Dm_diagonalised_ASD_odd}
\end{align}

\section{From thick to thin discs}
\label{sec:fromthicktothin}

\subsection{The collisionless case}
\label{sec:thicktothinFP}

In this Appendix, let us show how one can estimate the diffusion coefficients when the disc is too thin to use the continuous expression from equation~\eqref{g_ultra_continuous}. It will be shown how this second approach is consistent with that from equation~\eqref{g_ultra_continuous} and how the infinitely thin results from FPP15 are recovered.

As observed in equation~\eqref{g_ultra_continuous}, the use of the Riemann formula w.r.t. the index $k_{z}^{p}$ is only justified if the typical step distance ${ \Delta k_{z} \!\simeq\! \pi /h }$ from equation~\eqref{step_distance_kz} is sufficiently small compared to the scale of variation of the function ${ k_{z} \!\mapsto\! g_{s} (k_{z}) }$. In the limit of a thinner disc, ${ h \!\to\! 0 }$, so that ${ \Delta k_{z} \!\to\! + \infty }$. This approximation cannot be used anymore and the discrete sum over the quantised $k_{z}^{p}$ from equation~\eqref{g_continuous} should be kept. It is also within this framework that we may hope to recover in the razor-thin limit the known results from FPP15 for an infinitely thin stellar disc. Starting from equation~\eqref{g_continuous} for the symmetric diffusion coefficients, one rewrites equation~\eqref{Dm_autocorrelation_even} as
\begin{align}
D_{\bm{m}}^{\rm sym} & \, (\bm{J}) = \delta_{m_{z}}^{\rm even} \frac{1}{(2 h)^{2}} \nonumber
\\
& \!\!\!\!\! \, \times \, \sum_{n_{p} , n_{q}} \!\! \int \!\! \mathrm{d} k_{r}^{p} \, \mathcal{J}_{m_{r}} \!\!\bigg[\! \sqrt{\tfrac{2 J_{r}}{\kappa}} k_{r}^{p} \!\bigg] \mathcal{J}_{m_{z}} \!\!\bigg[\! \sqrt{\tfrac{2 J_{z}}{\nu}} k_{z}^{n_{p}} \!(k_{r}^{p}) \!\bigg] \frac{\alpha_{p}^{2}}{1 \!-\! \lambda_{p}} \nonumber
\\
& \!\!\!\!\! \, \times \, \int \!\! \mathrm{d} k_{r}^{q} \mathcal{J}_{m_{r}} \!\!\bigg[\! \sqrt{\tfrac{2 J_{r}}{\kappa}} k_{r}^{q} \!\bigg] \mathcal{J}_{m_{z}} \!\!\bigg[\! \sqrt{\tfrac{2 J_{z}}{\nu}} k_{z}^{n_{q}} \! (k_{r}^{q}) \!\bigg] \frac{\alpha_{q}^{2}}{1 \!-\! \lambda_{q}} \nonumber
\\
& \!\!\!\!\! \, \times \, \widehat{C}_{\psi^{\rm e}} [m_{\phi} , \bm{m} \!\cdot\! \bm{\Omega} , R_{\rg} , k_{r}^{p} , k_{r}^{q} , k_{z}^{n_{p}} \!(k_{r}^{p}) , k_{z}^{n_{q}} \!(k_{r}^{q}) ] \, ,
\label{Dm_autocorrelation_even_discrete}
\end{align}
where the autocorrelation of the external perturbation has been defined in equation~\eqref{definition_general_autocorrelation}. Starting from equation~\eqref{Dm_autocorrelation_odd}, the antisymmetric analog of equation~\eqref{Dm_autocorrelation_even_discrete} is straightforward to obtain, thanks to the substitution ${ \alpha_{p} \!\to\! \beta_{p} }$ and ${ \delta_{m_{z}}^{\rm even} \!\to\! \delta_{m_{z}}^{\rm odd} }$. However, as emphasised in equation~\eqref{definition_odd_vertical}, one should pay attention to the fact that it will involve an odd-restricted vertical Fourier transform of the potential perturbations. The next step, as in equation~\eqref{diagonal_autocorrelation}, is to diagonalise the autocorrelation of the external perturbation, taking into account that, contrary to the continuous case from equation~\eqref{Dm_autocorrelation_even}, the vertical frequencies ${ k_{z} \!=\! k_{z} (k_{r} , n) }$ are no longer a free variable but should be seen as functions of the associated $k_{r}$ and $n$. Following the same calculations as in Appendix~\ref{sec:appendixautocorrelation}, and using the shortening notation ${ k_{z}^{n_{1\!/\!2}} \!=\! k_{z}^{n_{1\!/\!2}} (k_{r}^{1\!/\!2}) }$, we may write
\begin{align}
\bigg<\! \widehat{\psi^{\rm e}}_{k_{r}^{1} , k_{z}^{n_{1}} } & \,  [R_{\rg} , \omega_{1}] \, \widehat{\psi^{\rm e}}^{*}_{k_{r}^{2} , k_{z}^{n_{2}} } [R_{\rg} , \omega_{2} ] \!\bigg> = 2 \pi \delta_{\rm D} (\omega_{1} \!-\! \omega_{2}) \,  \nonumber
\\
& \! \, \times \,  \delta_{\rm D} (k_{r}^{1} \!-\! k_{r}^{2}) \frac{1}{2} \!\! \int_{-2 h}^{2 h} \!\!\!\!\!\!\! \mathrm{d} v \, \widehat{\mathcal{C}} [v] \, G_{\rm sym} [k_{z}^{n_{1}} , k_{z}^{n_{2}} , v] \, .
\label{diag_auto_discrete}
\end{align}
Thanks to the Dirac delta ${ \delta_{\rm D} (k_{r}^{1} \!-\! k_{r}^{2}) }$, $k_{z}^{n_{1}}$ and $k_{z}^{n_{2}}$ are evaluated for the same $k_{r}$ so that equation~\eqref{step_distance_kz} gives
\begin{equation}
n_{1} \neq n_{2} \;\; \Rightarrow \;\; | k_{z}^{n_{1}} \!-\! k_{z}^{n_{2}} |  \gtrsim \pi / h \, ,
\label{equality_n1_n2}
\end{equation}
where for a thin disc, the quantised $k_{z}$ will therefore tend to be far apart. Given the approximation of $G_{\rm sym}$ obtained in equation~\eqref{approximation_Gs}, we may assume that equation~\eqref{diag_auto_discrete} is non-negligible only for ${ n_{1} \!=\! n_{2} }$, \text{i.e.} ${ k_{z}^{n_{1}} \!=\! k_{z}^{n_{2}} }$. One may rewrite equation~\eqref{diag_auto_discrete} similarly to equation~\eqref{diagonal_autocorrelation} as
\begin{align}
\bigg<\!\!  & \, \widehat{\psi^{\rm e}}_{\! m_{\phi}, k_{r}^{1} , k_{z}^{n_{1}}}  [R_{\rg} , \omega_{1}] \, \widehat{\psi^{\rm e}}^{*}_{\! m_{\phi} , k_{r}^{2} , k_{z}^{n_{2}}} [R_{\rg} , \omega_{2}] \!\!\bigg> \!=\!  2 \pi h \, \delta_{\rm D} (\omega_{1} \!-\! \omega_{2}) \nonumber
\\
& \;\;\;\;\;\;\;  \times \, \delta_{\rm D} (k_{r}^{1} \!-\! k_{r}^{2}) \, \delta_{n_{1}}^{n_{2}} \, \widehat{\mathcal{C}} [m_{\phi} , \omega_{1} , R_{\rg} , k_{r}^{1} , k_{z}^{n_{1}}] \, ,
\label{diagonal_autocorrelation_discrete}
\end{align}
where the presence of the Kronecker symbol $\delta_{n_{1}}^{n_{2}}$ should be noted. Thanks to this diagonalised autocorrelation, the discrete expression of the symmetric diffusion coefficients from equation~\eqref{Dm_autocorrelation_even_discrete} immediately takes the form
\begin{align}
D_{\bm{m}}^{\rm sym} (\bm{J}) \!= & \, \delta_{m_{z}}^{\rm even} \frac{1}{4 h} \sum_{n_{p}} \!\! \int \!\! \mathrm{d} k_{r}^{p} \, \mathcal{J}_{m_{r}}^{2} \!\!\bigg[\! \sqrt{\!\tfrac{2 J_{r}}{\kappa}} k_{r}^{p} \!\bigg] \mathcal{J}_{m_{z}}^{2} \!\!\bigg[\! \sqrt{\!\tfrac{2 J_{z}}{\nu}} k_{z}^{n_{p}} \!(k_{r}^{p}) \!\bigg] \nonumber
\\
& \times \, \bigg[\! \frac{\alpha_{p}^{2}}{1 \!-\! \lambda_{p}} \!\bigg]^{2} \widehat{\mathcal{C}} [m_{\phi} , \bm{m} \!\cdot\! \bm{\Omega} , R_{\rg} , k_{r}^{p} , k_{z}^{n_{p}} \!(k_{r}^{p})] \, .
\label{Dm_diagonalised_even_discrete}
\end{align}
This expression is the direct discrete equivalent of equation~\eqref{Dm_diagonalised_even}, and these two expressions are in full agreement. Indeed, starting from equation~\eqref{Dm_diagonalised_even_discrete}, the continuous expression on $k_{z}^{p}$ is recovered using the Riemann sum formula with, as given by equation~\eqref{step_distance_kz}, a step distance ${ \Delta k_{z} \!=\! \pi /h }$. Equation~\eqref{Dm_diagonalised_even} is then exactly recovered. As in equation~\eqref{Dm_diagonalised_ASD_even} for the continuous approach, the approximation of the small denominators can be used so as to write
\begin{align}
D_{\bm{m}}^{\rm sym} (\bm{J}) = & \, \delta_{m_{z}}^{\rm even} \frac{1}{4 h} \! \sum_{n_{p}} \! \Delta k_{r}^{n_{p}} \!\mathcal{J}_{m_{r}}^{2} \!\!\bigg[\! \sqrt{\!\tfrac{2 J_{r}}{\kappa}} k_{r,n_{p}}^{\rm max} \!\bigg] \mathcal{J}_{m_{z}}^{2} \!\!\bigg[\! \sqrt{\!\tfrac{2 J_{z}}{\nu}} k_{z , n_{p}}^{\rm max} \!\bigg] \nonumber
\\
& \, \!\!\!\!\!\! \times \bigg[\! \frac{(\alpha_{n_{p}}^{\rm max})^{2}}{1 \!-\! \lambda_{n_{p}}^{\rm max}} \!\bigg]^{2} \widehat{\mathcal{C}} [ m_{\phi} , \bm{m} \!\cdot\! \bm{\Omega} , R_{\rg} , k_{r , n_{p}}^{\rm max} , k_{z ,n_{p}}^{\rm max} ] \, ,
\label{Dm_diagonalised_ASD_even_discrete}
\end{align}
where for a given value of $n_{p}$, we considered the behaviour of the function ${ k_{r}^{p} \!\mapsto\! \lambda (k_{r}^{p} , k_{z}^{n_{p}} (k_{r}^{p})) }$. We assumed it reached a maximum $\lambda_{n_{p}}^{\rm max}$ for ${ k_{r} \!=\! k_{r , n_{p}}^{\rm max} }$ on a region of typical extension ${ \Delta k_{r}^{n_{p}} }$. Finally, we also used the shortening notation ${ k_{z , n_{p}}^{\rm max} \!=\! k_{z}^{n_{p}} (k_{r , n_{p}}^{\rm max}) }$.
The expression~\eqref{Dm_diagonalised_even_discrete} can straightforwardly be translated to the antisymmetric diffusion coefficients as
\begin{align}
D_{\bm{m}}^{\rm anti} (\bm{J}) = & \, \delta_{m_{z}}^{\rm odd} \frac{1}{4 h} \sum_{n_{p}} \!\! \int \!\! \mathrm{d} k_{r}^{p} \, \mathcal{J}_{m_{r}}^{2} \!\!\bigg[\! \sqrt{\!\tfrac{2 J_{r}}{\kappa}} k_{r} \!\bigg] \mathcal{J}_{m_{z}}^{2} \!\!\bigg[\! \sqrt{\!\tfrac{2 J_{z}}{\nu}} k_{z}^{n_{p}} \!(k_{r}^{p}) \!\bigg] \nonumber
\\
& \times \, \bigg[\! \frac{\beta_{p}^{2}}{1 \!-\! \lambda_{p}} \!\bigg]^{2} \widehat{\mathcal{C}} [m_{\phi} , \bm{m} \!\cdot\! \bm{\Omega} , R_{\rg} , k_{r}^{p} , k_{z}^{p}]  \,  ,
\label{Dm_diagonalised_odd_discrete}
\end{align}
where the antisymmetric quantised $k_{z}$ frequencies from equation~\eqref{quantisation_odd} should be considered. As emphasised in equation~\eqref{approximation_Ga}, one should also pay attention to the fact that in equation~\eqref{Dm_diagonalised_odd_discrete}, $\widehat{\mathcal{C}}$ contains an even-restricted vertical Fourier transform of the autocorrelation, despite the fact that one is considering antisymmetric diffusion coefficients. Similarly, the approximation of the small denominators from equation~\eqref{Dm_diagonalised_ASD_even_discrete} extends straightforwardly to the antisymmetric case.

Given the discrete diffusion coefficients from equation~\eqref{Dm_diagonalised_even_discrete}, we may now illustrate how in the infinitely thin limit, the diffusion coefficients obtained in FPP15 are recovered. As illustrated in figure~\ref{fig_quantisation}, notice that except for the fundamental symmetric frequency $k_{z, \rm s}^{1}$, one has ${ k_{z}^{n} \!>\! \pi / (2 h) }$. As a consequence, in the infinitely thin limit, for which ${ h \!\to\! 0 }$, one has ${ k_{z}^{n} \!\to\! + \infty }$, except for ${ k_{z , \rm s}^{1} }$. Recall that the dependence of ${ \widehat{\mathcal{C}} [k_{z}^{p}] }$ with $h$ is given by equation~\eqref{calcul_diag_auto_IV} and takes the form
\begin{equation}
\widehat{\mathcal{C}} [k_{z}^{p}] = \!\! \int_{-2h}^{2h} \!\!\!\!\! \mathrm{d} v \, \widehat{\mathcal{C}} [v] \, \cos [k_{z}^{p} v] \, .
\label{limit_C_thin}
\end{equation}
The following upper bound holds ${ |\widehat{\mathcal{C}} [k_{z}^{p}] | \!\leq\! 4 h \, \widehat{\mathcal{C}}_{\rm max} }$, which, in the razor-thin limit, will cancel the prefactor in ${ 1 / (4h) }$ present in equation~\eqref{Dm_diagonalised_even_discrete}.
Recalling the fact that ${ \forall n \!\geq\! 0 \, , \lim_{x \!\to\! + \infty} \mathcal{J}_{n} (x) \!=\! 0 }$, it follows straightforwarldy that
\begin{equation}
\lim\limits_{\rm thin} D_{\bm{m}}^{\rm anti} (\bm{J}) = 0 \, .
\label{limit_Dm_anti}
\end{equation}
Similarly, for the symmetric diffusion coefficients, the sum on $n_{p}$ from equation~\eqref{Dm_diagonalised_even_discrete} can be limited to the only fundamental term ${ n_{p} \!=\! 1 }$. In equation~\eqref{kz1_even}, we estimated that in the thin limit, one has the asymptotic behaviour ${ k_{z, \rm s}^{1} \!\simeq\! \sqrt{k_{r} / h} }$. Consequently, equation~\eqref{Dm_diagonalised_even_discrete} implies that as soon as ${ m_{z} \!\neq\! 0 }$, ${ \lim_{\rm thin} D_{\bm{m}}^{\rm sym} \!=\! 0 }$. As a conclusion, in the infinitely thin limit, only the symmetric diffusion coefficients associated with ${ m_{z} \!=\! 0 }$ will not vanish. Similarly, starting from equation~\eqref{Dm_diagonalised_even_discrete}, it is straightforward to note that one must have ${ J_{z} \!=\! 0 }$ so as to have a non vanishing symmetric diffusion coefficient.
Hence, in the razor-thin limit, for ${m_{z} \!=\! 0}$ and ${ J_{z} \!=\! 0 }$, 
\begin{align}
\lim\limits_{\rm thin} D_{\bm{m}}^{\rm sym} (\bm{J}) = \lim\limits_{\rm thin} & \, \frac{1}{4 h} \!\! \int \!\! \mathrm{d} k_{r}^{p} \, \mathcal{J}_{m_{r}}^{2} \!\bigg[\! \sqrt{\tfrac{2 J_{r}}{\kappa}} k_{r}^{p} \!\bigg] \bigg[\! \frac{\alpha_{1}^{2}}{1 \!-\! \lambda_{p}} \!\bigg]^{2} \nonumber
\\
& \, \times \, \widehat{\mathcal{C}} [m_{\phi} , \bm{m} \!\cdot\! \bm{\Omega} , R_{\rg} , k_{r}^{p} , k_{z , \rm s}^{1}] \, .
\label{limit_thin_sym}
\end{align}
Given the definition of $\alpha_{p}$ from equation~\eqref{alpha_even}, ${ \lim_{\rm thin} \alpha_{1} \!=\! 1 }$. Similarly, we have shown in equation~\eqref{limit_lambda_thin}, that for the fundamental symmetric mode, in the razor-thin limit, one has ${ \lim_{\rm thin} \lambda_{p} \!=\! \lambda_{p}^{\rm thin} }$. The last step of the calculation is to study the behaviour of ${ \widehat{\mathcal{C}} [k_{z , \rm s}^{1}] }$ in the razor-thin limit, as written in equation~\eqref{limit_C_thin}. Equation~\eqref{limit_C_thin} takes the form of an integral over an interval of length ${ 4 h }$ of a function oscillating at the frequency ${ k_{z , \rm s}^{1} \!\simeq\! \sqrt{k_{r} / h} }$. The number of oscillations of this function on this interval is of the order ${ k_{z , \rm s}^{1} h \!\sim\! \sqrt{k_{r} h} }$, so that in the razor-thin limit, the number of oscillations of the function ${ v \!\mapsto\! \cos [ k_{z , \rm s}^{1} v ] }$ tends to $0$. This allows us to perform the replacement ${ \cos [ k_{z , \rm s}^{1} v ] \!\to\! 1 }$. As a consequence, in the razor-thin limit, equation~\eqref{limit_C_thin} becomes
\begin{equation}
\lim\limits_{\rm thin} \widehat{\mathcal{C}} [k_{z , \rm s}^{1} ] = 4 h \, \widehat{\mathcal{C}} [v \!=\! 0] \, .
\label{limit_C_thin_final}
\end{equation}
When injected in equation~\eqref{limit_C_thin_final}, one finally obtains
\begin{align}
\lim\limits_{\rm thin} D_{\bm{m}}^{\rm sym} (\bm{J}) = & \, \!\! \int \!\! \mathrm{d} k_{r}^{p} \, \mathcal{J}_{m_{r}}^{2} \!\bigg[\! \sqrt{\tfrac{2 J_{r}}{\kappa}} k_{r}^{p} \!\bigg] \, \bigg[\! \frac{1}{1 \!-\! \lambda_{p}^{\rm thin}} \!\bigg]^{2} \nonumber
\\
& \, \times \, \widehat{\mathcal{C}}_{\rm thin} [m_{\phi} , \bm{m} \!\cdot\! \bm{\Omega} , R_{\rg} , k_{r}^{p}] \, ,
\label{limit_thin_final}
\end{align}
where ${ \widehat{\mathcal{C}}_{\rm thin} [m_{\phi} , \bm{m} \!\cdot\! \bm{\Omega} , R_{\rg} , k_{r}^{p} ] }$ stands for the local power spectrum of the external perturbations in the equatorial plane as computed in the infinitely thin case presented in FPP15. Hence equation~\eqref{limit_thin_final} is in complete agreement with the results obtained in that paper.

\subsection{The collisional case}
\label{sec:thicktothinBL}

Let us now show how one can estimate the collisional susceptibility coefficients in the case where the disc is too thin to use the continuous expressions from equation~\eqref{1/D_continuous}. We will especially show how this second approach starting from equation~\eqref{1/D_discrete} allows us to recover the razor-thin susceptibility coefficients derived in FPC15.

We observed in equation~\eqref{1/D_continuous} that the use of the Riemann sum formula w.r.t. the index $k_{z}^{n_{p}}$ is only justified if the typical step distance ${ \Delta k_{z} \!\simeq\! \pi / h }$ from equation~\eqref{step_distance_kz} is sufficiently small compared to the scale of variation of the function present in the r.h.s. of equation~\eqref{1/D_discrete}. In the limit of a thinner disc, for which ${ h \!\to\! 0 }$, one has ${ \Delta k_{z} \!\to\! + \infty }$. The approximation based on the Riemann sum formula cannot be used, and one should therefore stick with the discrete sum from equation~\eqref{1/D_discrete}. It is within this limit that one may recover the razor-thin results obtained in FPC15. As illustrated in figure~\ref{fig_quantisation}, one should note that except for the fundamental symmetric frequency ${ k_{z, \mathrm{s}}^{1} }$, one always has ${ k_{z}^{n_{p}} \!>\! \pi / (2 h) }$. As a consequence, in the razor-thin limit for which ${ h \!\to\! 0 }$, one has ${ k_{z}^{n_{p}} \!\to\! + \infty }$, except for $k_{z,\mathrm{s}}^{1}$. In equation~\eqref{1/D_discrete}, one should also note the presence of a prefactor in ${1/h}$. In the limit ${ h \!\to\! 0 }$, one therefore has to study the asymptotic behaviour of a term of the form
\begin{equation}
\frac{1}{h} \frac{1}{k_{r}^{2} \!+\! (k_{z}^{n_{p}})^{2}}   \underset{\rm thin}{\longrightarrow}
\begin{cases}
\begin{aligned}
\displaystyle & \frac{1}{k_{r}} & \text{if} & \;\;\;k_{z}^{n_{p}} \!=\! k_{z,\mathrm{s}}^{1} \, , 
\\
\displaystyle & 0 & \text{if} & \;\;\; k_{z}^{n_{p}} \!\neq\! k_{z,\mathrm{s}}^{1} \, .
\end{aligned}
\end{cases}
\label{asymptotic_h}
\end{equation}
As a consequence, in the razor-thin limit, since all the other terms appearing in equation~\eqref{1/D_discrete} are bounded, one has
\begin{equation}
\lim\limits_{\rm thin} \frac{1}{\mathcal{D}_{\bm{m}_{1},\bm{m}_{1}}^{\rm anti}} = 0 \, .
\label{limit_1/D_anti}
\end{equation}
Similarly, in the razor-thin limit, for the symmetric susceptibility coefficients from equation~\eqref{1/D_discrete}, the sum on $n_{p}$ can be limited to the only fundamental term ${ n_{p} \!=\! 1 }$. We recall the asymptotic behaviour in $0$ of the Bessel functions ${ \mathcal{J}_{n} (x) \!\sim\! (1/n!) (x/2)^{n} }$. As a consequence, since ${ k_{z,\mathrm{s}}^{1} \!\to\! 0 }$ in the razor-thin limit, as soon as ${ m_{1}^{z} \!\neq\! 0 }$, one has ${ \lim_{\rm thin} 1/\mathcal{D}_{\bm{m}_{1},\bm{m}_{1}}^{\rm sym} \!=\! 0 }$. Therefore, in the razor-thin limit, only diffusion associated with ${ m_{1}^{z} \!=\! 0 }$ will not vanish, and we may restrict ourselves to only considering this term. We also note that in order to have non-vanishing susceptiblity coefficients, one should restrict oneself to the case ${ J_{z}^{1} \!=\! 0 }$. In the razor-thin limit, one has ${\lim_{\rm thin}} \lambda_{p} \!=\! \lambda_{p}^{\rm thin}$, and thanks to equation~\eqref{alpha_even}, one has ${ \lim_{\rm thin} \alpha_{1} \!=\! 1 }$. As a consequence, in the razor-thin limit with ${ m_{1}^{z} \!=\! 0 }$ and ${ J_{z}^{1} \!=\! 0 }$, the symmetric susceptibility coefficients from equation~\eqref{1/D_discrete} may be approximated as
\begin{equation}
\lim\limits_{\rm thin} \frac{1}{\mathcal{D}_{\bm{m}_{1} , \bm{m}_{1}}^{\rm sym}} \sim \frac{1}{\mathcal{D}_{\bm{m}_{1} , \bm{m}_{1}}^{\rm thin}} \, \mathcal{J}_{0} \bigg[\! \sqrt{\tfrac{2 J_{z}^{2}}{\nu_{1}}} k_{z,\mathrm{s}}^{1} \!\bigg] \, ,
\label{limit_1/D_sym}
\end{equation}
where we introduced the razor-thin WKB susceptibility coefficients obtained in FPC15 as
\begin{align}
\frac{1}{\mathcal{D}_{\bm{m}_{1} , \bm{m}_{1}}^{\rm thin} (J_{\phi}^{1} , J_{r}^{1} , J_{\phi}^{1} , J_{r}^{2} , \omega)} = & \, \frac{G}{2\pi R_{1}} \!\! \int \!\! \rd k_{r} \, \frac{1}{1 \!-\! \lambda_{k_{r}}^{\rm thin} (R_{1} , \omega)} \nonumber
\\
& \!\!\!\!\!\!\!\! \!\!\!\!\!\!\!\! \!\!\!\!\!\!\!\! \!\!\! \times \, \mathcal{J}_{m_{1}^{r}} \bigg[\! \sqrt{\tfrac{2 J_{r}^{1}}{\kappa_{1}}} \, k_{r} \!\bigg] \, \mathcal{J}_{m_{1}^{r}} \bigg[\! \sqrt{\tfrac{2 J_{r}^{2}}{\kappa_{1}}} \, k_{r} \!\bigg] \, .
\label{razor_thin_1/D}
\end{align}
We start from equation~\eqref{final_drift} and estimate the drift coefficients in the razor-thin limit. We rewrite the thick system's DF from equation~\eqref{definition_DF_quasi_isothermal} as
\begin{equation}
F_{\rm thick} (J_{\phi}^{1} , J_{r}^{1} , J_{z}^{1}) \!=\! F_{\rm thin} (J_{\phi}^{1} , J_{r}^{1}) \frac{\nu_{1}}{2 \pi \sigma_{z}^{2}} \exp \!\bigg[\! - \frac{\nu_{1} J_{z}}{\sigma_{z}^{2}} \!\bigg] \, ,
\label{DF_thick_thin}
\end{equation}
where we introduced the razor-thin DF $F_{\rm thin}$ as
\begin{equation}
F_{\rm thin} (J_{\phi}^{1} , J_{r}^{1}) = \frac{\Omega_{\phi} \Sigma}{\pi \kappa_{1} \sigma_{r}^{2}} \, \exp \!\bigg[\! -\! \frac{\kappa_{1} J_{r}^{1}}{\sigma_{r}^{2}} \!\bigg] \, .
\label{definition_Fthin}
\end{equation}
To illustrate this straightforward calculation, we only consider the remaining dependences w.r.t. $J_{z}^{2}$ in equation~\eqref{final_drift}. One has to consider an expression of the form
\begin{align}
\frac{\nu_{1}}{2 \pi \sigma_{z}^{2}} \!\! \int \!\! \rd J_{z}^{2} \, \exp \bigg[\! - \frac{\nu_{1} J_{z}^{2}}{\sigma_{z}^{2}} \!\bigg] \, & \,  \mathcal{J}_{0}^{2} \bigg[\! \sqrt{\tfrac{2 J_{z}^{2}}{\nu_{1}}} \, k_{z , \mathrm{s}}^{1} \!\bigg] = \nonumber
\\
& \,  \frac{1}{2 \pi} \mathcal{I}_{0} \bigg[\! \frac{(k_{z,\mathrm{s}}^{1})^{2}}{\nu_{1}^{2}/\sigma_{z}^{2}} \!\bigg] \, \exp \!\bigg[\! -\! \frac{(k_{z,\mathrm{s}}^{1})^{2}}{\nu_{1}^{2} / \sigma_{z}^{2}} \!\bigg] \nonumber
\\
& \underset{\rm thin}{\longrightarrow} \frac{1}{2 \pi} \, ,
\label{shape_limit_drift}
\end{align}
where we used the formula ${6.615}$ from~\cite{Gradshteyn2007}, and also used equations~\eqref{kz1_even} and~\eqref{link_sigmaz_z0_nu}, so as to have in the razor-thin limit ${ (k_{z,\mathrm{s}}^{1} )^{2} / (\nu_{1}^{2} / \sigma_{z}^{2}) \!\sim\! h \!\to\! 0 }$. As a consequence, injecting equation~\eqref{shape_limit_drift} into the general expression~\eqref{final_drift} of the drift coefficients, one finally obtains
\begin{align}
\lim\limits_{\rm thin} A_{\bm{m}_{1}}^{\rm sym} (\bm{J}_{1}) = & \, - \frac{4 \pi ^{3} \mu}{(\bm{m}_{1} \!\cdot\! \bm{\Omega}_{1})'} \nonumber
\\
& \times \, \int \!\! \rd J_{r}^{2} \, \frac{\bm{m}_{1} \!\cdot\! \partial F_{\rm thin} / \partial \bm{J} (J_{\phi}^{1} , J_{r}^{2})}{|\mathcal{D}_{\bm{m}_{1} , \bm{m}_{1}}^{\rm thin} (J_{\phi}^{1} , J_{r}^{1} , J_{\phi}^{1} , J_{r}^{2} , \bm{m}_{1} \!\cdot\! \bm{\Omega}_{1})|^{2}} \, ,
\label{limit_drift}
\end{align}
where one has to restrict oneself to ${ m_{1}^{z} \!=\! 0 }$ and ${ J_{z}^{1} \!=\! 0 }$. Following the same approach, the razor-thin limit of the diffusion coefficients from equation~\eqref{final_diff} is straightforward to compute and reads
\begin{align}
\lim\limits_{\rm thin} D_{\bm{m}_{1}}^{\rm sym} (\bm{J}_{1}) = & \, \frac{4 \pi^{3} \, \mu}{(\bm{m}_{1} \!\cdot\! \bm{\Omega}_{1})'}  \nonumber
\\
& \times \, \int \!\! \rd J_{r}^{2} \, \frac{F_{\rm thin} (J_{\phi}^{1} , J_{r}^{2})}{| \mathcal{D}_{\bm{m}_{1} , \bm{m}_{1}}^{\rm thin} (J_{\phi}^{1} , J_{r}^{1} , J_{\phi}^{1} , J_{r}^{2} , \bm{m}_{1} \!\cdot\! \bm{\Omega}_{1}) |^{2}} \, .
\label{limit_diff}
\end{align}
The two razor-thin expressions from equations~\eqref{limit_drift} and~\eqref{limit_diff} are in full agreement with the expressions obtained in FPC15 where the razor-thin WKB limit of the inhomogeneous Balescu-Lenard equation was first presented.

\section{Perturbation autocorrelation}
\label{sec:appendixautocorrelation}

\balance

This Appendix shows how the hypothesis of quasi-stationarity from equation~\eqref{assumption_autocorrelation} leads to a diagonalisation of the autocorrelation w.r.t. $k_{r}$ and $k_{z}$ as expressed in equation~\eqref{diagonal_autocorrelation}. To shorten the notations, let us drop the index $m_{\phi}$ in equation~\eqref{diagonal_autocorrelation}, and use the notation ${ \psi \!=\! \psi^{\rm e} }$. Using the definitions of the local radial Fourier transform and the even-restricted vertical Fourier transform from equation~\eqref{definition_FT}, the l.h.s. of equation~\eqref{diagonal_autocorrelation} may be written as
\begin{align}
\!\!\!\!\!\! \bigg<\! \widehat{\psi}_{k_{r}^{1} , k_{z}^{1}} & \, [ R_{\rg} , \omega_{1}]  \, \widehat{\psi}^{*}_{k_{r}^{2} , k_{z}^{2}} [R_{\rg} , \omega_{2} ] \!\bigg> \!=\! \frac{1}{(2 \pi)^{2}} \!\!\! \int \!\!\! \mathrm{d} t_{1} \mathrm{d} t_{2} \mathrm{d} R_{1} \mathrm{d} R_{2} \mathrm{d} z_{1} \mathrm{d} z_{2} \nonumber
\\
 & \, \!\!\!\!\!\! \times g_{r} [R_{\rg} \!-\! R_{1}] \, g_{r} [R_{\rg} \!-\! R_{2}] \, \re^{-\ri (R_{1} - R_{\rg}) k_{r}^{1}} \, \re^{\ri (R_{2} - R_{\rg}) k_{r}^{2}}  \nonumber
\\
& \, \!\!\!\!\!\! \times \cos (k_{z}^{1} z_{1}) \cos (k_{z}^{2} z_{2}) \, \big<\! \psi [R_{1} , z_{1} , t_{1}] \, \psi^{*} [R_{2} , z_{2} , t_{z}] \!\big> \, ,
\label{calcul_diag_auto}
\end{align}
where we defined ${ g_{r} [R] \!=\! \re^{-R^{2}/(2 \sigma^{2})} }$ and the integrations on $z_{1}$ and $z_{2}$ have to be performed on ${ [-h ; h] }$. As in FPP15, one can perform the integrations on $t_{1}$, $t_{2}$, $R_{1}$ and $R_{2}$ to write
\begin{align}
& \, \bigg<\! \widehat{\psi}_{k_{r}^{1} , k_{z}^{1}} [ R_{\rg} , \omega_{1}] \, \widehat{\psi}^{*}_{k_{r}^{2} , k_{z}^{2}} [R_{\rg} , \omega_{2} ] \!\bigg> \!=\! 2 \pi \delta_{\rm D} (\omega_{1} \!-\! \omega_{2}) \delta_{\rm D} (k_{r}^{1} \!-\! k_{r}^{2}) \nonumber
\\
& \, \!\!\!\! \times \!\!\! \int \!\!\! \mathrm{d} z_{1} \mathrm{d} z_{2} \cos (k_{z}^{1} z_{1} \!) \cos(k_{z}^{2} z_{2} \!)  \, \widehat{\mathcal{C}} [\omega_{1} , \!R_{\rg} , \!k_{r}^{1} , \!z_{1} \!\!+\!\! z_{2} , \!z_{1} \!\!-\!\! z_{2} ] \, ,
\label{calcul_diag_auto_II}
\end{align}
where ${ \widehat{\mathcal{C}} [..., R_{\rg} , k_{r}^{1}, ...] }$ stands for the local radial Fourier transform of the function ${ r \!\mapsto\! \widehat{\mathcal{C}} [ ... , R_{\rg} , r , ... ] }$ in the neighbourhood of ${ r \!=\! 0 }$ at the frequency $k_{r}^{1}$, on a scale ${ \sigma' \!=\! \sqrt{2} \sigma }$ as defined in equation~\eqref{definition_FT} (see FPP15). In equation~\eqref{calcul_diag_auto_II} to compute the remaining integrals on $z_{1}$ and $z_{2}$, one performs the change of variables ${ u \!=\! z_{1} \!+\! z_{2} }$ and ${ v \!=\! z_{1} \!-\! z_{2} }$. Keeping only the remaining dependences on $u$ and $v$ and writing ${ k_{1/2} \!=\! k_{z}^{1/2} }$, the second line of equation~\eqref{calcul_diag_auto_II} reads
\begin{equation}
\frac{1}{2} \!\! \int_{- 2h}^{2h} \!\!\!\!\!\!\!\! \mathrm{d} v \! \int_{-2h + |v|}^{2h - |v|} \!\!\!\!\!\!\!\!\!\!\! \mathrm{d} u \, \cos \!\bigg[\! \frac{u \!+\! v}{2} k_{1} \!\bigg] \cos \!\bigg[\! \frac{u \!-\! v}{2} k_{2}  \!\bigg] \, \widehat{\mathcal{C}} [u , v] \, .
\label{calcul_diag_auto_III}
\end{equation}
Let us now assume that on the scale $h$ on which the external perturbations are considered, the function ${ u \!\mapsto\! \widehat{\mathcal{C}} [u , v] }$ slowly depends on $u$, so that we may perform the replacement ${ \widehat{\mathcal{C}} [u , v] \!\to\! \widehat{\mathcal{C}} [ 0 ,v ] }$. As a consequence, in equation~\eqref{calcul_diag_auto_III}, the integral on $u$ may be computed. It reads
\begin{align}
& \int_{-2 h + |v|}^{2 h - |v|} \!\!\!\!\!\!\!\!\!\!\!\! \mathrm{d} u \, \cos \!\bigg[\! \frac{u \!+\! v}{2} k_{1} \!\bigg]  \cos \!\bigg[\! \frac{u \!-\! v}{2} k_{2} \!\bigg] = \, \frac{2}{(k_{1} \!-\! k_{2}) (k_{1} \!+\! k_{2})}  \nonumber
\\
& \times \, \bigg\{ k_{1} \cos [ k_{2} (h \!-\! |v|)]  \sin [h k_{1}] \!+\! k_{1} \cos [h k_{2}] \sin [k_{1} (h \!-\! |v|) ] \nonumber 
\\
& \!-\! k_{2} \cos [ k_{1} ( h \!-\! |v|)] \sin [h k_{2}]  \!-\! k_{2} \cos [h k_{1}] \sin [k_{2} (h \!-\! |v|)] \bigg\} \nonumber
\\
& \, =  G_{\rm sym} [k_{1} , k_{2} , v] \, .
\label{definition_Gs_diag_auto}
\end{align}
The next step of the calculation is to approximate the function $G_{\rm sym}$, so as to diagonalise it w.r.t. to $k_{1}$ and $k_{2}$. To a given pair ${ (k_{1} , k_{2}) }$, let us associate the coordinates ${ ( \tilde{k} , \tilde{\delta k} )}$ defined as
\begin{equation}
\tilde{k} = \frac{k_{1} \!+\! k_{2}}{2} \;\;\; ; \;\;\; \tilde{\delta k} = \frac{k_{1} \!-\! k_{2}}{2} \, .
\label{definition_tilde}
\end{equation}
Let us then assume that $G_{\rm sym}$ follows the ansatz
\begin{equation}
G_{\rm sym} (\tilde{k} \!+\! \tilde{\delta k} , \tilde{k} \!-\! \tilde{\delta k} , v) = H_{\rm sym} (\tilde{k} , v) \, \delta_{\rm D} (\tilde{\delta k}) \, . 
\label{ansatz_Gs}
\end{equation}
The constraint which has to be satisfied by $H_{\rm sym}$ is then given by
\begin{align}
H_{\rm sym} (\tilde{k} , v) & \, = \!\! \int_{- \infty}^{+ \infty} \!\!\!\!\!\!\!\!\! \mathrm{d} \tilde{\delta k} \, H_{\rm sym} (\tilde{k} , v) \, \delta_{\rm D} (\tilde{\delta k}) \nonumber
\\
& \, = \!\! \int_{- \infty}^{+ \infty} \!\!\!\!\!\!\!\!\! \mathrm{d} \tilde{\delta k} \, G_{\rm sym} (\tilde{k} \!+\! \tilde{\delta k} , \tilde{k} \!-\! \tilde{\delta k} , v) \nonumber
\\
& \, = \pi  \cos (\tilde{k} v) \, .
\label{constraint_Hs}
\end{align}
We may therefore use the approximation
\begin{equation}
G_{\rm sym} (k_{1} , k_{2} , v) = 2 \pi \, \delta_{\rm D} (k_{1} \!-\! k_{2}) \, \cos \!\bigg[\! \frac{k_{1} \!+\! k_{2}}{2} v \!\bigg] \, ,
\label{approximation_Gs}
\end{equation}
where the factor $2$ comes from the property ${ 2 \delta_{\rm D} \!(\tilde{\delta k}) \!=\! \delta_{\rm D} \!(\tilde{\delta k}/2) }$.
Equation~\eqref{calcul_diag_auto_III} then leads to
\begin{align}
\eqref{calcul_diag_auto_III} & \, = \pi \delta_{\rm D} (k_{z}^{1} \!-\! k_{z}^{1}) \!\! \int_{-2h}^{2h} \!\!\!\!\!\! \mathrm{d} v \, \widehat{\mathcal{C}} [u \!=\! 0 , v] \, \cos [k_{z}^{1} v] \nonumber
\\
& \, = \pi \delta_{\rm D} (k_{z}^{1} \!-\! k_{z}^{2}) \, \widehat{\mathcal{C}} [k_{z}^{1}] \, , 
\label{calcul_diag_auto_IV}
\end{align}
where the first line of equation~\eqref{calcul_diag_auto_IV} could be seen as a local even-restricted vertical Fourier transform of the function ${ v \!\mapsto\! \widehat{\mathcal{C}} [0,v] }$ on the interval ${ [-2 h ; 2 h] }$ as defined in equation~\eqref{definition_FT}, and we wrote ${ \widehat{\mathcal{C}} [k_{z}^{1}] \!=\!\widehat{\mathcal{C}} [ u\!=\! 0 , k_{z}^{1}] }$ for simplicity. As a conclusion, injecting this result into equation~\eqref{calcul_diag_auto_II} yields
\begin{align}
\!\!\!\!\! \bigg<\! \widehat{\psi}_{k_{r}^{1} , k_{z}^{1}} & [ R_{\rg} , \omega_{1}] \, \widehat{\psi}^{*}_{k_{r}^{2} , k_{z}^{2}} [R_{\rg} , \omega_{2} ] \!\bigg> \!=\! 2 \pi^{2} \delta_{\rm D} (\omega_{1} \!-\! \omega_{2}) \nonumber
\\
& \,  \!\!\!\!\! \times \delta_{\rm D} (k_{r}^{1} \!-\! k_{r}^{2}) \, \delta_{\rm D} (k_{z}^{1} \!-\! k_{z}^{2}) \, \widehat{\mathcal{C}} [m_{\phi} , \omega_{1} , R_{\rg} , k_{r}^{1} , k_{z}^{1}] \, ,
\label{final_diag_auto}
\end{align}
so as to recover the diagonalised autocorrelation from equation~\eqref{diagonal_autocorrelation}.

When considering the antisymmetric diffusion coefficients, as underlined in equation~\eqref{Dm_autocorrelation_odd}, the autocorrelation of the external perturbation involves the odd-restricted vertical Fourier transformed potential perturbations defined in equation~\eqref{definition_odd_vertical}. As a consequence, for antisymmetric perturbations, the diagonalisation of the autocorrelation as started in equation~\eqref{calcul_diag_auto} only requires to make the change ${\text{``${\cos}$"} \!\to\! \text{``${\sin}$"}}$. In the antisymmetric case, while the diagonalisations w.r.t. $\omega$ and $k_{r}$ remain the same, equation~\eqref{calcul_diag_auto_III} now requires to evaluate
\begin{equation}
\frac{1}{2} \!\! \int_{- 2h}^{2h} \!\!\!\!\!\!\!\! \mathrm{d} v \! \int_{-2h + |v|}^{2h - |v|} \!\!\!\!\!\!\!\!\!\!\! \mathrm{d} u \, \sin \!\bigg[\! \frac{u \!+\! v}{2} k_{1} \!\bigg] \sin \!\bigg[\! \frac{u \!-\! v}{2} k_{2} \!\bigg] \, \widehat{\mathcal{C}} [u , v] \, .
\label{calcul_diag_auto_III_odd}
\end{equation}
Using the same assumption as in equation~\eqref{calcul_diag_auto_III}, let us assume that the function ${ \widehat{\mathcal{C}} [u,v] }$ slowly depends on $u$, so that equation~\eqref{definition_Gs_diag_auto} becomes
\begin{align}
& \int_{-2 h + |v|}^{2 h - |v|} \!\!\!\!\!\!\!\!\!\!\!\! \mathrm{d} u \, \sin \!\bigg[\! \frac{u \!+\! v}{2} k_{1} \!\bigg]  \sin \!\bigg[\! \frac{u \!-\! v}{2} k_{2} \!\bigg] = \, - \frac{2}{(k_{1} \!-\! k_{2}) (k_{1} \!+\! k_{2})}  \nonumber
\\
& \times \, \bigg\{ k_{1} \cos [ k_{1} (h \!-\! |v|)]  \sin [h k_{2}] \!+\! k_{1} \cos [h k_{1}] \sin [k_{2} (h \!-\! |v|) ] \nonumber 
\\
& \!-\! k_{2} \cos [ k_{2} ( h \!-\! |v|)] \sin [h k_{1}]  \!-\! k_{2} \cos [h k_{2}] \sin [k_{1} (h \!-\! |v|)] \bigg\} \nonumber
\\
& \, =  G_{\rm anti} [k_{1} , k_{2} , v] \, .
\label{definition_Ga_diag_auto}
\end{align}
As in equation~\eqref{ansatz_Gs}, $G_{\rm anti}$ should follow the ansatz
\begin{equation}
G_{\rm anti} (\tilde{k} \!+\! \tilde{\delta k} , \tilde{k} \!-\! \tilde{\delta k} , v) = H_{\rm anti} (\tilde{k} , v) \, \delta_{\rm D} (\tilde{\delta k}) \, .
\label{ansatz_Ga}
\end{equation}
Again following equation~\eqref{constraint_Hs}, $H_{\rm anti}$ can be computed as
\begin{align}
H_{\rm anti} (\tilde{k} , v) & \, = \!\! \int_{- \infty}^{+ \infty} \!\!\!\!\!\!\!\!\! \mathrm{d} \tilde{\delta k} \, G_{\rm anti} (\tilde{k} \!+\! \tilde{\delta k} , \tilde{k} \!-\! \tilde{\delta k} , v) \nonumber
\\
& \, = \pi \cos (\tilde{k} v)
\label{constraint_Ha}
\end{align}
Therefore, in the antisymmetric case, as in equation~\eqref{approximation_Gs}, $G_{\rm anti}$ can be approximated by
\begin{equation}
G_{\rm anti} (k_{1} , k_{2} , v) = 2 \pi \delta_{\rm D} (k_{1} \!-\! k_{2}) \, \cos \!\bigg[\! \frac{k_{1} \!+\! k_{2}}{2} v \!\bigg] \, .
\label{approximation_Ga}
\end{equation}
As a conclusion, for antisymmetric contributions, the diagonalised autocorrelation takes the exact same form as the symmetric one obtained in equation~\eqref{final_diag_auto}. It involves an even-restricted vertical Fourier transform of the perturbation autocorrelation, as defined in equation~\eqref{calcul_diag_auto_IV}. Therefore, from equations~\eqref{approximation_Gs} and~\eqref{approximation_Ga}, whatever the symmetry of the basis elements considered, the diffusion coefficients are always sourced by the even component of the autocorrelation power spectrum.

\label{lastpage}
\end{document}